\documentclass[fleqn,usenatbib]{mnras}

\usepackage{newtxtext,newtxmath}

\usepackage[T1]{fontenc}
\usepackage{ae,aecompl}

\usepackage{graphicx}	
\usepackage{amsmath}	
\usepackage{amssymb}	
\usepackage[utf8]{inputenc}
\usepackage{tabularx}

\newcommand{\bfk}{\mathbf{k}}
\renewcommand{\bvarepsilon}{\mathbf{\varepsilon}}
\renewcommand{\d}{\operatorname{d}}
\newcommand{\fNL}{f_\mathrm{NL}}
\newcommand{\inv}{^{-1}}
\newcommand{\x}{\mathbf{x}}

\title[Eliminating systematic modes from the galaxy power]
{A map-based method for eliminating systematic modes from galaxy clustering power spectra with application to BOSS}

\author[B. Kalus et al.]{B. Kalus,$^{1}$\thanks{E-mail: benedict.kalus@icc.ub.edu}
W. J. Percival,$^{2,3,4}$
D. J. Bacon,$^{4}$
E.-M. Mueller,$^{4}$
L. Samushia,$^{5,6,4}$\newauthor
L. Verde,$^{1,7}$
A. J. Ross,$^{8}$
and J. L. Bernal$^{1,9}$
\\
$^{1}$ICC, University of Barcelona, IEEC-UB, Mart\'i i Franqu\`es, 1, 08028 Barcelona, Spain\\
$^{2}$Department of Physics and Astronomy, University of Waterloo, 200 University Ave W, Waterloo, ON N2L 3G1, Canada\\
$^{3}$Perimeter Institute for Theoretical Physics, 31 Caroline St. North, Waterloo, ON N2L 2Y5, Canada\\
$^{4}$Institute of Cosmology \& Gravitation, University of Portsmouth, Dennis Sciama Building, Portsmouth, PO1 3FX, UK\\
$^{5}$Department of Physics, Kansas State University, 116 Cardwell Hall, Manhattan, KS, 66506, USA\\
$^{6}$National Abastumani Astrophysical Observatory, Ilia State University, 2A Kazbegi Ave., GE-1060 Tbilisi, Georgia\\
$^{7}$ICREA, Pg. Llu\'is Companys 23, 08010 Barcelona, Spain\\
$^{8}$Center for Cosmology and AstroParticle Physics, The Ohio State University, Columbus, OH 43210, USA\\
$^{9}$Dept. de F\'isica Qu\`antica i Astrof\'isica, Universitat de Barcelona, Mart\'i i Franqu\`es 1, 08028 Barcelona, Spain\\
}

\date{Accepted XXX. Received YYY; in original form ZZZ}

\pubyear{2018}

\begin{document}
\label{firstpage}
\pagerange{\pageref{firstpage}--\pageref{lastpage}}
  \maketitle

\begin{abstract}
We develop a practical methodology to remove modes from a galaxy survey power spectrum that are associated with systematic errors. We apply this to the BOSS CMASS sample, to see if it removes the excess power previously observed beyond the best-fit $\Lambda$CDM model on very large scales. We consider several possible sources of data contamination, and check whether they affect the number of targets that can be observed and the power spectrum measurements. We describe a general framework for how such knowledge can be transformed into template fields. Mode subtraction can then be used to remove these systematic contaminants at least as well as applying corrective weighting to the observed galaxies, but benefits from giving an unbiased power. Even after applying templates for all known systematics, we find a large-scale power excess, but this is reduced compared with that observed using the weights provided by the BOSS team. This excess is at much larger scales than the BAO scale and does not affect the main results of BOSS. However, it will be important for the measurement of a scale-dependent bias due to primordial non-Gaussianity. The excess is beyond that allowed by any simple model of non-Gaussianity matching Planck data, and is not matched in other surveys. We show that this power excess can further be reduced but is still present using "phenomenological" templates, designed to consider further potentially unknown sources of systematic contamination. As all discrepant angular modes can be removed using "phenomenological" templates, the potentially remaining contaminant acts radially.

\end{abstract}




\section{Introduction}

Galaxy surveys allow measurements that are crucial
for our understanding of the Universe. For instance, 
Baryon Acoustic
Oscillation (BAO) observations provide a standard ruler measurement
that we can use to study our Universe's expansion 
history, while Redshift Space Distortion (RSD) measurements
test the theory of gravity that governs structure growth.
Furthermore, a full measurement of the shape of the galaxy power spectrum provides additional information about the
total matter density $\Omega_\mathrm{m}h^2$, the baryon density $\Omega_\mathrm{b}h^2$, the
neutrino mass density $\Omega_\mathrm{\nu}h^2$, and the local primordial non-Gaussianity parameter
$f_\mathrm{NL}$ \citep*[e.g.][]{Slosar:2008hx,Ross:2012sxfNL,Leistedt:2014zqa}.

While the field of density fluctuations revealed by galaxy surveys contains a lot of cosmological signal, this is contaminated by various foreground and instrumental sources.
A simple model of such contaminations is that the true density field in Fourier space
\begin{equation}
	D(\bfk)=F(\bfk)-\varepsilon f(\bfk),
	\label{eq:Fassum}
\end{equation}
is a linear combination of the measured density field $F(\bfk)$ and the contaminant field that
can be written in terms of a template $f(\bfk)$
with unknown amplitude $\varepsilon$. In practice, we do not know the exact 
shape of $f(\bfk)$ and the data is affected by more than one contaminant. This can be 
accounted for by extending Eq. \eqref{eq:Fassum} to include a range of different templates
$f_A(\bfk)$ with unknown amplitudes $\varepsilon_A$:
\begin{equation}
	D(\bfk)=F(\bfk)-\sum_A\varepsilon_A
    f_A(\bfk).
    \label{eq:FassumMulti}
\end{equation}
Mode deprojection 
\citep{Rybicki:1992jz} offers an elegant way of mitigating contaminants that can be modelled as
in Eq. \eqref{eq:FassumMulti} by analytically marginalising over all $\varepsilon_A^\mathrm{(true)}$.
This approach can directly be implemented in any covariance based estimator for the power 
spectrum, a natural choice being the Quadratic Maximum 
Likelihood \citep[QML]{Tegmark:1996qtQML} Estimator whose estimates are unbiased and optimal for any covariance matrix.
This technique has been applied by 
\citet*{Slosar:2004fr} in angular power
spectrum measurements from the Wilkinson Microwave 
Anisotropy Probe (WMAP) at low
multipoles $\ell$, in investigating the Integrated Sachs-Wolfe (ISW) effect as a function of
redshift \citep*{Ho:2008bz}, and by \citet{Pullen:2012rd} to the SDSS quasar sample. 
Furthermore, \citet{Leistedt:2013gfa} 
identified, using mode deprojection, a previously found
large-scale excess in the angular auto- and cross
power spectra of the catalogue of photometric
quasars from the Sixth Data Release (DR6) of SDSS
as being due to systematics.

The computational cost of mode deprojection can be 
reduced by identifying the most important templates 
before marginalising 
over them as described by 
\citet{Leistedt:2014wia}. A complementary strategy to improve the
computational efficiency of mode deprojection is to incorporate it into fast, but sub-optimal
power spectrum estimators, such as the 
\citet*[FKP]{Feldman:1993ky} Estimator for the 3D power
spectrum, as described in \citet{Kalus:2016cno}. This work was subsequently extended to the Pseudo-$C_\ell$ \citep{Hivon:2001jp}
Estimator for the angular power spectrum, by \citet*{Elsner:2016bvs}.

In this article, we consider how systematic templates can
be produced for common systematic issues, using the Northern Galactic Cap (NGC)
of the SDSS-III BOSS CMASS sample as our example data set. We derive a set of templates for this 
sample, and use the Mode Subtraction technique of \citet{Kalus:2016cno} to remove these modes 
when making a power spectrum measurements of these data. The MOde 
Subtraction code to Eliminate Systematic contamination in
galaxy clustering power spectrum measurements (MOSES)
is available at 
\url{https://github.com/KalusB/Moses}.

The outline of this paper is as follows: In Sec.
\ref{sec:modesub}, we review the Mode Subtraction 
technique and develop a method
to generate templates for a
given contaminant and survey. We use this
method to generate templates for known contaminants driven by: foreground stars
(Sec. \ref{sec:stellardensitycounts}), seeing (Sec.
\ref{sec:seeing}), airmass variations (Sec. 
\ref{sec:airmass}), galactic extinction (Sec. 
\ref{sec:extinction}) and the SDSS scanning strategy 
(Sec. \ref{sec:stripes}). We employ these templates in
power spectrum measurements from the SDSS-III BOSS CMASS
NGC sample in Sec. \ref{sec:application}.

\section{Methods for removing contaminants}
\label{sec:methods}

We start by reviewing the key results of 
\citet{Kalus:2016cno}, which discussed two related methods of
mitigating systematic templates: mode deprojection and
mode subtraction. The two were shown to be mathematically equivalent if we allow the templates to be matched, modulo a final normalisation step, although this can be added in making the techniques identical. In concept however, they are quite different as explained below.

\subsection{Mode Deprojection}
\label{sec:modedeproj}

Mode deprojection \citep{Rybicki:1992jz}
works by updating the
mode-by-mode covariance matrix $\mathbf{C}_{\alpha\beta}$
in a covariance based estimator as
\begin{equation}
	\tilde{\mathbf{C}}_{\alpha\beta}\equiv 
    \mathbf{C}_{\alpha\beta}
    +\lim_{\sigma\rightarrow\infty}\sigma f(\bfk_\alpha)
    f^\ast(\bfk_\beta).
    \label{eq:modedeprojection}
\end{equation}
Thus contaminated modes are excluded from the analysis at the likelihood stage, and consequently the result is an unbiased estimate of the underlying power spectrum.

Following \citep{Tegmark:1996qtQML}'s QML approach, the
power spectrum is estimated as 
\begin{equation}
	\widehat{P}(k_i)=\sum_{j}
    \mathbf{N}_{ij}^{-1}\mathbf{p}_j,
\end{equation}
where 
\begin{equation}
	\mathbf{p}_j=-\sum_{\alpha\beta}
    F^\ast(\bfk_\alpha)
    \frac{\partial\mathbf{C}^{-1}_{\alpha\beta}}
    {\partial P(k_j)}F(\bfk_\beta)
    \label{eq:pj}
\end{equation} 
is a covariance 
weighted two-point function of the contaminated density
field and 
\begin{equation}
	\mathbf{N}_{ij}=\operatorname{tr}
    \left\lbrace\mathbf{C}^{-1}
    \frac{\partial\mathbf{C}}
    {\partial P(k_i)}\mathbf{C}^{-1}
    \frac{\partial\mathbf{C}}
    {\partial P(k_j)}\right\rbrace
\end{equation}
normalises and optimises the estimator. After updating 
the covariance as prescribed by Eq. 
\eqref{eq:modedeprojection} and, for clarity,
assuming that all modes
are uncorrelated, Eq. \eqref{eq:pj} reads 
\citep{Kalus:2016cno}
\begin{equation}
	\mathbf{p}_j=\sum_{\bfk_\alpha\text{ in bin $j$}}
    \frac{\vert F(\bfk_\alpha)-
    \frac{S}{R}f(\bfk_\alpha)\vert^2}{P^2(k_j)}
    \label{eq:pjupdate}
\end{equation}
with
\begin{equation}
	R=\sum_\alpha\frac{\vert f(\bfk_\alpha)\vert^2}
    {P(k_\alpha)}\text{ and }S=\sum_\alpha
    \frac{F^\ast(\bfk_\alpha) f(\bfk_\alpha)}
    {P(k_\alpha)},
    \label{eq:RSsingledef}
\end{equation}
as well as a prior model power spectrum $P(k)$. The
numerator in Eq. \eqref{eq:pjupdate} has the same form
as Eq. \eqref{eq:Fassum} and we shall indeed see in the
next section that $\varepsilon^\mathrm{(BF)}=\frac{S}{R}$
can be interpreted as a best-fitting estimate of the
contamination amplitude.

\subsection{Mode Subtraction}
\label{sec:modesub}

Mode subtraction instead works by finding the best-fit amplitude of the contaminant, assuming
that $D(\bfk)$ is Gaussian distributed, and removing that directly from the data. Formally, for
a single contaminant, we can find the best-fit amplitude
\begin{equation}
	\varepsilon^\mathrm{(BF)}=\frac{S}{R},
\end{equation}
thus, we find the same result as with mode deprojection in the QML approach prior to 
normalisation and optimisation. Extending this result to several contaminants, we find the
best-fit amplitude
for a vector of contaminants as required in Eq.~(\ref{eq:FassumMulti}) as
\begin{equation}
	\bvarepsilon^\mathrm{(BF)}=\mathbf{R}\inv\mathbf{S},
    \label{eq:epsbfmulticont}
\end{equation}
where 
\begin{equation}		
	\mathbf{R}_{AB}=
  \sum_{\mu}\frac{f_A^\ast(\bfk_\mu)
  f_B(\bfk_\mu)}{P(k_\mu)}\text{ and }\mathbf{S}_A=\sum_\alpha
    \frac{F^\ast(\bfk_\alpha) f_A(\bfk_\alpha)}
    {P(k_\alpha)}
    \label{eq:Rdef}
\end{equation}
are generalisations of Eq. \eqref{eq:RSsingledef}.

Na\"ively inserting Eq. \eqref{eq:epsbfmulticont} into Eq. \eqref{eq:FassumMulti} and then
applying the FKP Estimator provides a biased estimate of the power spectrum because 
$\bvarepsilon^\mathrm{(BF)}$ is correlated with the data \citep*{Elsner:2015aga}. However, we
can debias the contribution of each mode by modifying the estimator as \citep{Kalus:2016cno}
\begin{equation}
	\widehat{P}(k_i)=\frac{1}{N_i}\sum_{\bfk_\alpha\text{ in bin $i$}}
    \frac{\vert F(\bfk_\alpha)-\sum_{AB}
    \mathbf{R}\inv_{AB}\mathbf{S}_B f_A(\bfk_\alpha)\vert^2}{1-\sum_{AB}
    \frac{f_A(\bfk_\alpha)\mathbf{R}\inv_{AB}f^\ast_B(\bfk_\alpha)}{P(k_i)}},
    \label{eq:unbiasedFKPmodesub}
\end{equation}
where $N_i$ is the number of modes in bin $i$. Throughout this article, we assume an 
isotropic power spectrum as there is no evidence
for anisotropy at large scales \citep{Pullen:2010zy}.

Mode subtraction is commonly performed by introducing 
additional weights $w$, which can either be
assigned to individual galaxies (to account for effects
such as redshift failures, fibre collisions, etc.) or which
directly or indirectly depend on positions on the sky
(such as for seeing, airmass, stellar density, etc.).
For example, one template might be the inverse of the change in galaxy density as a function of the seeing: by weighting galaxies by these numbers, we are subtracting this template from the field. This will become more clear in the following subsection.

\subsection{A practical approach to decontamination: single contaminant}
\label{sec:practicalapprsingle}

We base our analysis on the Fourier-based framework of 
\citet{Feldman:1993ky}, which we adjust
to include the removal of systematics. We start by defining the 
contaminated field, where the contaminants are multiplicative and are accounted for
by the systematic weight $w(\x)$:
\begin{equation}
	D(\x)=w_\mathrm{FKP}(\x)\frac{w(\x)n_\mathrm{g}(\x)-\alpha n_\mathrm{r}(\x)}{\sqrt{I_2}},
    \label{eq:Ddefwithweights}
\end{equation}
where $n_\mathrm{g}(\x)$ is the galaxy density, and $\alpha n_\mathrm{r}(\x)$ the expected density derived from a random catalogue. Ignoring geometrical effects, the power in the field $D(\x)$ gives an unbiased estimate of the power spectrum after subtracting off a shot-noise term, if the normalisation is 
\begin{equation}
	I_2=\int\d^3\x\;\bar{n}^2(\x)w_\mathrm{FKP}^2(\x).
\end{equation}
The optimal weights are
\begin{equation}
	w_\mathrm{FKP}(\x)=\frac{1}{1+\bar{n}(\x)P_\mathrm{FKP}},
\end{equation}
where $P_\mathrm{FKP}$ is 
a typical value of the power spectrum,
that maximises the signal-to-noise of the power spectrum 
estimate at the desired scales.

The aim of this section is to translate Eq.  
\eqref{eq:Ddefwithweights} into the
debiased mode subtraction framework, i.e. writing $D$ in terms of
$F$ and $f$, and thence, identifying how $F$ and $f$ are related to
the weight $w(\x)$ and the observed galaxy and random counts, $n_\mathrm{g}$ and
$n_\mathrm{r}$, respectively. Eq. 
\eqref{eq:Ddefwithweights} reflects the fact
that most known contaminants affect the observed galaxy density
multiplicatively. To give an example, a bright star obscures a 
fraction of the targets in its angular vicinity, i.e. the number of
targets that are not observed depends on the number of targets that 
actually exist. In spite of that, the underlying assumption behind 
mode subtraction is given by Eq. \eqref{eq:Fassum}, i.e. that a
template of the contaminant $f$ can be 
subtracted from the observation $F$ to
obtain a ``clean'' density field $D$. 
\citet{Jasche:2013lwa} lift this apparent 
contradiction by adopting a different data model 
which is implemented in the Algorithm for  
Reconstruction and Sampling 
\citep[ARES]{Jasche:2009hx,Jasche:2017rze}. Here, we
introduce a framework that allows us to directly 
transform the corrective weights into contaminant
templates that can be applied within the simpler 
model of Eq. \eqref{eq:Fassum} and, hence, the 
methods discussed in \citet{Kalus:2016cno}. To do 
so, we move the weights from acting on the observed 
galaxy density to letting their inverse act on the 
random catalogue. Mathematically speaking, we divide 
both the numerator and denominator 
of the fraction in Eq. \eqref{eq:Ddefwithweights}
by the weights and obtain
\begin{equation}
	D(\x)=w_\mathrm{FKP}^\prime(\x)\frac{n_\mathrm{g}(\x)-\alpha w^{-1}(\x)n_\mathrm{r}(\x)}{\sqrt{I_2}},
\end{equation}
where $w_\mathrm{FKP}^\prime(\x)\equiv w_\mathrm{FKP}(\x)w(\x)$ is an
updated FKP weight.
We know the parameter $\alpha$ well as it has been chosen to 
match the random catalogue to the galaxy catalogue. The amplitude of the contaminant is 
unknown and is something we wish to determine, so we split the second term into a part without weights
and into another with weights, such that we can introduce 
a free parameter $\varepsilon\approx\alpha$ that we can
marginalise over:
\begin{equation}
	D(\x)=w_\mathrm{FKP}^\prime(\x)\frac{n_\mathrm{g}(\x)-\alpha^\prime n_\mathrm{r}(\x)-\varepsilon[w^{-1}(\x)-1]n_\mathrm{r}(\x)}{\sqrt{I_2}}.
   \label{eq:Dwitheps}
\end{equation}
In order to normalise Eq. \eqref{eq:Dwitheps} to give zero expected over-density, we need to use
a revised value of $\alpha^\prime$ matching the galaxy and revised random catalogues
\begin{equation}
	\alpha^\prime=\frac{\int\d^3\x\;[n_\mathrm{g}(\x)
    -\varepsilon[w^{-1}(\x)-1]n_\mathrm{r}(\x)]}
    {\int\d^3\x\; n_\mathrm{r}(\x)}.
\end{equation}
Recalling that in the case of not including weights, we have
\begin{equation}
	\alpha_\mathrm{FKP}=\frac{\int\d^3\x\; 
    n_\mathrm{g}(\x)}{\int\d^3\x\; n_\mathrm{r}(\x)},
\end{equation}
we can split $\alpha$ into two terms: one independent of $\varepsilon$, and one 
proportional to $\varepsilon$: 
\begin{equation}
	\alpha^\prime
    =\alpha_\mathrm{FKP}-\varepsilon
    \left[\frac{\int\d^3\x w^{-1}(\x) n_\mathrm{r}(\x)}
    {\int\d^3\x n_\mathrm{r}(\x)}-1\right].
    \label{eq:alpha}
\end{equation}
Then we can write Eq. \eqref{eq:Dwitheps} as 
\begin{align}
	D(\x)=&w_\mathrm{FKP}^\prime(\x)\frac{n_\mathrm{g}(\x)-\alpha_\mathrm{FKP} n_\mathrm{r}(\x)}{\sqrt{I_2}}\nonumber\\&-\varepsilon w_\mathrm{FKP}^\prime(\x)
    \frac{\left[w^{-1}(\x)-\frac{\int\d^3\x\; w^{-1}(\x) 
    n_\mathrm{r}(\x)}
    {\int\d^3\x\; n_\mathrm{r}(\x)}\right]n_\mathrm{r}(\x)}{\sqrt{I_2}}.
    \label{eq:Dmodesub}
\end{align}
In the mode subtraction framework, we
can identify
\begin{equation}
	F(\x)=w_\mathrm{FKP}^\prime(\x)\frac{n_\mathrm{g}(\x)-\alpha_\mathrm{FKP} n_\mathrm{r}(\x)}{\sqrt{I_2}}
    \label{eq:Farbitrary}
\end{equation}
and
\begin{equation}
	f(\x)=\frac{w_\mathrm{FKP}^\prime(\x)n_\mathrm{r}(\x)}{\sqrt{I_2}}\left[w^{-1}(\x)-
    \frac{\int\d^3\x\; 
    w^{-1}(\x) n_\mathrm{r}(\x)}{\int\d^3\x\;  n_\mathrm{r}(\x)}\right].
	\label{eq:farbitrary}
\end{equation}
Thus, the uncorrected field $F$ is similar to the FKP field without
systematic weights, but with a modified FKP weight. 
The template is the expected correction that has to be 
subtracted based on expectation of the galaxy number density from
the random catalogue and the systematic weight.

\subsection{A practical approach to decontamination: multiple contaminants}
\label{sec:practicalapprmulti}

One big advantage of the mode subtraction framework is that it can be 
easily extended to $N_\mathrm{cont}$ different contaminant templates.
Different contaminants can be included in the traditional weighting
scheme by just multiplying $n_\mathrm{g}$ with a weight for each
contaminant one can imagine. Formally, $w(\x)$ has to be 
known exactly for each mode to be subtracted. In practice, if the functional form of the weight is not exactly 
known, the mode subtraction framework allows us to include more than 
one template for each contaminant: we simply need a set of templates that span the region of uncertainty. Having a free parameter for each
template then naturally mitigates the templates that are supported
by the data with the correct amplitude. When dealing with more
than one template, we write the effect of each contaminant $E_A(\x)$ that we define such that
$\langle E_A(\x)\rangle=0$ for all contaminants $A$. We model 
the total weight in terms of
\begin{equation}
	w_\mathrm{FKP}^\prime(\x)=\frac{w_\mathrm{FKP}(\x)}{1+\sum_{A=1}^{N_\mathrm{cont}}E_A(\x)}.
    \label{eq:multcontweights}
\end{equation}
Eq. \eqref{eq:Ddefwithweights} then reads
\begin{equation}
	D(\x)=w_\mathrm{FKP}^\prime(\x)\frac{n_\mathrm{g}(\x)-\alpha\left(1+\sum_{A=1}^{N_\mathrm{cont}}E_A(\x)\right)n_\mathrm{r}(\x)}{\sqrt{I_2}}.
\end{equation}
As stated, we introduce free parameters $\varepsilon_A$ for each 
contaminant to take the uncertainties of each of their amplitudes 
into account. We then have
\begin{equation}
	D(\x)=w_\mathrm{FKP}^\prime(\x)\frac{n_\mathrm{g}(\x)-\alpha^\prime
    n_\mathrm{r}(\x)-\sum_{A=1}^{N_\mathrm{cont}}\varepsilon_A E_A(\x)n_\mathrm{r}(\x)}{\sqrt{I_2}}.
\end{equation}
To ensure again that the expected overdensity field is zero, we need
\begin{equation}
	\alpha^\prime=\alpha_\mathrm{FKP}-\sum_{A=1}^{N_\mathrm{cont}}\varepsilon_A
    \frac{\int\d^3\x\; E_A(\x)n_\mathrm{r}(\x)}
    {\int\d^3\x\; n_\mathrm{r}(\x)}.
\end{equation}
Recollecting all $\varepsilon_A$ terms yields
\begin{align}
	D(\x)=&w_\mathrm{FKP}^\prime(\x)\frac{n_\mathrm{g}(\x)-\alpha_\mathrm{FKP}n_\mathrm{r}(\x)}{\sqrt{I_2}}
    \nonumber\\
    &-\sum_{A=1}^{N_\mathrm{cont}}\varepsilon_A\frac{E_A(\x)-
    \frac{\int\d^3\x\; E_A(\x)n_\mathrm{r}(\x)}
    {\int\d^3\x\; n_\mathrm{r}(\x)}}{\sqrt{I_2}}w_\mathrm{FKP}^\prime(\x)n_\mathrm{r}(\x),
\end{align}
where we can read off 
\begin{equation}
	F(\x)=w_\mathrm{FKP}^\prime(\x)\frac{n_\mathrm{g}(\x)-\alpha_\mathrm{FKP}n_\mathrm{r}(\x)}{\sqrt{I_2}}
\end{equation}
and 
\begin{equation}
	f_A(\x)=w_\mathrm{FKP}^\prime(\x)\frac{E_A(\x)-
    \frac{\int\d^3\x\; E_A(\x)n_\mathrm{r}(\x)}
    {\int\d^3\x\; n_\mathrm{r}(\x)}}{\sqrt{I_2}}n_\mathrm{r}(\x)
    \label{eq:multconttemp}
\end{equation}
in the same way as we did to obtain Eq. \eqref{eq:Farbitrary} and 
\eqref{eq:farbitrary}. The field $F$ 
is again similar to the FKP field. Each $E_A(\x)$ describes how 
contaminant $A$ affects the number of galaxies 
in a certain region around the point $\x$. Although the effect of most
contaminants is expected to be relative to $F$, this section has 
shown how absolute templates $f$ can be constructed using the expected 
 number of galaxies from the random catalogue $n_\mathrm{r}$.
Each template is an estimate of the
absolute number density that has to
be added or subtracted to correct for the contaminant in question. 
The following sections will
show how the $E_A$ are obtained in practice for specific contaminants.

\subsection{Methodology for Templates of Known Sources of
Contamination}
\label{sec:knowntemps}

In order to be able to compare the results using mode 
subtraction to the results using the weights as in 
\cite{Rossinprep}, we generate our templates in a similar 
way as their weights. We start with a map of the 
contaminant $n_\mathrm{c}$ in Hierarchical Equal Area 
isoLatitude Pixelization of a sphere \citep[HEALPix
\footnote{\url{http://healpix.sf.net}}, cf. maps in 
Appendix \ref{sec:maps}]{Gorski:2004by}.
To obtain a template to mitigate against the
contaminant in question, we pixelise the galaxy and 
random catalogues in the same way as the contaminant map.
We assign each cell to a bin according to 
the degree of contamination in the respective cell. For
each bin, we average over the ratio of observed and 
expected galaxy number count $\frac{n_\mathrm{g}}{\langle 
n_\mathrm{g}\rangle}$.
We estimate the expected number density as 
\begin{equation}
	\langle n_\mathrm{g}\rangle=\alpha n_\mathrm{r}.
    \label{eq:ngexp}
\end{equation} 
As random catalogues are usually constructed not to
contain any clustering information, Eq. \eqref{eq:ngexp} 
is a biased estimate of the galaxy number density in a 
cluster or void. However, assuming that the distribution 
of foreground stars and the actual galaxy number density 
are uncorrelated, we
expect the average over all $\frac{n_\mathrm{g}}{\langle 
n_\mathrm{g}\rangle}$ in each contaminant bin to
equal unity. A significant deviation from one is an 
indication of the contaminant affecting 
the observed galaxy number density. Following 
\citet{Rossinprep}'s procedure to obtain
weights for a given galaxy, we use this information and
fit a linear regression line 
\begin{equation}
	\frac{n_\mathrm{g}}{\langle 
    n_\mathrm{g}\rangle}^{(1)}(n_\mathrm{c})\equiv 
    C_0+C_1 n_\mathrm{c}
    \label{eq:linearcontfit}
\end{equation}
through the $\frac{n_\mathrm{g}}{\langle 
n_\mathrm{g}\rangle}$ data points. 
The weight for the $i$th galaxy in 
the survey at right ascension $\alpha_i$ and $\delta_i$ 
is then given by the inverse of the fitting function
evaluated at the contaminant level in the pixel which 
contains the galaxy:
\begin{equation}
	w_i=\frac{1}{\frac{n_\mathrm{g}}{\langle n_\mathrm{g}\rangle}^{(1)}(n_\mathrm{c}(\alpha_i,\delta_i))}.
\end{equation}
To obtain a single template for the contaminant, we 
simply have to insert this weight into Eq. 
\eqref{eq:farbitrary}.

For the case that one does not want to assume a linear 
relation between the contaminant and its effect on the
observed number of galaxies, one can fit higher order 
polynomials
\begin{equation}
	\frac{n_\mathrm{g}}{\langle 
    n_\mathrm{g}\rangle}^{(N)}
    (n_\mathrm{c})\equiv\sum_{A=0}^N 
    C_A n_\mathrm{c}^A
\end{equation}
and build multiple templates that can cover a range of fits to this trend using Eq.
\eqref{eq:multcontweights}-\eqref{eq:multconttemp}. At linear order, the contaminant has a
significant effect if $\frac{n_\mathrm{g}}{\langle n_\mathrm{g}\rangle}$ is significantly 
different from 1, thus we define
\begin{equation}
	E_1(z,\alpha,\delta)\equiv\frac{n_\mathrm{g}}{\langle n_\mathrm{g}\rangle}^{(1)}(n_\mathrm{c}(\alpha,\delta))-1.
\end{equation}
In order to fulfil Eq. \eqref{eq:multcontweights}, we define 
\begin{equation}
	E_A(z,\alpha,\delta)\equiv\frac{n_\mathrm{g}}{\langle n_\mathrm{g}\rangle}^{(A)}(n_\mathrm{c}(\alpha,\delta))-\frac{n_\mathrm{g}}{\langle n_\mathrm{g}\rangle}^{(A-1)}(n_\mathrm{c}(\alpha,\delta))
\end{equation}
for higher orders. In this way, correlations between the
$E_A$ are reduced. Therefore, templates 
that correspond to expansion orders that
are actually not in the data obtain naturally 
negligible best-fitting values
of $\varepsilon^\mathrm{(BF)}$. 

\section{Application to BOSS}
\label{sec:application}

We use data from the Final  
\citep{Alam:2015mbd}
SDSS-III \citep{Eisenstein:2011sa}
BOSS \citep{Dawson:2012va} Data Release
that has been obtained using
the BOSS spectrograph \citep{Smee:2012wd} on the
Sloan Foundation 2.5-meter Telescope \citep{Gunn:2006tw}.

BOSS galaxies were
selected for spectroscopic observation 
from photometric SDSS-I and -II data. 
BOSS observed two spectroscopic galaxy samples, the Low 
Redshift (LOWZ) sample consisting of 361,762 LRGs at
$0.16<z<0.36$, and the Constant Mass (CMASS) sample, 
that includes both LRGs and fainter blue galaxies at 
$0.43<z<0.7$. By combining the two red and blue 
populations into one single sample the
shot-noise in the measured density field is reduced. The
total number of galaxies in CMASS amounts to 777,202, 
of which 568,776 are in the Galactic North and the 
remaining 208,426 galaxies are in the Galactic South 
\citep{Reid:2015gra}. There are also 13,290 ``known''
galaxy spectra from SDSS-II that fulfil the selection 
criteria of CMASS and are therefore also included. The
number of ``known'' spectra for the LOWZ sample is
much larger, with 153,517 ``known'' galaxies, mainly 
SDSS-II LRGs.
The final footprint of BOSS covers 
9329 square degrees and can 
be seen e.g. in Fig. \ref{fig:starsNSIDE256}.
As the CMASS
sample probes a larger volume than the LOWZ sample and
as it is more affected by large-scale systematics 
\citep{Rossinprep}, we have chosen it as the test sample
for mode subtraction.

The colour criterion used in the selection process for 
the CMASS sample is dominated by limits on the 
parameter
\begin{equation}
	d_\perp \equiv r_\mathrm{mod}-i_\mathrm{mod}-\frac{g_\mathrm{mod}-r_\mathrm{mod}}{8},
\end{equation}
where $g_\mathrm{mod}$, $r_\mathrm{mod}$ and $i_\mathrm{mod}$ are the model $g,r$ and $i$-band magnitudes adopting either a de Vaucouleurs or an exponential luminosity profile, depending on which of the two fits better in the $r$-band \citep{Stoughton:2002ae}. 
Other important quantities in the selection process are
the model $i$-band magnitude $i_\mathrm{cmod}$
calculated from the best-fitting linear combination 
of the de Vaucouleurs and exponential luminosity profiles
\citep{Abazajian:2004aja}, and the $i$-band magnitude 
within a $2^{\prime\prime}$
aperture radius $i_\mathrm{fib2}$.
The requirements on CMASS galaxies are then given by
\begin{align}
	17.5<i_\mathrm{cmod}<19.9\nonumber\\
    r_\mathrm{mod}-i_\mathrm{mod}<2\nonumber\\
    d_\perp>0.55\nonumber\\
    i_\mathrm{fib2}<21.5\nonumber\\
    i_\mathrm{cmod}<19.86+1.6(d_\perp-0.8).
    \label{eq:CMASSselection}
\end{align}
Isolated stars can be distinguished from galaxies as
they have profiles closer to that of the point spread 
function
(PSF). After fitting the magnitudes $i_\mathrm{psf}$ and
$z_\mathrm{psf}$ to the point spread function, one can define
further criteria to avoid targeting stars:
\begin{align}
	i_\mathrm{psf}&>4.2+0.98i_\mathrm{mod}\nonumber\\
    z_\mathrm{psf}&>9.125+0.54z_\mathrm{mod}.
    \label{eq:starcut}
\end{align}
Several observational and instrumental effects, such as
the presence of foreground stars, the Earth's atmosphere,
interstellar dust or the surveys scanning strategy,
alter the magnitudes of objects depending on the 
contaminant along the LOS to each objects, potentially
affecting the selection as described in Eq. 
\eqref{eq:CMASSselection}. We therefore generate 
templates according to our recipe in \ref{sec:knowntemps}
and apply them to the 
BOSS CMASS NGC sample in the following Subsections.

\subsection{Stellar Density Counts}
\label{sec:stellardensitycounts}

The presence of foreground stars 
affects galaxy clustering measurement through
obscuration (we cannot observe galaxies behind a foreground star that is brighter than the 
target), selection bias (photometric measurements needed for the target selection of 
spectroscopic survey are more inaccurate close to foreground stars) and confusion (a star is
misclassified as a galaxy).
For current spectroscopic surveys, we expect confusion to be
negligible. Hence, the higher the 
stellar density is, the lower is the number of galaxies
we observe, 
as found by \citet{Ross:2011cz}. It has been 
confirmed by \citet{Rossinprep} that foreground stars 
cause the strongest 
systematic error in BOSS CMASS data. The foreground stars are 
within our own Galaxy, which can be described as a half-sky mode in 
Fourier space. Thus, the foreground stars add large-scale power to 
the actual galaxy power spectrum in a very similar way as a positive 
$\fNL$ signal \citep{Ross:2012sxfNL}. As stellar densities have been 
reported as the main source of systematic error in BOSS CMASS data 
\cite{Ross:2012sxfNL}, it is the first systematic we want to confront using the 
mode subtraction technique. 

As described in Sec. \ref{sec:knowntemps}, we 
start by creating a stellar number count map in HEALPix (cf. Fig. 
\ref{fig:starsNSIDE256}) of the contaminant which, in 
this case, we take from the SDSS DR8 star catalogue. 
Both the BOSS data and the catalogue of 
stars that we use thus come from the same survey, hence 
having similar footprints (there are additional stripes 
running through the Milky Way in the catalogue of stars 
that are masked out in Fig. \ref{fig:starsNSIDE256}) and 
instrumental systematics. The advantage of using HEALPix 
is that the number count in each pixel is proportional to 
the angular density of stars, because all pixels cover 
equally sized areas. Another advantage of HEALPix in 
general is that the resolution of a map can be easily 
changed due to the hierarchical ordering of the cells. 
The resolution can be identified by the ``number of 
pixels per side'' ($N_\mathrm{side}$), which is related 
to the total number of pixels on the sphere
$N_\mathrm{pix}=12 N_\mathrm{side}^2$.  The resolution in 
Fig. \ref{fig:starsNSIDE256} is $N_\mathrm{side}=256$. 
\citet{Rossinprep} use $N_\mathrm{side}=128$ 
to reduce the shot noise in the stellar data. We reduce 
the resolution to $N_\mathrm{side}=64$, which is 
sufficient to cover the angular positions of our FKP grid
with $128^3$ grid points. 

The number count and the number density of stars are only 
proportional to each other in cells that are entirely 
within the survey footprint.
In the original stellar number count map, we could see prominent edge
effects, because HEALPix cells on the edges are only 
partially filled. Considering that
each cell has exactly four adjacent cells,
we reduce this effect by assuming the following for
the completeness of each HEALPix-pixel: pixels in 
the $N_\mathrm{side}=256$-map that
have only non-zero
neighbours are complete, and for every neighbour that is
zero, we assume that the pixel in question is 25 per cent
less complete, such that cells, whose neighbours are all 
empty, are also empty.
We generate a HEALPix map with these 
completeness values and reduce its resolution to 
$N_\mathrm{side}=64$ in the same way as the stellar number count map. 
We divide the number count of each partially filled pixel 
by this resulting completeness map, such that we obtain a 
map whose entries are proportional to the number 
density of stars.

\begin{figure}
 \centering
\includegraphics[width=\columnwidth]{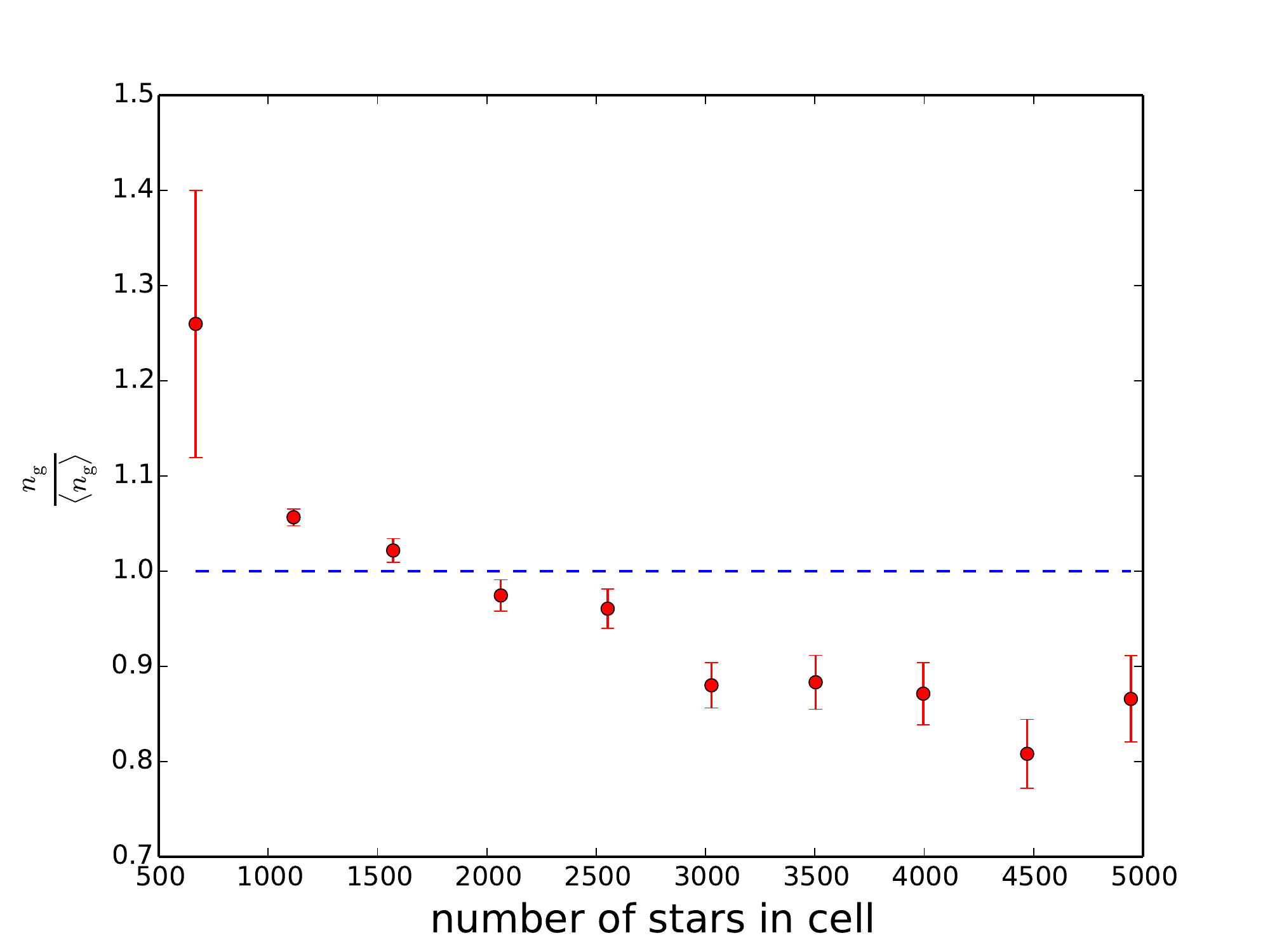}
\caption{The average ratio of observed galaxies to expected galaxies from the random catalogue in cells with given numbers of stars. 
\label{fig:nbarbynstars}}
\end{figure}

The objects in the BOSS DR12 CMASS galaxy and random 
catalogues are assigned to HEALPix cells as described in
Sec. \ref{sec:knowntemps}. The randoms are a catalogue of Poisson sampled positions from the
expected background density (that is without clustering) under the same
spatial selection function as the actual galaxy catalogue. The first step in the creation of the 
random catalogue is to pick random angular positions distributed according to the completeness 
of BOSS, which, in a given sector $i$, is estimated as
\begin{equation}
	C_{\mathrm{BOSS},i}\equiv \frac{N_{\mathrm{star},i}+
    N_{\mathrm{gal},i}+N_{\mathrm{fail},i}+
    N_{\mathrm{cp},i}}{N_{\mathrm{star},i}+
    N_{\mathrm{gal},i}+N_{\mathrm{fail},i}+
    N_{\mathrm{cp},i}+N_{\mathrm{missed},i}},
    \label{eq:completeness}
\end{equation}
where $N_{\mathrm{star},i}$, $N_{\mathrm{gal},i}$, 
$N_{\mathrm{fail},i}$, $N_{\mathrm{cp},i}$,
$N_{\mathrm{missed},i}$ are the numbers of objects
spectroscopically confirmed to be stars, objects that 
were spectroscopically confirmed a galaxies, objects
whose classification failed, 
close-pair objects of which
no spectra could be taken due to fibre collision but 
with at least one more object in the same target class,
and all other objects without spectra 
\citep{Reid:2015gra}. The next step is to apply veto masks to account for
regions where spectra cannot be taken for, e.g., the area around the 
central bolts of the tiles, the area around targets that 
have higher targeting priority, areas around bright stars with magnitudes smaller than 11.5.
Finally, each object that is still in the random
catalogue after vetoing is assigned a random redshift
that follows the distribution of the (weighted) galaxies. The 
random catalogues provided by the BOSS collaboration
contain 50 times more random galaxies than there are
in the galaxy catalogue.
After averaging over all 
$\frac{n_\mathrm{g}}{\langle 
n_\mathrm{g}\rangle}$ in a stellar number count bins, we
see a significant deviation from one (cf. Fig. 
\ref{fig:nbarbynstars}), as has been found by 
\cite{Ross:2011cz,Rossinprep} before.
In Fig. \ref{fig:nbarbynstars} we see that in pixels 
containing less than 1500 stars, we
observe more galaxies than we expect from the random 
catalogue, whereas in pixels with more than 2000 stars, 
we seem to miss galaxies in the observations.

\begin{figure}
 \centering
\includegraphics[width=\columnwidth]{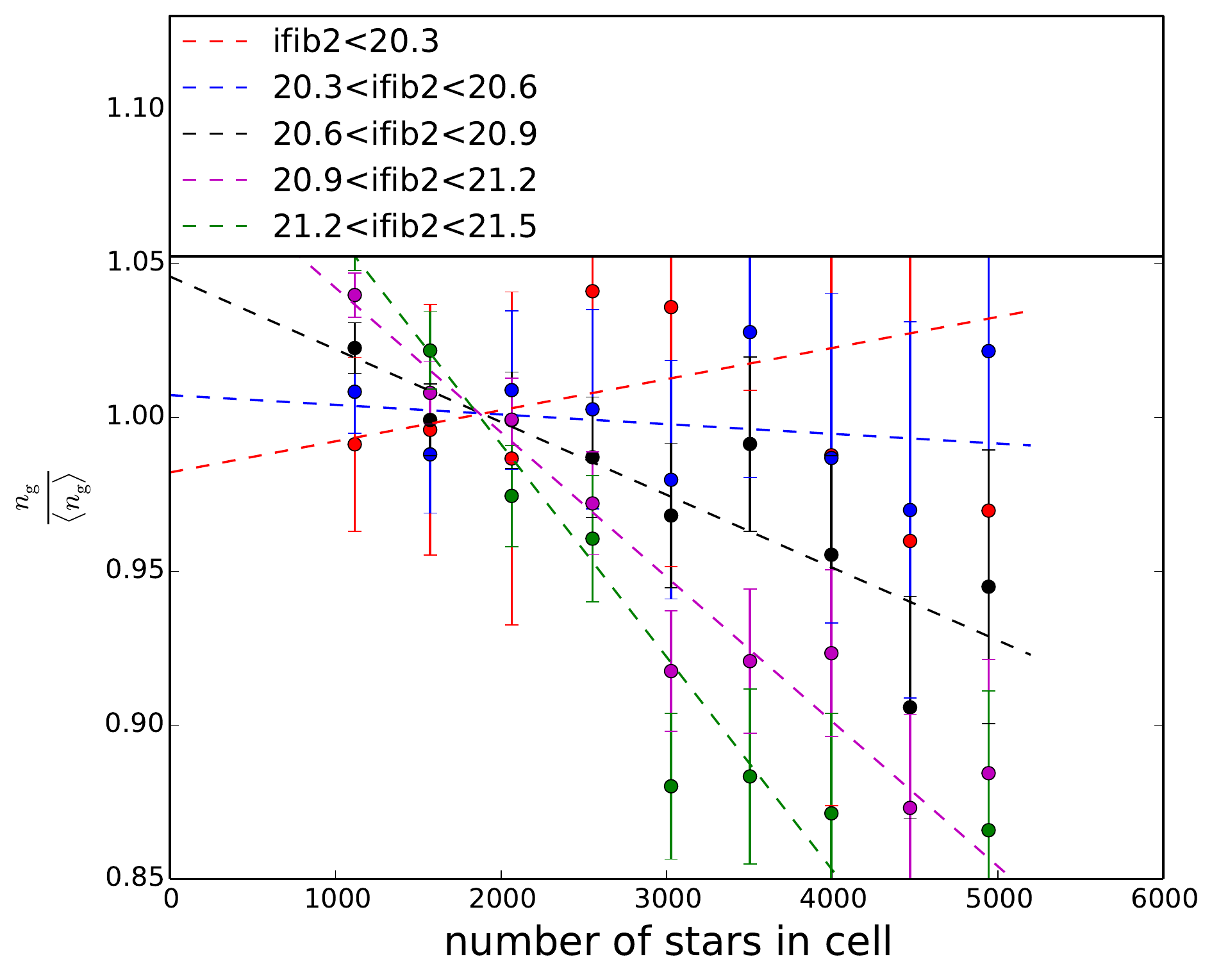}
\caption{Plot similar to Fig. \ref{fig:nbarbynstars}, but points in different colours are for different subsamples of galaxies with different $i_\mathrm{fib2}$ ranges. The dashed lines are the best-fitting lines through the data points. 
\label{fig:chi2fit}}
\end{figure}

\begin{figure}
 \centering
\includegraphics[width=\columnwidth]{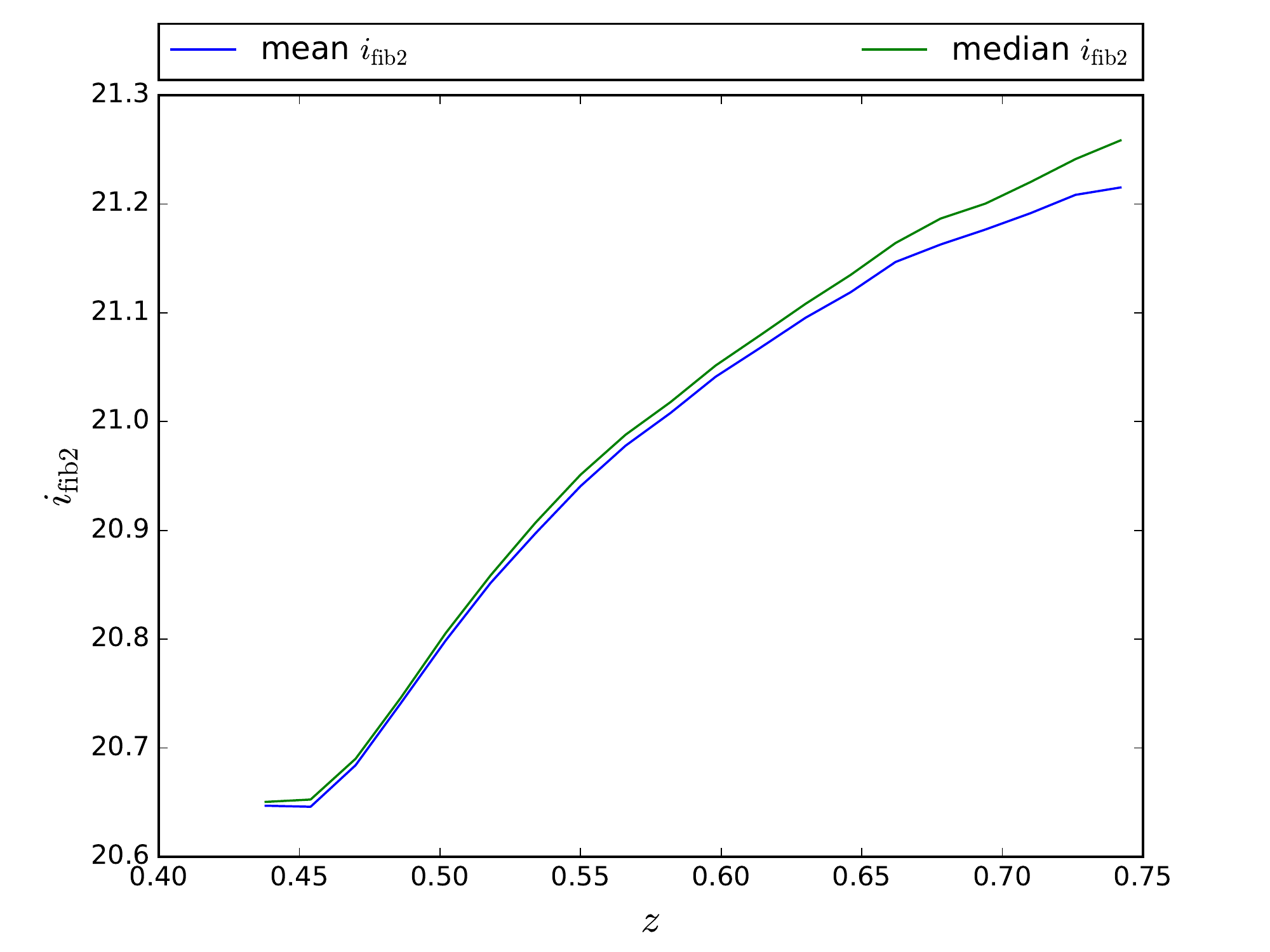}
\caption{Mean and median values of the 
	$i_\mathrm{fib2}$-magnitudes of BOSS 
    DR12 CMASS galaxies at given redshifts 
    $z$. 
\label{fig:ifibvsredplot}}
\end{figure}

For the effect of obscuration it is reasonable to assume 
that galaxies with different magnitudes are affected differently by 
foreground stars. \citet{Ross:2012sxfNL,Rossinprep}
therefore made plots similar 
to Fig. \ref{fig:nbarbynstars}, but with the galaxies split into 
sub-samples by their $i$-band magnitudes within a
$2^{\prime\prime}$ aperture radius $i_\mathrm{fib2}$. We also follow 
that procedure to generate Fig. \ref{fig:chi2fit}, where we can see
that galaxy 
observations are affected very differently according 
to their surface brightnesses: for galaxies with 
$i_\mathrm{fib2}<20.6$ we see no significant 
deviation between the expected and observed
number of galaxies. For fainter (in terms of surface 
brightness) galaxies, the best-fitting $\frac{n_
\mathrm{g}}{\langle n_\mathrm{g}\rangle}$-lines are 
negative and are steeper the larger the galaxies'
magnitudes (i.e. the fainter
they are). This meets our expectation,
because part of the stellar
contamination effect is due to obscuration.
To obtain \citet{Rossinprep}'s
weights for a given galaxy, Eq. \eqref{eq:linearcontfit} is extended by making the fitting
coefficients $C_0(i_\mathrm{fib2})$ and $C_1(i_\mathrm{fib2})$ magnitude dependent:
\begin{equation}
	\frac{n_\mathrm{g}}{\langle n_\mathrm{g}\rangle}^{(1)}(n_\mathrm{stars},i_\mathrm{fib2})\equiv C_0(i_\mathrm{fib2})+C_1(i_\mathrm{fib2})n_\mathrm{stars}.
    \label{eq:linearcontfitstars}
\end{equation}
In \citet{Rossinprep}, the stellar density weight for the $i$th galaxy in the survey
with magnitude $i_\mathrm{fib2}$ at right ascension $\alpha_i$ and $\delta_i$ is obtained by
evaluating Eq. \eqref{eq:linearcontfitstars} at its magnitude and at the number of stars in the 
pixel where it is situated: 
\begin{equation}
	w_i=\frac{1}{\frac{n_\mathrm{g}}{\langle n_\mathrm{g}\rangle}^{(1)}(n_\mathrm{stars}(\alpha_i,\delta_i),i_\mathrm{fib2})}.
\end{equation}
A template based technique, however, requires a field value for $i_\mathrm{fib2}$ in either 
configuration space or Fourier space, so one cannot generate 
the template using the $i_\mathrm{fib2}$-values of individual 
galaxies. Instead, we average $i_\mathrm{fib2}$ in redshift slices 
(cf. Fig. \ref{fig:ifibvsredplot}), because farther
galaxies tend to have smaller surface brightness,
and we assign the averages to template grid cells 
according to their redshifts. Apart 
from this, the weights entering Eq. \eqref{eq:farbitrary} are obtained in the 
same way as \citet{Rossinprep}'s weights:
\begin{equation}
	w(\x)=w(z,\alpha,\delta)=\frac{1}{\frac{n_\mathrm{g}}{\langle n_\mathrm{g}\rangle}^{(1)}(n_\mathrm{stars}(\alpha,\delta),\langle i_\mathrm{fib2}\rangle(z))}.
\end{equation}

\begin{figure}
 \centering
\includegraphics[width=\columnwidth]{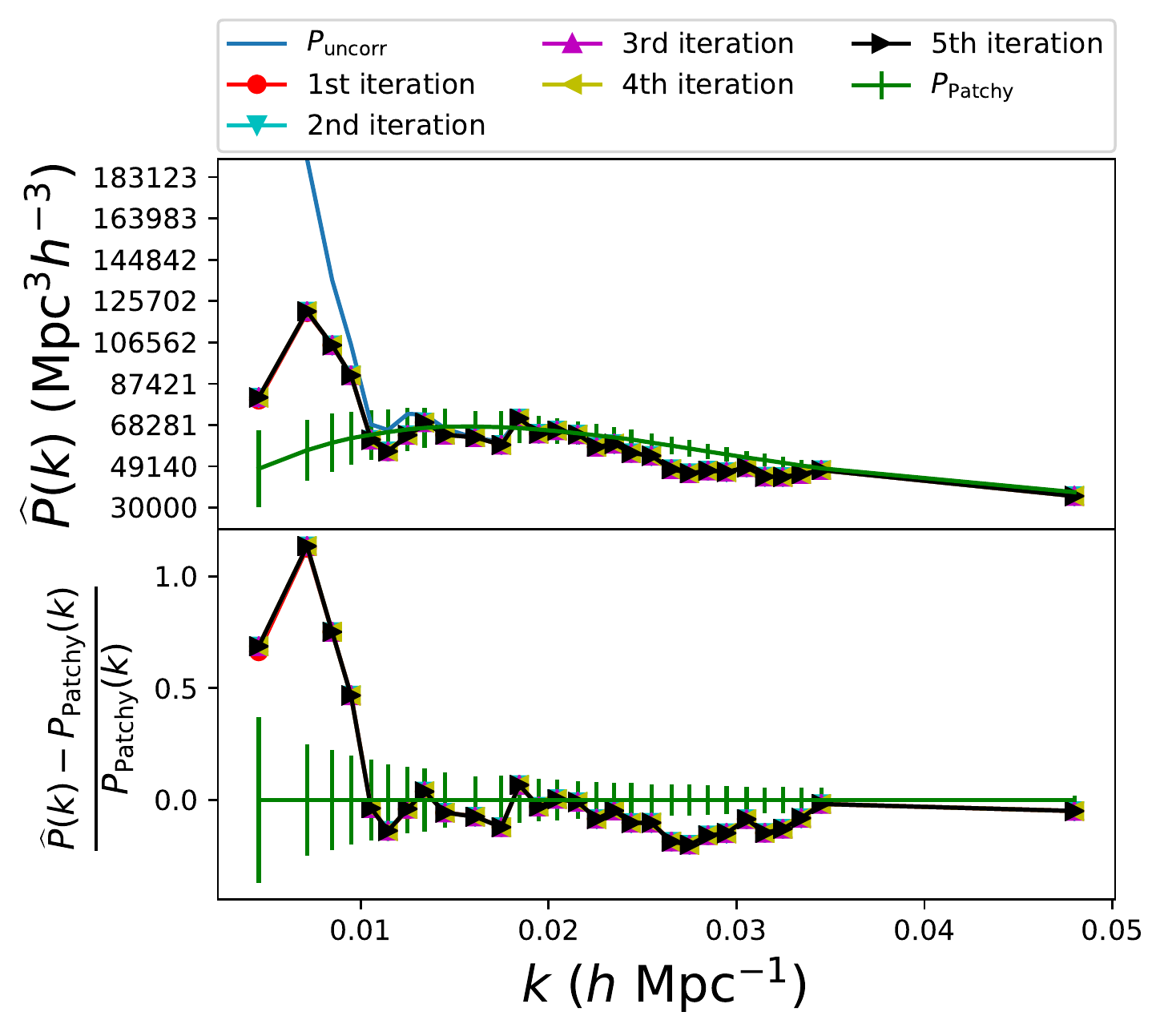}
\caption{The power spectra of the uncorrected BOSS DR12 CMASS NGC
galaxies $P_\mathrm{uncorr}$ and after 1 to 5 iterations of the
debiased mode subtraction
procedure, compared to the average power spectrum of the MultiDark-Patchy 
mocks $P_\mathrm{Patchy}$.
\label{fig:P_withoutsdcweightsPlotbyIteration}}
\end{figure}

We compute the BOSS DR12 CMASS NGC power spectrum using 
MOSES. We additionally apply the FKP, 
fibre-collision, redshift-failure and seeing weights 
that are provided in the catalogue files. The 
resulting power spectrum is shown in Fig. 
\ref{fig:P_withoutsdcweightsPlotbyIteration}. We 
also compute the power 
spectra of 2048 MultiDark-Patchy mock catalogues
\citep{Kitaura:2015uqa}
generated using the Perturbation Theory Catalog Generator of Halo and Galaxy 
Distributions \citep[PATCHY]{Kitaura:2014mja}. The mocks are generated using 
\citet{Kitaura:2013cwa}'s Augmented Lagrangian Perturbation Theory and a non-linear bias 
stochastic scheme. The bias parameters are fitted to the clustering of the BigMultiDark Planck 
simulation for each redshift snapshot \citep{Klypin:2014kpa}\footnote{www.multidark.org}. 
The mass assignment to halos was done with the Halo mAss Distribution ReconstructiON (HADRON)
code \citep{Zhao:2015jga}, that takes the local dark matter density, the cosmic web environment 
and the halo-exclusion effect into account. Finally, light-cones are obtained using the SUrvey
GenerAtoR \citep[SUGAR]{Rodriguez-Torres:2015vqa} code.

We use the MultiDark-Patchy mock power spectra for three different purposes:
\begin{itemize}
	\item We estimate the error on our power spectrum measurements as the sample variance of the 
    	MultiDark-Patchy power spectra. 
    \item We use the average of the mock power spectra to compare the power spectrum estimates
    	from data against. This has the advantage that the comparison is free of any kind of
        window effects even though the window changes the large scales a lot.
	\item We also use the average mock power spectrum as the prior power spectrum needed for 	
    	MOSES in the debiasing step of Eq. \eqref{eq:unbiasedFKPmodesub}. 
\end{itemize}
To check the stability of this choice of prior power 
spectrum, we use, for the first run, the average of 
the MultiDark-Patchy power spectra as the input prior 
power spectrum of the error mitigation procedure, 
and then we iterate by rerunning MOSES with the previous 
output power spectrum as the prior for the next run. We 
cannot see any 
significant difference between the power spectra of the five runs 
plotted in Fig. \ref{fig:P_withoutsdcweightsPlotbyIteration}. 
Furthermore, all of the five spectra agree well with the power 
spectrum obtained by mitigating the stellar density contamination 
using \citet{Rossinprep}'s weights (cf. Fig. 
\ref{fig:P_withoutsdcweightsPlotbyIteration}). This 
shows that MOSES can successfully remove the stellar contamination to first 
order. On the other hand, we also observe a significant 
discrepancy between the average MultiDark-Patchy power and our result. One might argue that, 
due to this discrepancy, the MultiDark-Patchy mocks are not suitable to calculate the 
covariance of this measurement. 
However, we are not testing any alternative model to the $\Lambda$CDM model, but any
possible deviations from this caused by systematics. Thus, in the absence of any alternative 
model that we could use to generate a different set of mock catalogues, the errors of the 
MultiDark-Patchy mocks and their underlying $\Lambda$CDM model describe well the error on our
expectation.

\subsubsection{Higher Order Stellar Templates}
\label{sec:higherorder}

\begin{figure}
 \centering
\includegraphics[width=\columnwidth]{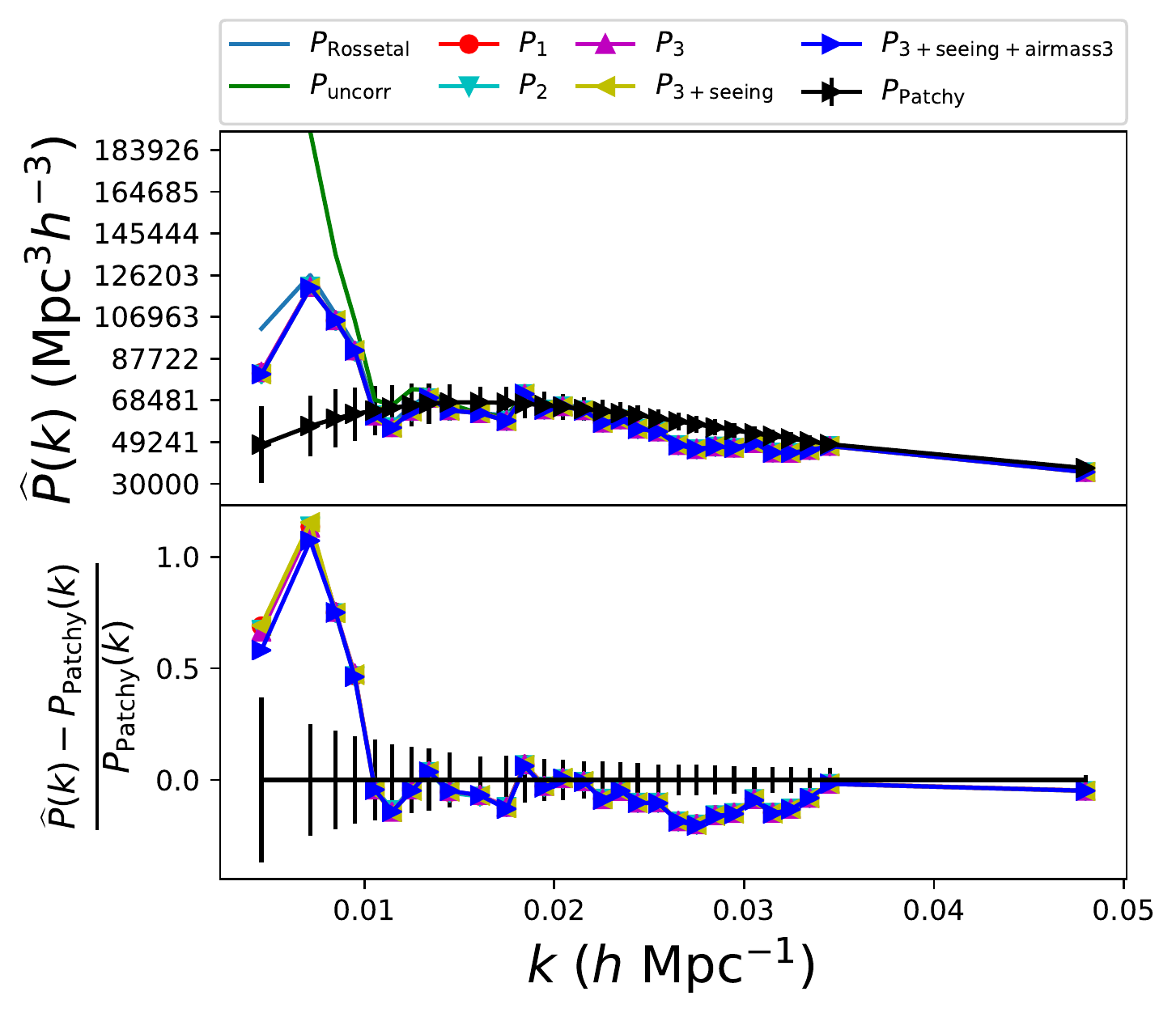}
\caption{The power spectra of the uncorrected BOSS DR12 CMASS NGC
galaxies and after applying debiased mode 
subtraction with a first order stellar template 
$P_1$, an additional second order template $P_2$,
and a third order template $P_3$. We also show the cases of using one seeing template $P_{3+\mathrm{seeing}}$, and of using 3 airmass templates (first, second and third order)  
$P_\mathrm{3+seeing+airmass3}$. We compare to the 
average power spectrum of the MultiDark-Patchy 
mocks. 
\label{fig:P_withoutsdcweightsPlotbyOrders}}
\end{figure}

A linear fit is not the only possible relationship
between observed galaxies and the number of stars. As
discussed in Sec. \ref{sec:practicalapprmulti}, one big advantage of our 
template based method is that we can add more 
templates for any form of contamination we have a reason to include.
To liberate ourselves from the linear assumption, we follow the steps described in Sec.
\ref{sec:knowntemps} but, as in the previous subsection, we allow for 
$i_\mathrm{fib2}$-dependent fitting coefficients and generate our higher order templates with
the redshift dependence of the average magnitude in mind.

We use Eq. \eqref{eq:epsbfmulticont}
to find the best-fitting amplitude of each template, which are 
listed in Tab. \ref{tab:epsilonbf3stars}. The amplitude for the 
first and third order templates, $\varepsilon_1^\mathrm{(BF)}$
and $\varepsilon_3^\mathrm{(BF)}$,
respectively, do not change significantly
when other templates are fitted at the same time. The second order
amplitude changes, but it is always at least one order of
magnitude less than $\varepsilon_1^\mathrm{(BF)}$ and 
$\varepsilon_3^\mathrm{(BF)}$. The fourth
order amplitude $\varepsilon_4^\mathrm{(BF)}$ is also much smaller, which 
suggests that the true relationship between observed 
number of stars and galaxies is an odd function. 
We compute
the debiased mode subtracted power spectra, which we plot in Fig.
\ref{fig:P_withoutsdcweightsPlotbyOrders}. We observe that, even 
though $\varepsilon_3^\mathrm{(BF)}$ is almost as large as 
$\varepsilon_1^\mathrm{(BF)}$,
including the third order stellar contamination 
template, or in fact any other higher order template, 
does not change the 
resulting power spectrum significantly. This is because
the field values of the third order template are two
orders of magnitude smaller than those of the first order
template. Therefore, for the third order template to have
an effect, we would need $\varepsilon_3^\mathrm{(BF)}
\gg\varepsilon_1^\mathrm{(BF)}$. This is similar for
other higher order templates.

\begin{table}
 \centering
 \begin{minipage}{\linewidth}
  \caption{Best-fitting contamination amplitudes $\varepsilon^\mathrm{(BF)}$ for a power
  	spectrum measurement using different numbers of stellar templates.}
  \label{tab:epsilonbf3stars}
  \begin{tabularx}{\textwidth}{Xrrrr}
 \hline
 order & \#templates: 1 & 2 & 3 & 4 \\ 
 \hline
 $1^\mathrm{st}$ & 0.0071 & 0.0071 & 0.0072 & 0.0073\\
 $2^\mathrm{nd}$ & & 0.0008 & 0.0001 & 0.0008\\
 $3^\mathrm{rd}$ & & &  0.0055 & 0.0054 \\
 $4^\mathrm{th}$ & & & & -0.0001 \\
\hline
\end{tabularx}
\end{minipage}
\end{table}

\subsubsection{Sub-Sampling the Stars by Magnitude}
\label{sec:magnitudesplit}

As the distribution of faint and bright stars is
different on the sky, we split the SDSS star 
catalogue into sub-samples
according to the stellar magnitudes. 

\begin{figure}
 \centering
 \includegraphics[width=0.9\linewidth]{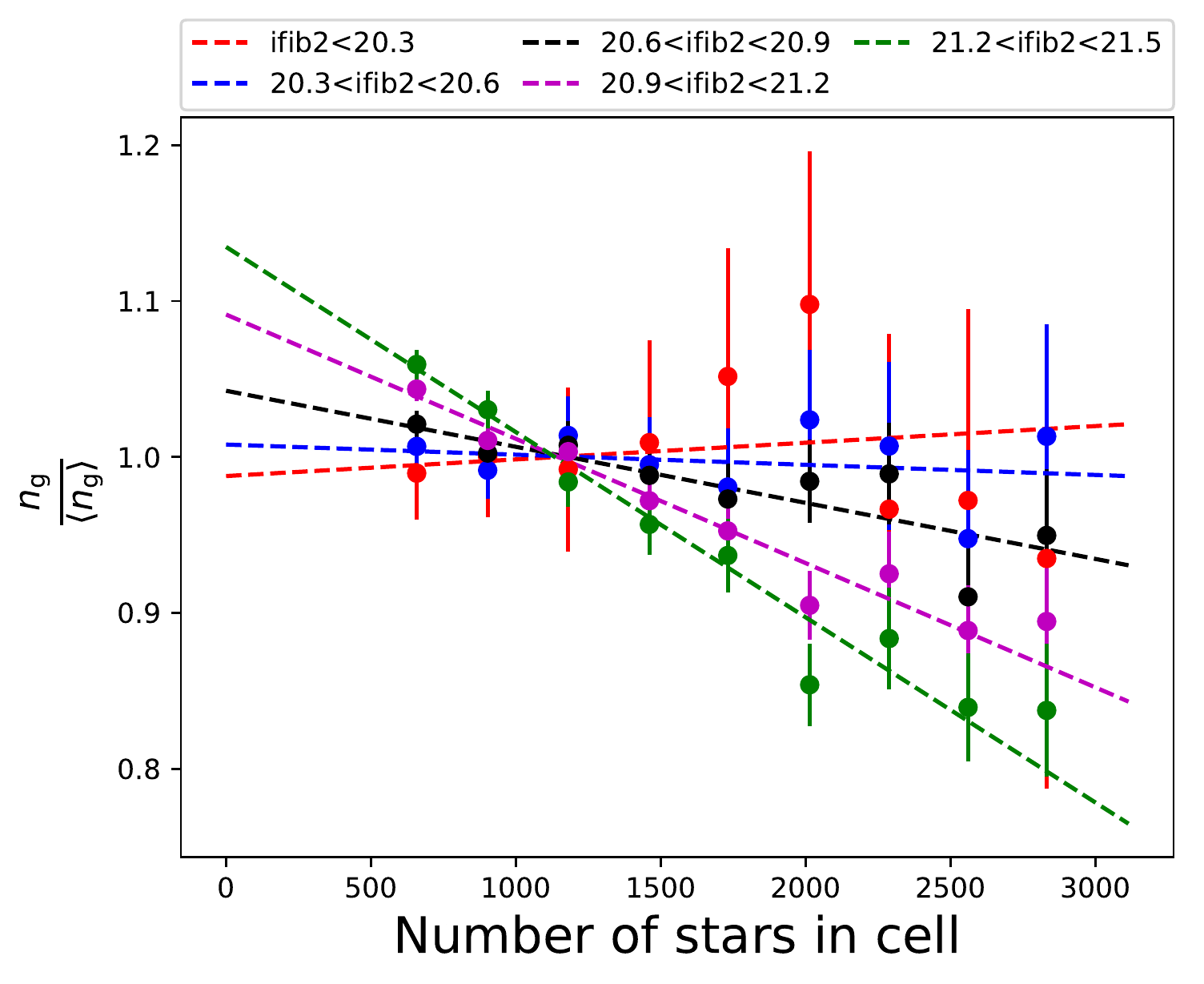}
 \includegraphics[width=0.9\linewidth]{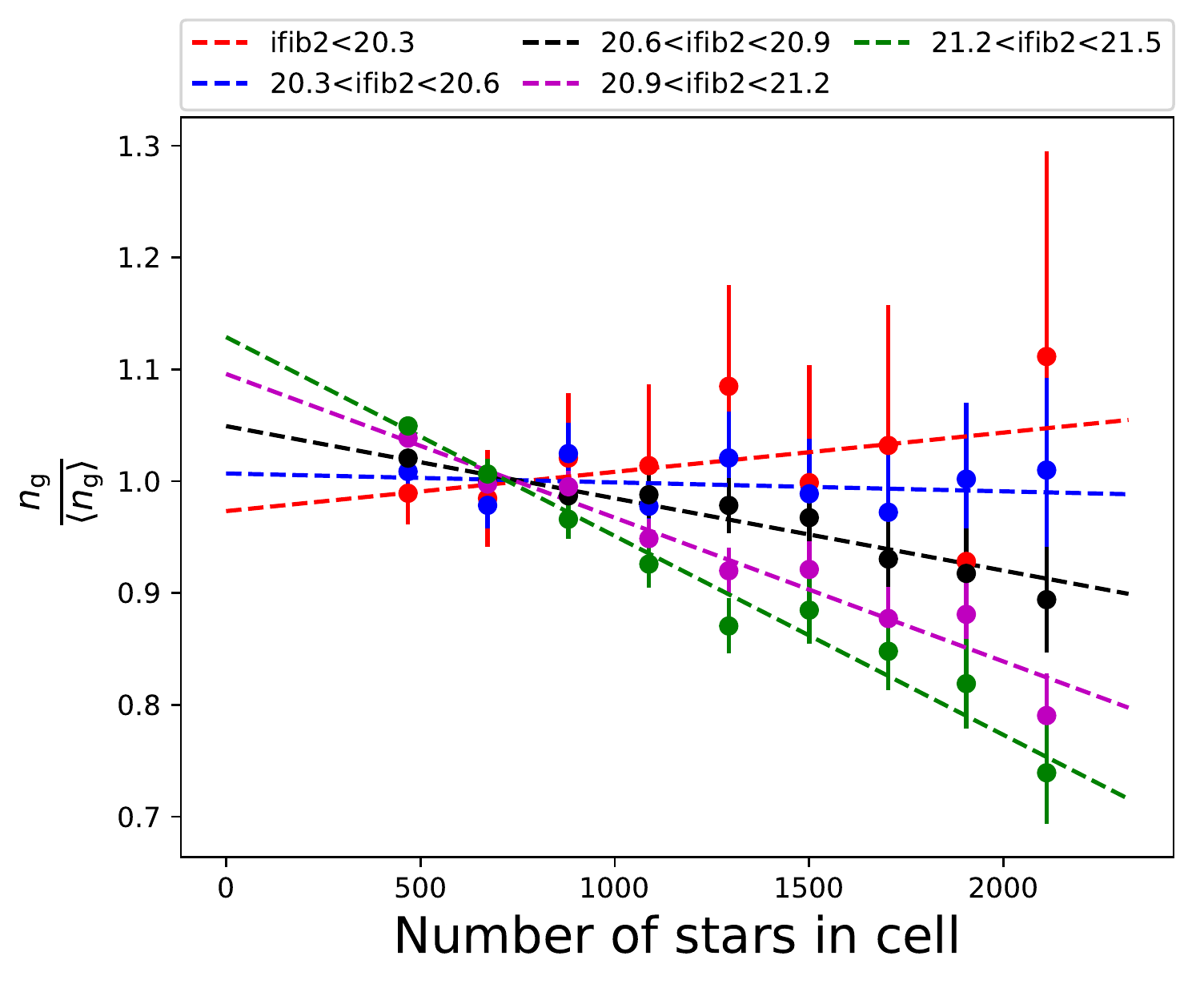}
	\caption{The relationship between observed galaxy density and the number of bright stars ($i<18.7$, upper panel) and faint stars ($i>18.7$, lower panel).}
 \label{fig:starmagsplit2chi2fit}
\end{figure}

\begin{figure}
 \centering
\includegraphics[width=\columnwidth]{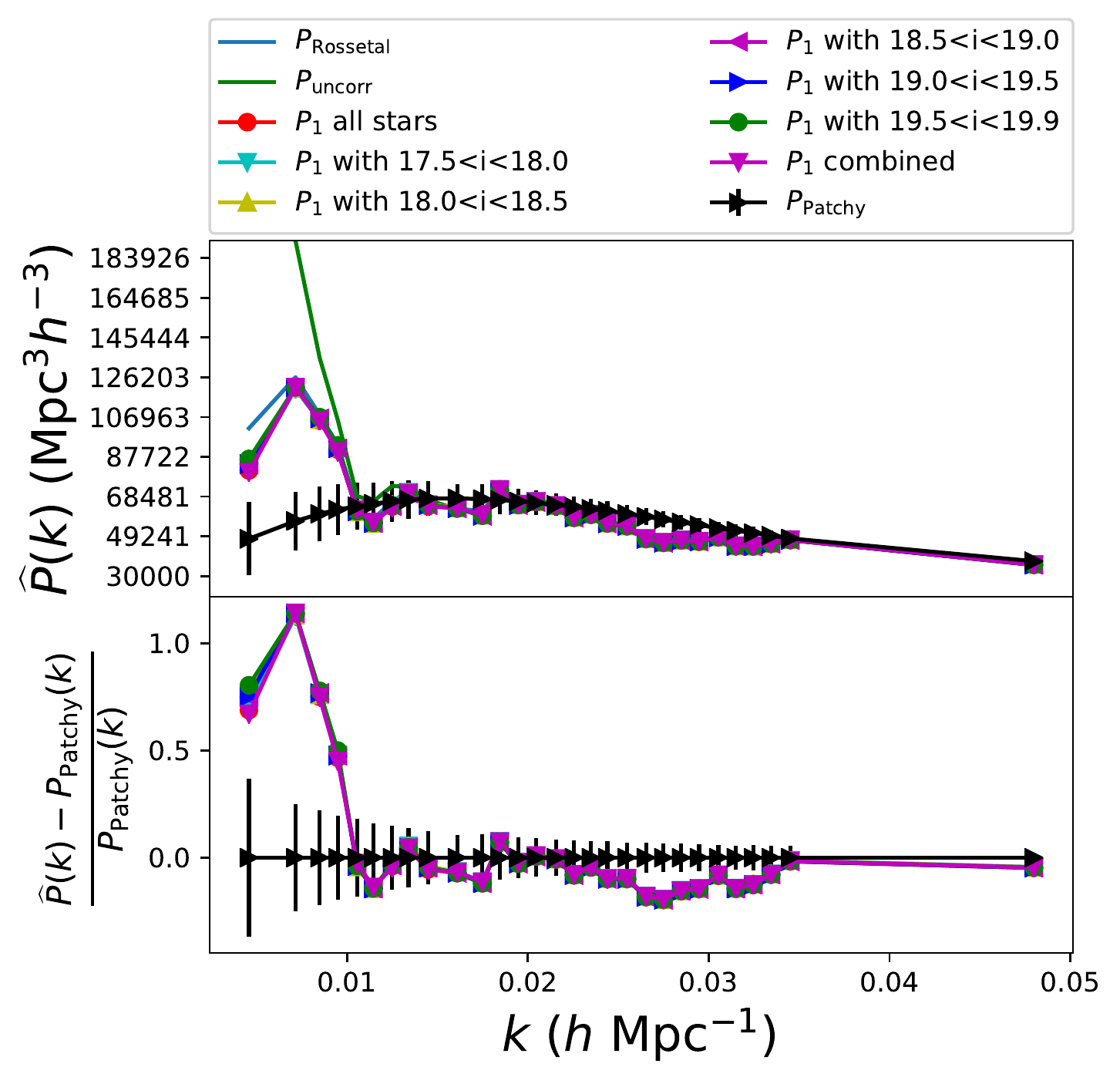}
\caption{The power spectra of the uncorrected BOSS DR12 CMASS NGC
galaxies (green), and those after mode subtraction using five different templates for stars with different magnitudes, compared to the average power spectrum of the MultiDark-Patchy
mocks (black) and the power spectrum using the Ross \textit{et al.} 
weights (blue). 
\label{fig:P_withoutsdcweightsPlotwithMagnitudeSplit}}
\end{figure}

First, we split the star sample into 
two sub-samples at
the central $i$-band magnitude value of $i=18.7$. In
Fig. \ref{fig:starmagsplit2} one can see that the two 
sub-samples also have different spatial distributions:
Bright stars are more likely to be found close to the 
Galactic plane, whereas faint stars are more spread out.
The $n_\mathrm{g}/\langle n_\mathrm{g}\rangle$ diagrams 
in Fig. \ref{fig:starmagsplit2chi2fit}
do not look very different, though.
We refine the magnitude
split of the stars and split them
into five magnitude bins,
each with a width of 0.5, except for the last bin with
$19.5<i<19.9$. By comparing the 
masked maps of each sample 
(Fig. \ref{fig:starmagsplit5}), one can see that the 
differences in the distribution of stars are only prominent
in regions close to the galactic plane from where no
galaxies enter BOSS. Therefore, the templates are all strongly correlated, which
we can also see in Tab. 
\ref{tab:epsilonbfmagsplit}. When fitted separately,
all templates have roughly the same amplitude, and each 
template alone can remove the whole stellar contamination
signal, suggesting that they contain mostly the same 
information. When combined, their amplitudes differ, but
the resulting power spectrum does not change. The 
resulting power spectra are plotted in Fig.
\ref{fig:P_withoutsdcweightsPlotwithMagnitudeSplit}.
We therefore
conclude that at different magnitudes, the effect of 
stars on the galaxy power spectrum is very similar and we
therefore do not have to sub-sample the stars by 
magnitude when mitigating against stellar effects. 

\begin{table}
 \centering
 \begin{minipage}{\linewidth}
  \caption{Best-fitting contamination amplitudes for a power
  	spectrum measurement using five templates for different
    magnitude ranges of the stars.
    The values on the left hand side are obtained by fitting
    only one template at a time, whereas those on the right
    have been obtained in a simultaneous fit.}
  \label{tab:epsilonbfmagsplit}
  \begin{tabularx}{\textwidth}{Xrr}
 \hline
 magnitude range & separate fit & simultaneous fit\\ 
 \hline
 $17.5<i<18.0$ & 0.007 & 0.013 \\
 $18.0<i<18.5$ & 0.006 & -0.009 \\
 $18.5<i<19.0$ & 0.006 & -0.004 \\
 $19.0<i<19.5$ & 0.007 & 0.009 \\
 $19.5<i<19.9$ & 0.006 & -0.003 \\ 
\hline
\end{tabularx}
\end{minipage}
\end{table}

\subsubsection{Number Count versus Integrated Magnitude}
\label{sec:brightness}

\begin{figure}
 \centering
 \includegraphics[width=\linewidth]{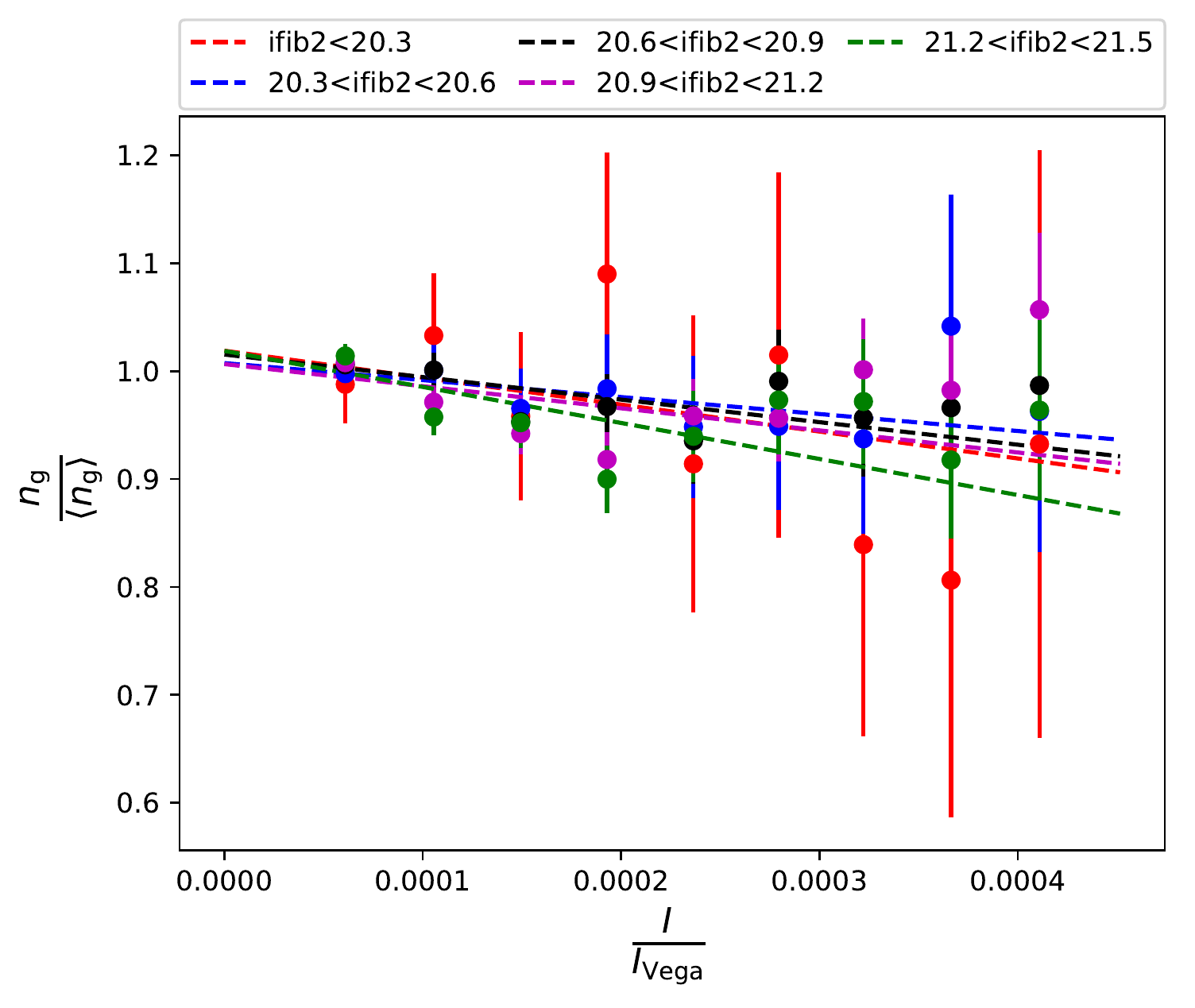}
	\caption{The relationship between 
    	observed galaxy density and the 
        integrated stellar foreground 
        brightness in units of the
        brightness of Vega.}
 \label{fig:intchi2fit}
\end{figure}

So far, all our stellar templates are based on the number
count of stars in regions of the sky. As the effect of the
stars is due to their light, this is not the only plausible
way of making templates and we explore, as an alternative,
basing the templates on the stellar foreground brightness $I(\alpha,\delta)$ in each HEALPix cell. The astronomical
magnitude $m$ of an object is defined through the decimal 
logarithm of its brightness $I$ in units of the brightness of
a reference object $I_\mathrm{ref}$:
\begin{equation}
	m-m_\mathrm{ref}\equiv -2.5\log_{10}\left(\frac{I}{I_\mathrm{ref}}\right).
\end{equation}
Given the $i$-band magnitudes provided in the
star catalogue file and used in the previous
section, we can obtain the stellar foreground 
brightness as the sum over the brightness of all stars in 
a HEALPix cell around the coordinates $(\alpha,\delta)$:
\begin{equation}
	I(\alpha,\delta)\propto\sum_{\mathrm{stars \in cell}} 10^{-i/2.5}.
\end{equation}
The distribution of the stellar 
foreground brightness, mapped in Fig.
\ref{fig:intmap}, is very similar to the 
distribution of the number of foreground stars
(cf. Fig. \ref{fig:starsNSIDE256}). However,
in Fig. \ref{fig:intchi2fit} we see that the
relationship between observed galaxy density 
and the foreground brightness
$n_\mathrm{g}/\langle 
n_\mathrm{g}\rangle(I(\alpha,\delta),i_\mathrm
{fib2})$ looks very different compared to
the same plot for the number counts (cf. Fig.
\ref{fig:chi2fit}), but this can be explained by 
the fact that the number count and the brightness are 
approximately logarithmically related. A linear fit does not
agree well with the data and the errorbars are so large that we could fit almost any shape 
with almost any slope. In this case, a more thorough treatment of the error of the template
would be needed; however, as the quality of fit of the number count based templates is much
better, we leave this for future work.

\begin{figure}
	\centering
    \includegraphics[width=\linewidth]{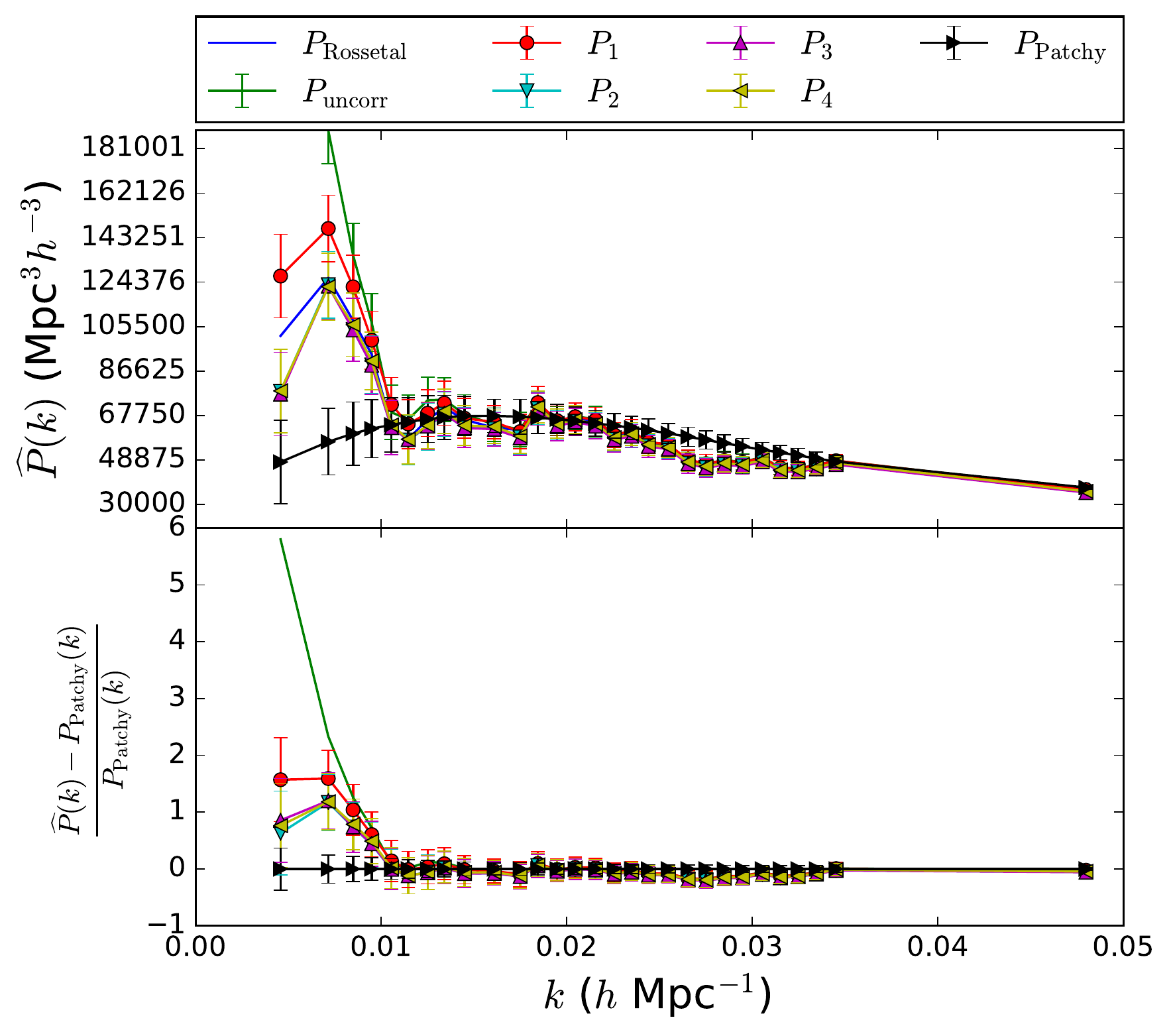}
    \caption{Plot similar to Fig. \ref{fig:P_withoutsdcweightsPlotbyOrders} but with 
    templates based on the integrated brightness of stars rather than their number counts.}
    \label{fig:P_withoutsdcweightsPlotbyOrdersInt}
\end{figure}

The first order template for integrated brightness of stars also does worse in removing the contamination than the first order
template based on the number counts, as the plot of the
resulting power spectrum in Fig.
\ref{fig:P_withoutsdcweightsPlotbyOrdersInt} shows. 
Introducing higher order templates results in power spectra
that are similar to the power spectra obtained in the 
sections before. It shows that the method of introducing
templates based on a series expansion of the expected
contaminant is working if more than one template is significant, and 
if they are uncorrelated. On the other hand, it also shows that there is
no improvement by constructing the templates on the 
integrated brightness rather than on the number count of
the foreground stars.

\subsection{Seeing}
\label{sec:seeing}
The light travelling to ground based 
telescopes has to travel through
the Earth's atmosphere. Due to 
turbulence in the atmosphere, its 
refractive index changes on short 
time scales. This blurs the image
of an astronomical object and the 
flux of the object is spread out.
This causes an increased magnitude 
error, and hence, makes it 
problematic to distinguish between 
galaxies and stars in the target
selection process \citep{Ross:2011cz},
because the star-galaxy 
separation cut relies entirely on magnitudes.
This can cause spurious fluctuations in the 
observed density field of galaxies \citep{Ross:2011cz}. 
The seeing can 
be quantified by measuring
the apparent diameter of a point source. In all 
power spectrum estimates in the previous
subsection, we took seeing into account by applying the
seeing weights that are provided in the 
galaxy catalogue file and which are
mapped in Fig. \ref{fig:Seeingmap}, which is 
inconsistent with the way we mitigated against the 
effects of foreground stars. Here, we use it as a 
test case of the equivalence of templates and 
corrective weights that was
outlined in Sec. \ref{sec:practicalapprsingle}.

As the effect is purely
angular and does not depend 
on intrinsic properties of the 
galaxies, we can build our templates by directly 
inserting \citet{Rossinprep}'s
weights into Eq. \eqref{eq:farbitrary}. The yellow lines
and left facing triangles in Fig.
\ref{fig:P_withoutsdcweightsPlotbyOrders} represent the 
power spectrum we obtained after replacing the direct 
application of the seeing weights with seeing templates, 
and using three templates for the stellar contamination. 
The plot shows that there is no difference between the 
results obtained using the weights and those obtained 
using templates based on the same weights. This shows 
that MOSES works and that the discrepancy between 
the measured and
theoretical power spectra is not due to using
different error mitigation techniques 
inconsistently. 

\begin{table}
 \centering
 \begin{minipage}{\linewidth}
  \caption{Best-fitting contamination amplitudes for a power
  	spectrum measurement using three stellar templates and seeing
    weights (left) and replacing the seeing weights by seeing 
    templates (right).}
  \label{tab:epsilonbf3stars1seeing}
  \begin{tabularx}{\textwidth}{Xrr}
 \hline
 template & only stellar & + 1 seeing template\\ 
 \hline
 stars $1^\mathrm{st}$ order & 0.00719 & 0.00739 \\
 stars $2^\mathrm{nd}$ order & 0.00009 & -0.00002 \\
 stars $3^\mathrm{rd}$ order & 0.00552 & 0.00576 \\ 
 seeing & & -0.00237 \\
\hline
\end{tabularx}
\end{minipage}
\end{table}

\subsection{Airmass}
\label{sec:airmass}

\begin{figure}
 \centering
\includegraphics[width=\columnwidth]{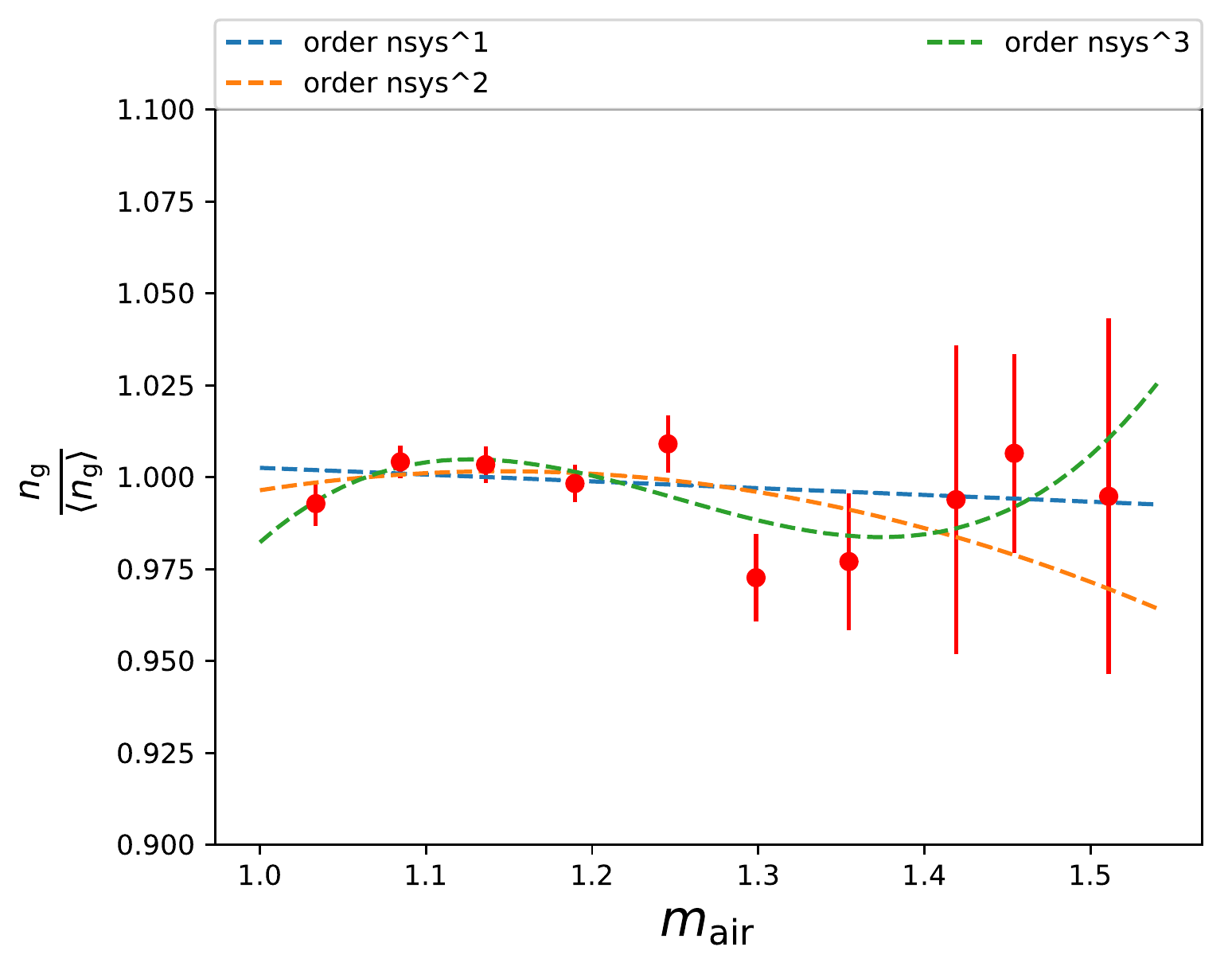}
\caption{The relationship between observed galaxy density and airmass (cf. Eq. \eqref{eq:airmass}).}
\label{fig:airmasschi2fit}
\end{figure}

Another variation in astronomical
observations due to the Earth's atmosphere arises because light 
coming from a source close to the horizon has
to travel through more atmosphere than the light coming from a source
close to the zenith. The effect is quantified by the airmass
\begin{equation}
	m_\mathrm{air}\equiv \frac{\int\d s\; \rho}
    {\int \d s_\mathrm{zenith}\;\rho},
    \label{eq:airmass}
\end{equation}
which is the column density, i.e., the integral over the mass density of the atmosphere $\rho$,
along
the line of sight $s$, divided by the zenith column density. The
mass density depends on time-varying quantities such as the 
temperature and other weather phenomena. Furthermore, the angle 
between the zenith and the line of sight changes with the seasons.
Hence, the amount of photons to be scattered or
absorbed varies with both position and observing times, effectively
varying the depth of the survey and the magnitude 
error. Information about the airmass is provided in the random
catalogue. A map can be found in Fig. \ref{fig:Airmassmap}.
It prominently shows the drift scanning strategy of SDSS. The airmass 
does not change significantly along SDSS scanning stripes, as the telescope
remains stationary along a great circle, but there are 
sharp leaps from stripe to stripe, which can cause spurious
fluctuations in the density field. A plot
similar to Fig. \ref{fig:nbarbynstars} that relates $n_\mathrm{g}/
\langle n_\mathrm{g}\rangle$ to the airmass is shown in Fig.
\ref{fig:airmasschi2fit}, where the
data points are consistent with
$n_\mathrm{g}/\langle n_\mathrm{g}\rangle=1$ for 
almost all
values of airmass. The linear fit $n_\mathrm{g}/\langle
n_\mathrm{g}\rangle^\mathrm{(1)}$ through Fig.
\ref{fig:airmasschi2fit} 
is almost constantly equal to one. The quadratic fit
$n_\mathrm{g}/\langle
n_\mathrm{g}\rangle^\mathrm{(2)}$ shows a slight negative trend at
larger airmasses and the cubic fit $n_\mathrm{g}/\langle
n_\mathrm{g}\rangle^\mathrm{(3)}$ looks like an over-fit. 
\citet{Rossinprep} made a 
similar analysis including a $\chi^2$ null test. 
Based on that test, they state that corrections for 
such a marginally significant effect are ill-advised. 
However, they recommend to 
reconsider this choice for any future
studies of the clustering of BOSS galaxies at the largest 
scales.

We proceed as in Sec. \ref{sec:knowntemps}. We fit the three 
polynomials
\begin{equation}
	\frac{n_\mathrm{g}}{\langle n_\mathrm{g}\rangle}^{(N)}(m_\mathrm{air})=\sum_{i=0}^N C_i m_\mathrm{air}^i
\end{equation}
to the data that we have plotted in Fig. \ref{fig:airmasschi2fit}. We
define
\begin{equation}
	E_{\mathrm{am},N}(\alpha,\delta)\equiv \frac{n_\mathrm{g}}{\langle n_\mathrm{g}\rangle}^{(N)}(m_\mathrm{air}(\alpha,\delta))-\frac{n_\mathrm{g}}{\langle n_\mathrm{g}\rangle}^{(N-1)}(m_\mathrm{air}(\alpha,\delta)),
\end{equation}
which we insert into Eq. \eqref{eq:multcontweights} and
\eqref{eq:multconttemp} to obtain templates to mitigate the effect
of the airmass.

We perform the mode subtraction method 
and find the best-fitting template amplitudes given in Tab.
\ref{tab:epsilonbf3stars3airmass}. The third order template indeed 
is not favoured by the data and obtains a very small amplitude,
suggesting that the third order describes noise rather than an
actual effect of the airmass on the observed galaxy density. The first
order is almost constant and equal to one, so it cannot be expected to
significantly change the resulting power spectrum. The second order
template, however, has the largest amplitude coefficient. Yet, including all templates into the power spectrum measurement does only
lead to
minor corrections in the result, as the blue line in Fig. \ref{fig:P_withoutsdcweightsPlotbyOrders} shows.

\begin{table}
 \centering
 \begin{minipage}{\linewidth}
  \caption{Best-fitting contamination amplitudes for a power
  	spectrum measurement using three stellar templates
    (left) and additionally three airmass templates
    (right).}
  \label{tab:epsilonbf3stars3airmass}
  \begin{tabularx}{\textwidth}{Xrr}
 \hline
 template & only stellar & + airmass templates\\ 
 \hline
 stars $1^\mathrm{st}$ order & 0.0072 & 0.0061 \\
 stars $2^\mathrm{nd}$ order & 0.0001 & -0.0013 \\
 stars $3^\mathrm{rd}$ order & 0.0055 & 0.0042 \\ 
 airmass $1^\mathrm{st}$ order & & -0.0014 \\
 airmass $2^\mathrm{nd}$ order & & 0.0202 \\
 airmass $3^\mathrm{rd}$ order & & -0.0003 \\
\hline
\end{tabularx}
\end{minipage}
\end{table}

\subsection{Galactic Extinction}
\label{sec:extinction}

The interstellar medium 
within our Galaxy causes Galactic Extinction which can be mapped. As blue light is
more affected by scattering, extinction causes the light to become 
redder, and extinction is usually quantified as the difference
between the observed (obs) and intrinsic (int) $\mathrm{B-V}$ colour
\begin{equation}
	E_\mathrm{B-V}=(\mathrm{B-V})_\mathrm{obs}-(\mathrm{B-V})_\mathrm{int},
\end{equation}
where B stands for
the filter sensitive to blue light and V is sensitive to visible 
green-yellow light.

\begin{figure}
 \centering
\includegraphics[width=\columnwidth]{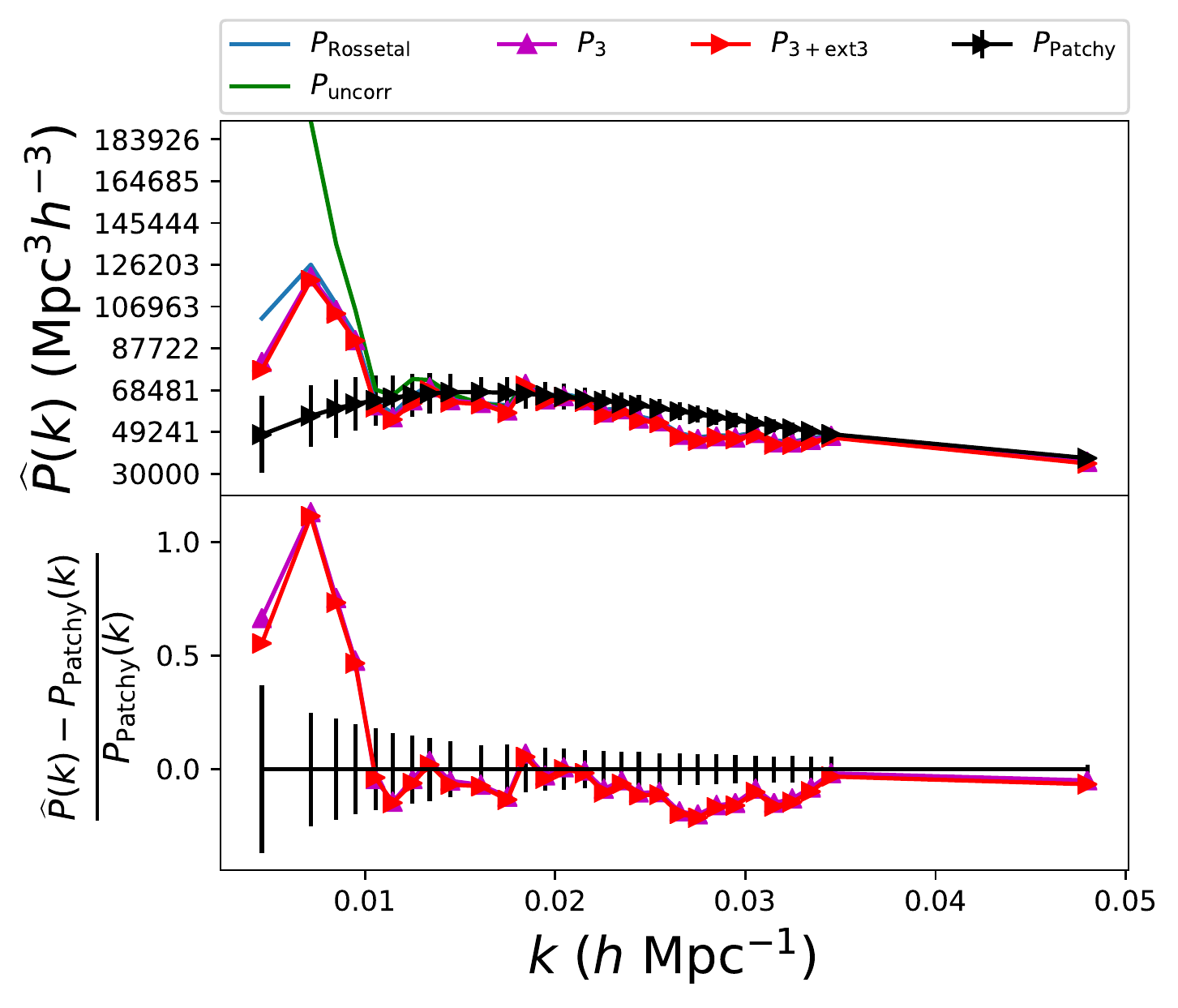}
\caption{The power spectra of the uncorrected BOSS DR12 CMASS NGC
galaxies (green), and those after mode subtraction using 3 stellar templates (magenta) and 3 stellar and 3 extinction templates (red), compared to the average power spectrum of the MultiDark-Patchy
mocks (black) and the power spectrum using the Ross \textit{et al.} 
weights (blue). 
\label{fig:P_withoutsdcweightsPlot_ext3}}
\end{figure}

The photometric magnitudes used in the BOSS target selection were
corrected using the dust map by 
\citet*[SFD]{Schlegel:1997yv}. \citet{Schlafly:2010dz}
found that, using a more 
accurate reddening law, the SFD map 
$E_\mathrm{B-V,SFD}$ has to be
recalibrated such that \citep{Schlafly:2010dz}
\begin{equation}
	E_\mathrm{B-V}=0.86E_\mathrm{B-V,SFD}.
\end{equation}
Due to the recalibration, there might be a 
colour-dependent shift in the target 
density. A similar $\chi^2$ null-test by 
\citet{Rossinprep} led to a similar conclusion as 
for the airmass test: extinction weights
do not significantly change the clustering 
statistics at BAO scales, 
but one should be prudent at large 
scales. 
For that reason, we test whether including
extinction templates changes our power spectrum at
large scales. The SFD values of $E_\mathrm{E-V,SFD}$
used in the BOSS targeting and listed in the catalogue files are mapped in Fig. 
\ref{fig:Extinctionmap}. There, one can see that 
extinction mostly affects the SGC part of the BOSS
footprint, which we do not analyse in this work.
Extinction in NGC occurs mostly in the regions close
to the Galactic disk, similar to the stars in Fig.
\ref{fig:starsNSIDE256}. We therefore might expect 
some
correlation between the stellar and extinction
templates, as their best-fitting amplitudes 
$\varepsilon^\mathrm{(BF)}$, listed in Tab.
\ref{tab:epsilonbf3stars3ext}, also suggest. The 
amplitudes of the first and third 
order stellar templates is slightly smaller when fitted
at the same time as the extinction templates. The 
amplitudes of all
extinction templates are less than all stellar template 
amplitudes,
explaining why the power spectrum does not change much 
when extinction templates are included (cf. Fig.
\ref{fig:P_withoutsdcweightsPlot_ext3}).

\begin{table}
 \centering
 \begin{minipage}{\linewidth}
  \caption{Best-fitting contamination amplitudes for a power
  	spectrum measurement using three stellar templates
    (left) and additionally three extinction templates
    (right).}
  \label{tab:epsilonbf3stars3ext}
  \begin{tabularx}{\textwidth}{Xrr}
 \hline
 template & only stellar & + extinction templates\\ 
 \hline
 stars $1^\mathrm{st}$ order & 0.0072 & 0.0070 \\
 stars $2^\mathrm{nd}$ order & 0.0001 & 0.0023 \\
 stars $3^\mathrm{rd}$ order & 0.0055 & 0.0043 \\ 
 \multicolumn{2}{l}{extinction $1^\mathrm{st}$ order}
 & -0.0009 \\
 \multicolumn{2}{l}{extinction $2^\mathrm{nd}$ order}
 & 0.0016 \\
 \multicolumn{2}{l}{extinction $3^\mathrm{rd}$ order}
 & 0.0020 \\
\hline
\end{tabularx}
\end{minipage}
\end{table}

\subsection{Scanning Stripes}
\label{sec:stripes}
\begin{figure}
 \centering
\includegraphics[width=\columnwidth]{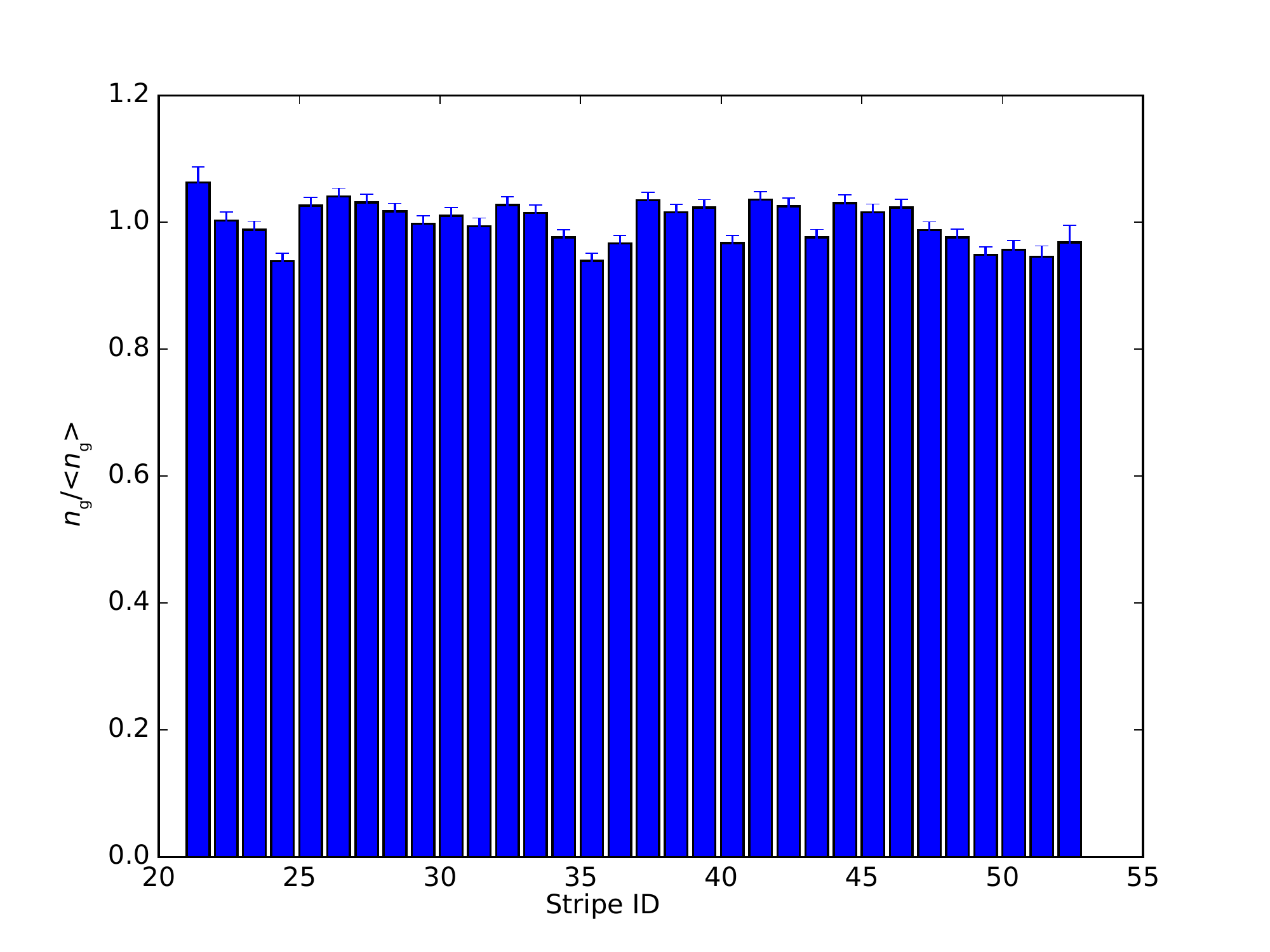}
\caption{$n_\mathrm{g}/\langle n_\mathrm{g}\rangle$ in the different scanning
stripes.
\label{fig:GalvsStripe}}
\end{figure}

Another possible source of data contamination is the 
instrument itself rather than astronomical or atmospheric
foregrounds. For example, the telescope might have a calibration 
offset between different nights.
Furthermore, one can see in Fig. \ref{fig:Seeingmap}
and \ref{fig:Airmassmap} that time-varying systematics 
are mostly exposing the drift scanning strategy of
SDSS. In fact, Fig. \ref{fig:GalvsStripe}
shows that the observed number of galaxies
in certain stripes can be significantly 
different from the number that is expected 
from the random catalogue.

\begin{figure}
	\centering
    \includegraphics[width=\columnwidth]
    {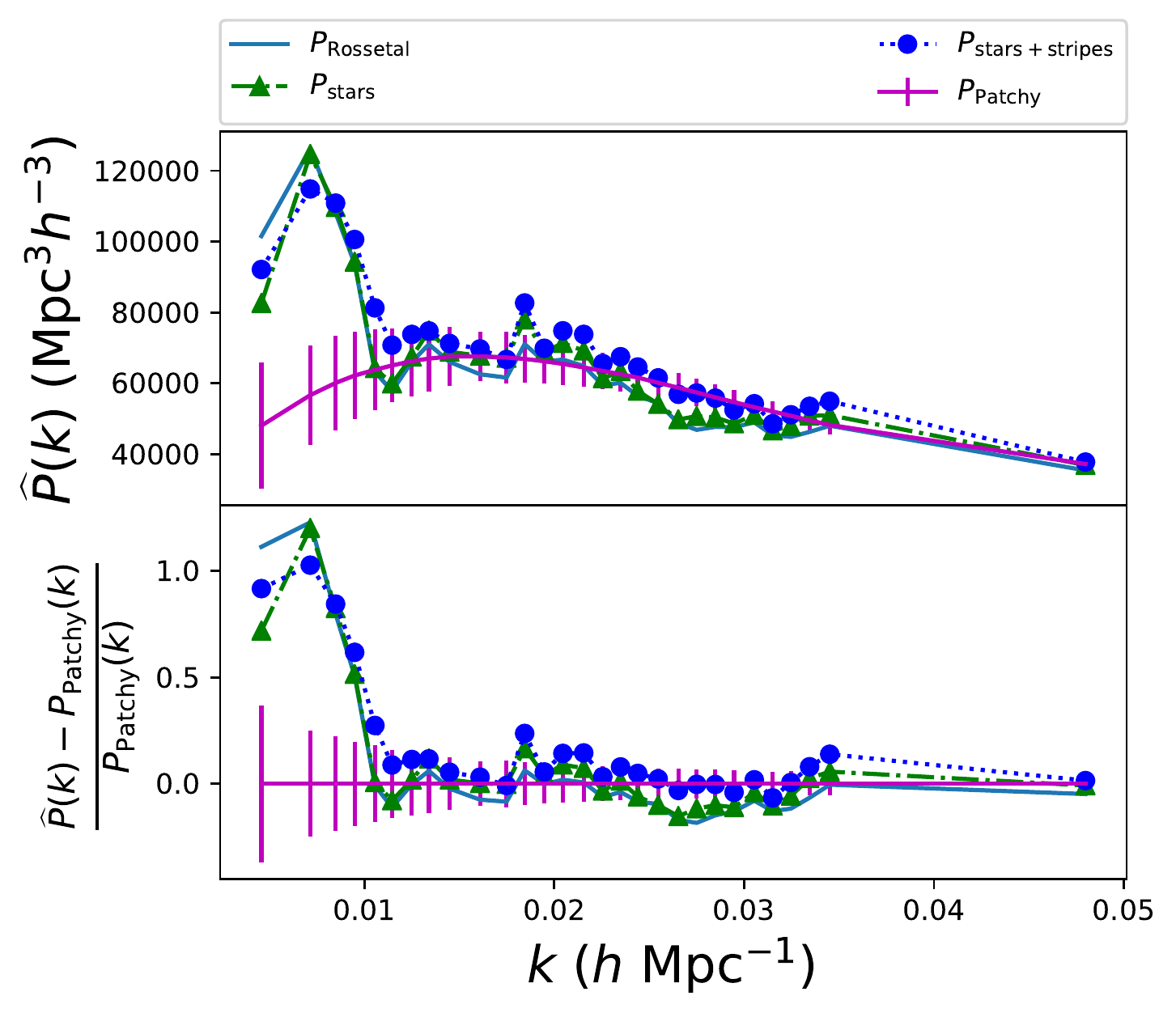}
    \caption{Power spectra of BOSS DR 12 
    CMASS NGC data, using mode subtraction to mitigate
    the effect of foreground
    stars (green), as well as stars and possible stripe
    dependent effects (dotted blue), compared to
the power using \citet{Rossinprep}'s stellar weights (solid blue) and the
	    average MultiDark-Patchy power spectrum (magenta).}
    \label{fig:P_starsandstripes}
\end{figure}

We use Eq. \eqref{eq:multconttemp} to build templates for each scanning stripe $\eta_A$ where
everything within the scanning stripe can be mitigated against, but not between stripes, i.e.
\begin{equation}
	E_A(\x)=\begin{cases}
    	1\text{, if }\x\in\eta_A,\\
		0\text{, else.}
	\end{cases}
\end{equation}
Applying these templates causes a smoothing of the
power spectrum (cf. Fig. \ref{fig:P_starsandstripes}) that can be explained by the fact that the 
stripe templates affect short 
scales perpendicular to the scanning stripes and long scales along the stripes.
The changes are less than the sample variance of the mock 
power spectra and therefore leave us with the large scale excess. The stripe templates also remove a dip
in the power spectrum compared to the MultiDark-Patchy power
spectrum at around $k\approx 0.27\; h\; \mathrm{Mpc}
^{-1}$ that is also present e.g. in the power 
spectrum monopole used for the redshift space 
distortion measurements by 
\citet{Gil-Marin:2015sqa}. Still, most of their 
signal comes from scales that are not accessible 
with the coarse grid that we use here, thus, their
results are likely to be unaffected by the stripe 
templates.

\section{Phenomenological Templates}
\label{sec:pheno}

The large-scale offset of the BOSS DR12 CMASS NGC power spectrum compared to what we expect from
the MultiDark-Patchy mocks could be due to new or not fully understood physics. However, it is at least as
likely instead to
be due to unknown systematics. In this Section, we explore phenomenological templates that we
generate without any particular source of systematic data contamination in mind.

\subsection{Templates Based on Spherical Harmonics Analyses}
\label{sec:alm}

\begin{figure}
	\centering
    \includegraphics[width=\columnwidth]{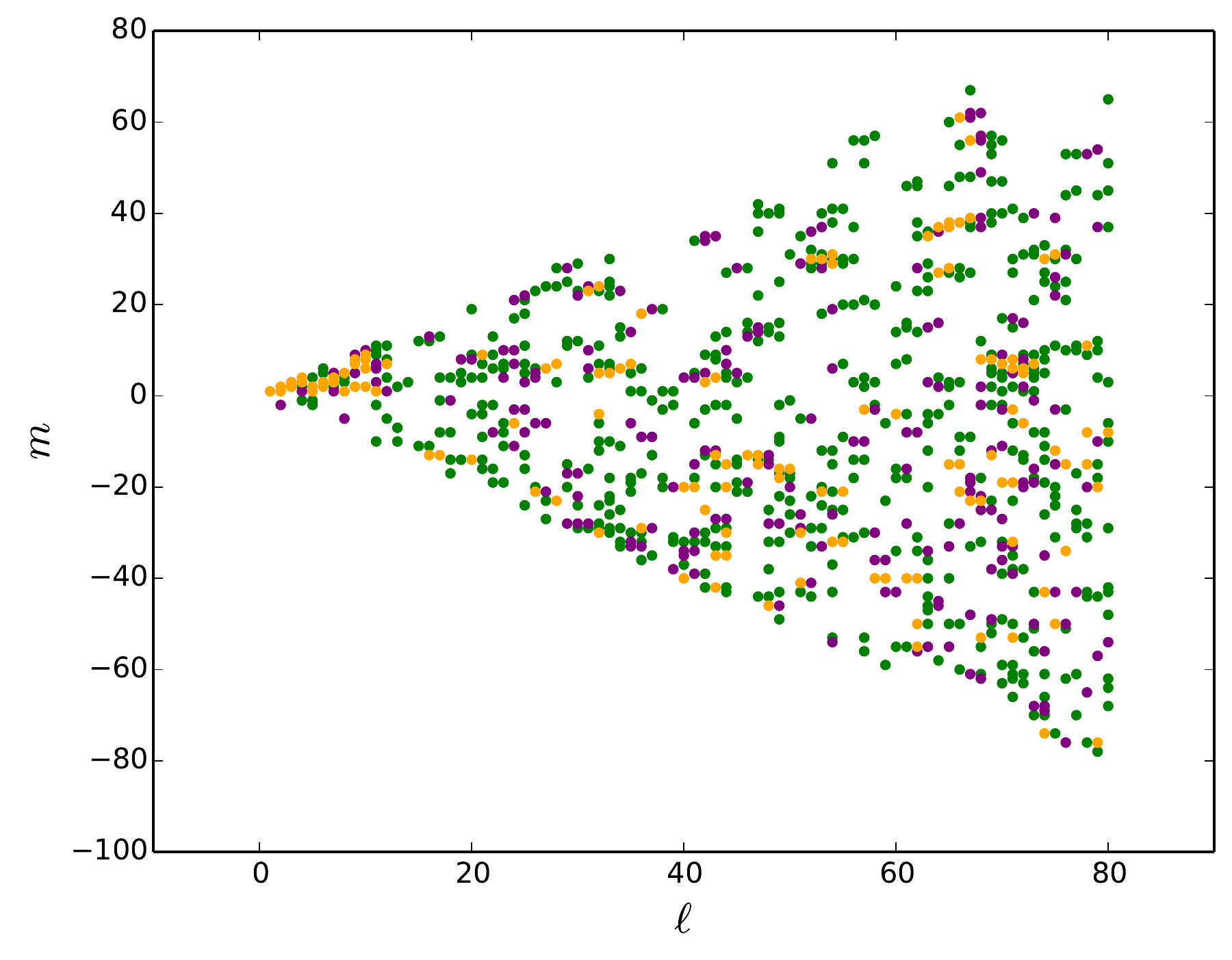}
    \caption{Graphical representation of multipoles 
    $(\ell,m)$ whose measured values of $a_{\ell m}$ is
    with $2\sigma$ (green), $3\sigma$ (blue) or more than 
    $4\sigma$ (red) discrepant 
    with the average value obtained
    from MultiDark-Patchy mocks.}
    \label{fig:almplot}
\end{figure}

As most systematics are expected to affect the data only 
in different angular directions, i.e.,
not radially, we start with a spherical harmonics decomposition of the data and the mocks. We 
average the density field along each line-of-sight (LOS) to obtain a density map
\begin{equation}
	F_\mathrm{map}(\delta,\alpha)\equiv 
    \frac{1}{r_\mathrm{max}-r_\mathrm{min}}\int_{r_\mathrm{min}}^{r_\mathrm{max}}\d r\;
    F(r,\delta,\alpha),
\end{equation}
where $F(r,\delta,\alpha)$ is the density field at distance $r$, declination $\delta$ and 
right ascension $\alpha$. This map can then be decomposed as
\begin{equation}
	F_\mathrm{map}(\delta,\alpha)=\sum_{\ell=0}^\infty\sum_{m=-\ell}^\ell a_{\ell m}Y_{\ell m}
    (\delta,\alpha)
\end{equation}
using normalised spherical harmonics $Y_{\ell m}(\delta,\alpha)$ and coefficients $a_{\ell m}$
that we estimate using the HEALPix software package as the 
following sum over all $N_\mathrm{pix}=12 N_\mathrm{side}^2$ 
HEALPix-pixels \citep{Gorski:2004by}:
\begin{equation}
	a_{\ell m}=\frac{4\pi}{N_\mathrm{pix}}\sum_{p=0}^{N_\mathrm{pix}-1}
    Y_{\ell m}^\ast(p) F_\mathrm{map}(p).
    \label{eq:alm}
\end{equation}
We compute Eq. \eqref{eq:alm} for both the data (after applying stellar templates)
and all MultiDark-Patchy mocks. This allows us to identify 
multipoles at which the data $a_{\ell m}$ is discrepant
with the distribution of the respective mock 
$a_{\ell m}$. These multipoles are represented by a dot
in Fig. \ref{fig:almplot}. At small scales ($\ell\geq 
12$),
we see that these appear randomly distributed, whereas at large 
scales ($\ell<12$), we see a large concentration of
multipoles that are more than $4\sigma$ away from the
expected $a_{\ell m}$ from the MultiDark-Patchy mocks. 
Interestingly, these are all at positive $m$.

\begin{figure}
	\centering
    \includegraphics[width=\columnwidth]{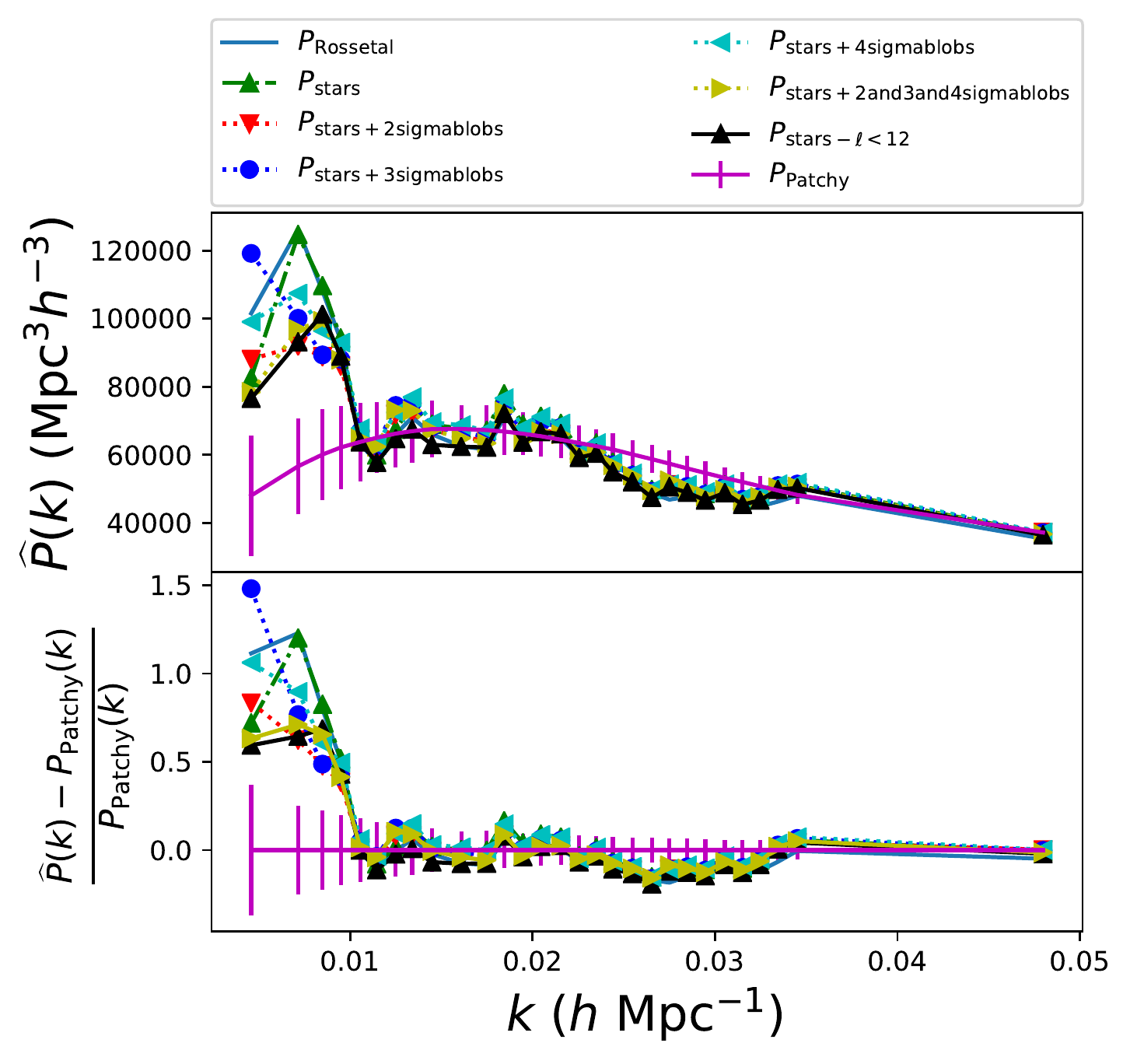}
    \caption{Power spectra of BOSS DR 12 
    CMASS NGC data, using 
    mode subtraction to mitigate the effect of foreground
    stars (green), as well as stars and the phenomenological templates of Eq. 
    \eqref{eq:blobmap} with a $2\sigma$ (red), $3\sigma$ (blue) or $4\sigma$ (cyan) threshold,
    as well as all templates combined (yellow), and a template that removes all modes at 
    $\ell<12$. We compare these with 
    the power using \citet{Rossinprep}'s stellar weights
    (solid blue) and to the Patchy power (magenta).}
    \label{fig:P_stars+blobs}
\end{figure}

We use the information contained in Fig. 
\ref{fig:almplot} to generate our first phenomenological 
templates: choosing a significance threshold,
we do not include 
multipoles for which the significance of the discrepancy
between data and mocks is less than the threshold, but
we include the ones exceeding the threshold by inserting
\begin{equation}
	E_A(z,\alpha,\delta)=
    \sum_{\ell m\; \mathrm{significant}} 
    \left(a_{\ell m}^\mathrm{(data)}-
    \left\langle a_{\ell m}^\mathrm{(mocks)}
    \right\rangle\right)
    Y_{\ell m}(\delta,\alpha)
    \label{eq:blobmap}
\end{equation}
into Eq. \eqref{eq:multconttemp}. Fig. \ref{fig:blobmaps}
shows the maps corresponding
to such a contaminant for a $2\sigma$, $3\sigma$ and
$4\sigma$ threshold.  All maps show that the centre of the survey footprint is over-dense 
compared to the mocks. In the $4\sigma$-map, an under-dense ring around the edge becomes more
prominent which could be due to unknown galactic effects, or might hint that our treatment of
stars could be improved, e.g. by a more thorough 
treatment of the error on the templates. In Fig. 
\ref{fig:P_stars+blobs},
we plot the data power spectrum after applying these
phenomenological templates. With the $2\sigma$-template,
the power offset in the first two bins halves, but for 
the $3\sigma$- and $4\sigma$-templates, this is not the
case. Applying all 3 phenomenological templates, the
power in the first bin is further reduced compared to
just applying the $2\sigma$-template, however, in the
third bin, we almost see the same power spectrum as if
we do not apply the phenomenological templates.

Furthermore, we generate templates 
\begin{equation}
	E_A(z,\alpha,\delta)=
    \sum_{\ell=0}^{11}\sum_{m=-\ell}^\ell
    \left(a_{\ell m}^\mathrm{(data)}-
    \left\langle a_{\ell m}^\mathrm{(mocks)}
    \right\rangle\right)
    Y_{\ell m}(\delta,\alpha)
    \label{eq:lowltemp}
\end{equation}
that remove all 
angular modes at $\ell<12$. Applying this template 
yields almost the same power spectrum as applying all
three previous phenomenological templates. This suggests
that something alters the power spectrum along the LOS because Eq. \eqref{eq:lowltemp} removes
all angular modes at multipoles where the data and the mocks are inconsistent with each other.
This could be new or not well understood physics, or a new type of unknown systematic, which
would require rethinking the common assumption that foregrounds effects are purely angular.

\subsection{Cross-correlating Redshift Shells}
\label{sec:shells}

\begin{figure}
	\centering
    \includegraphics[width=\columnwidth]{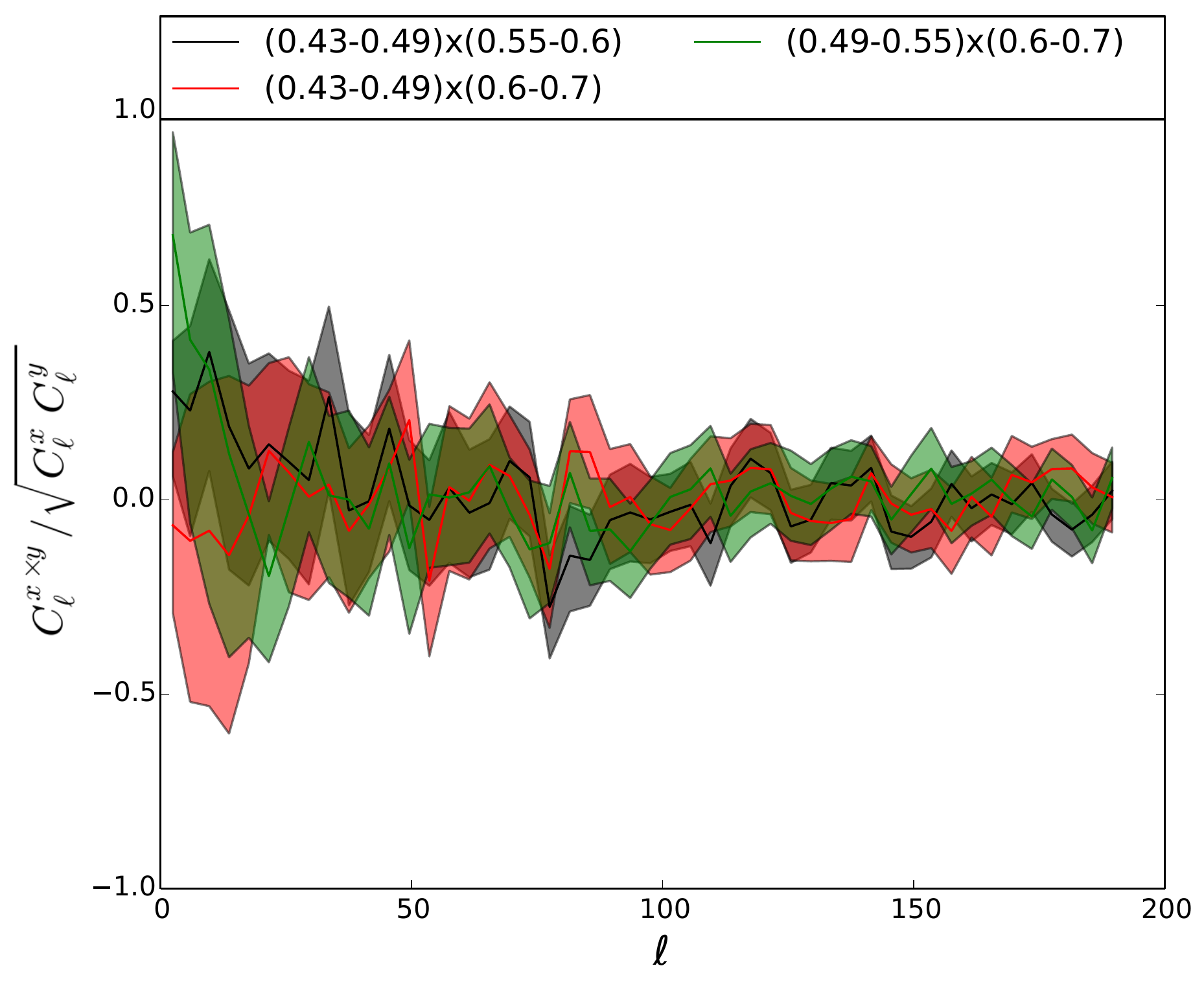}
    \caption{Cross-correlations between BOSS CMASS redshift shells. We 
    applied bandpowering with width 
    $\Delta\ell=4$. The error bars were obtained by cross-correlating
    the same redshift shells in all MultiDark-Patchy mocks.}
    \label{fig:Cl_BOSS_RedshiftXCorr}
\end{figure}

One can access the information encoded in the radial modes by only considering those modes when computing the power spectrum. In order to not lose information, here we use the projected angular power spectrum of the cross-correlation of non-overlapping redshift bins. Given a wide enough separation between the two subsamples (so that the density correlations are negligible), the only physical correlation arises due to magnification effects. Nonetheless, radial variations in observing conditions or foreground contaminants may also correlate the subsamples. Therefore, in this section we test
whether the remaining
offset in the power spectrum is a foreground angular 
contamination, or whether it is a 
cosmic signal. A foreground contaminant would affect
all redshift slices in a similar way and we would therefore see a 
strong correlation between different shells at the 
same scales. Therefore, as suggested e.g. by 
\citet{Ho:2012vy,Pullen:2012rd} and \citet{Agarwal:2013ajb}, the angular cross power 
spectrum
\begin{equation}
	C_\ell^{(xy)}\equiv \frac{1}{2\ell+1}\sum_{m=-\ell}^\ell 
    \left(a_{\ell m}^{(x)}\right)^\ast a_{\ell m}^{(y)}
\end{equation}
between redshift shells $x$ and $y$
can be used to characterise unknown systematics.
In Fig. \ref{fig:Cl_BOSS_RedshiftXCorr}, we quantify 
cross-correlations between redshift shells by
the correlation coefficient
\begin{equation}
	C_\ell^{(xy)}/\sqrt{C_\ell^{(xx)} C_\ell^{(yy)}}.
\end{equation}
To estimate by-chance
correlations, one can do the same cross-correlation
studies to the mock catalogues. We choose 4 radial bins in a way that they contain the same 
number of objects, thus the first bin extends from redshifts of
0.43 to 0.49, the second until 0.55, the third until 0.6 and the
fourth up to 0.7. Fig. \ref{fig:Cl_BOSS_RedshiftXCorr} shows that, as 
expected, at scales ($\ell\geq 12$), there is no significant cross-correlation between 
non-adjacent shells. However, the second and fourth shells are strongly correlated at large
scales and the first and third only mildly. We do not see any evidence of 
cross-correlation between the first and fourth shell.

\begin{figure}
	\centering
    \includegraphics[width=\columnwidth]{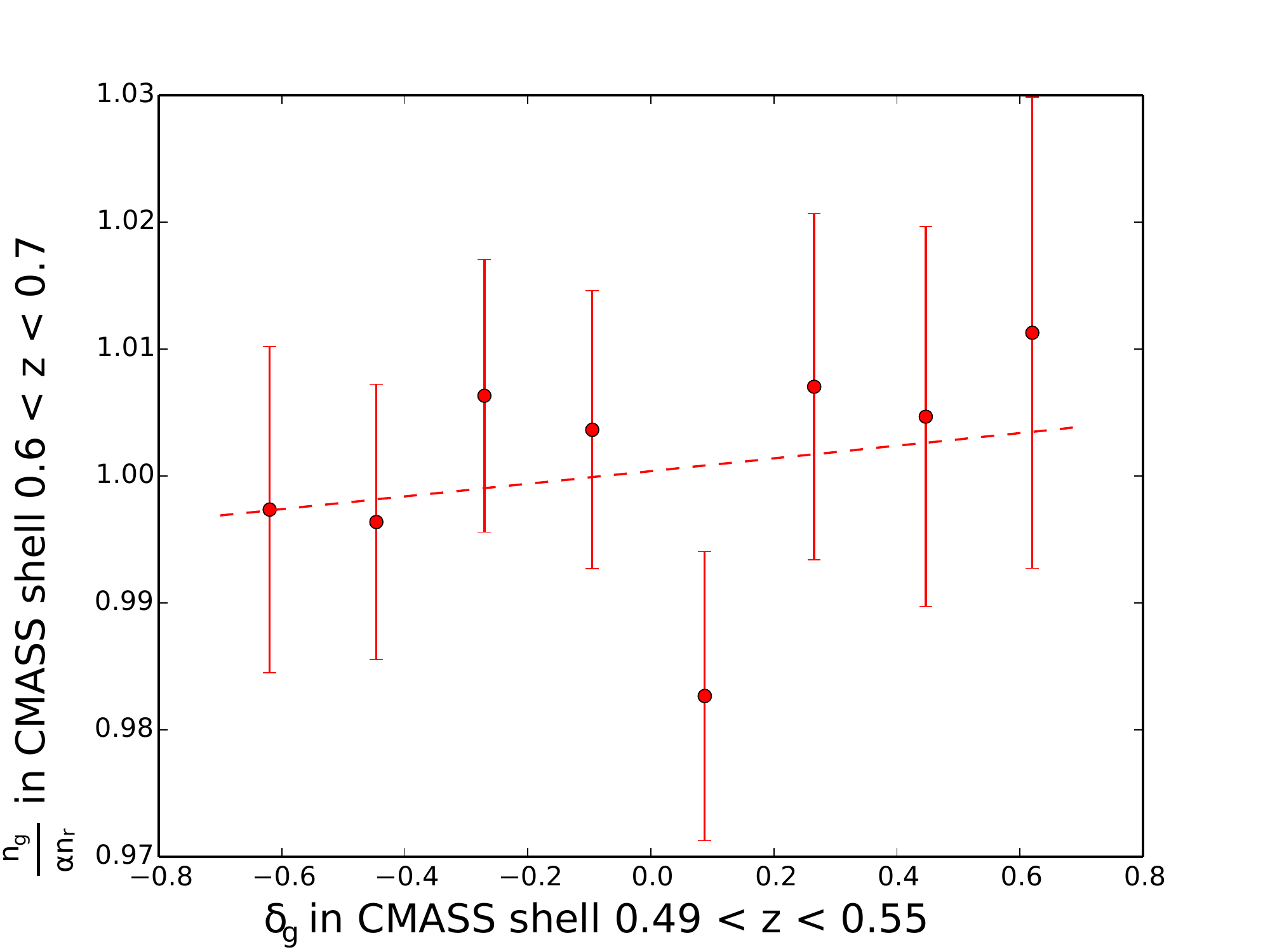}
    \caption{Plot similar to Fig. \ref{fig:nbarbynstars} but with the
    galaxy over-density $\delta_\mathrm{g}$ in the redshift shell 
    between $0.49<z<0.55$ as the foreground and 
    considering only galaxies between $0.6<z<0.7$ for 
    $n_\mathrm{g}/\langle n_\mathrm{g}\rangle$.}
    \label{fig:nbarbynCMASS}
\end{figure}

As we saw a strong correlation between the second and fourth radial bins, we test how the
ratio of observed versus expected galaxy number density 
$n_\mathrm{g}/\langle n_\mathrm{g}\rangle$ in the fourth shell changes with respect to the 
foreground galaxy over-density $\delta_\mathrm{g}$ (cf. Fig. \ref{fig:nbarbynCMASS}). For 
almost all values of $\delta_\mathrm{g}$ in the second bin, we see the expected amount of 
objects in the fourth bin. A template built in this way is therefore not significantly different 
from zero and, therefore, does not influence the power
spectrum measurement significantly.

\subsection{Cross-correlating LOWZ and CMASS}

After all the mode subtraction discussed above, we still find no angular contaminant which causes the remaining excess in the measured power spectrum at very large scales with respect to the one computed from the mocks. Surprisingly, the scales at which this excess is located coincide with the largest radial scales of the volume covered by the CMASS catalogue. This is why it is possible that such deviation appears only in radial modes. Therefore, and as we have checked that there is no consistent correlation between different combinations of sub-samples of CMASS separated by redshift, here we test if this excess can be explained with cosmic magnification.

Cosmic magnification due to foreground galaxies affects background galaxy number counts in two competing ways. On one hand, the space behind the lens is stretched, so the background number density decreases. On the other hand, as background sources are magnified, faint galaxies may surpass the detection threshold, which otherwise would have remained undetected. The net effect is then accounted for in the magnification bias, which depends on the specific background sample. Although the 
magnification signal does not strongly depend on the 
redshift of the background sources, it may have affected the targeting strategy and contaminate the selection procedure for CMASS, including more galaxies than expected in the faint end of the galaxy population, which is most probably the galaxies with the highest redshifts. Galaxy mocks assume Newtonian gravity but magnification is a relativistic effect, hence we can test if the CMASS sample has a significant amount of magnified galaxies which would not have been targeted otherwise by comparing the cross-correlation of CMASS galaxies with foreground catalogues with the corresponding mocks. In order to avoid the introduction of different assumptions or systematics in this analysis, we choose the BOSS LOWZ sample (spectroscopic as well) as our foreground sample.

\begin{figure}
	\centering
    \includegraphics[width=\columnwidth]{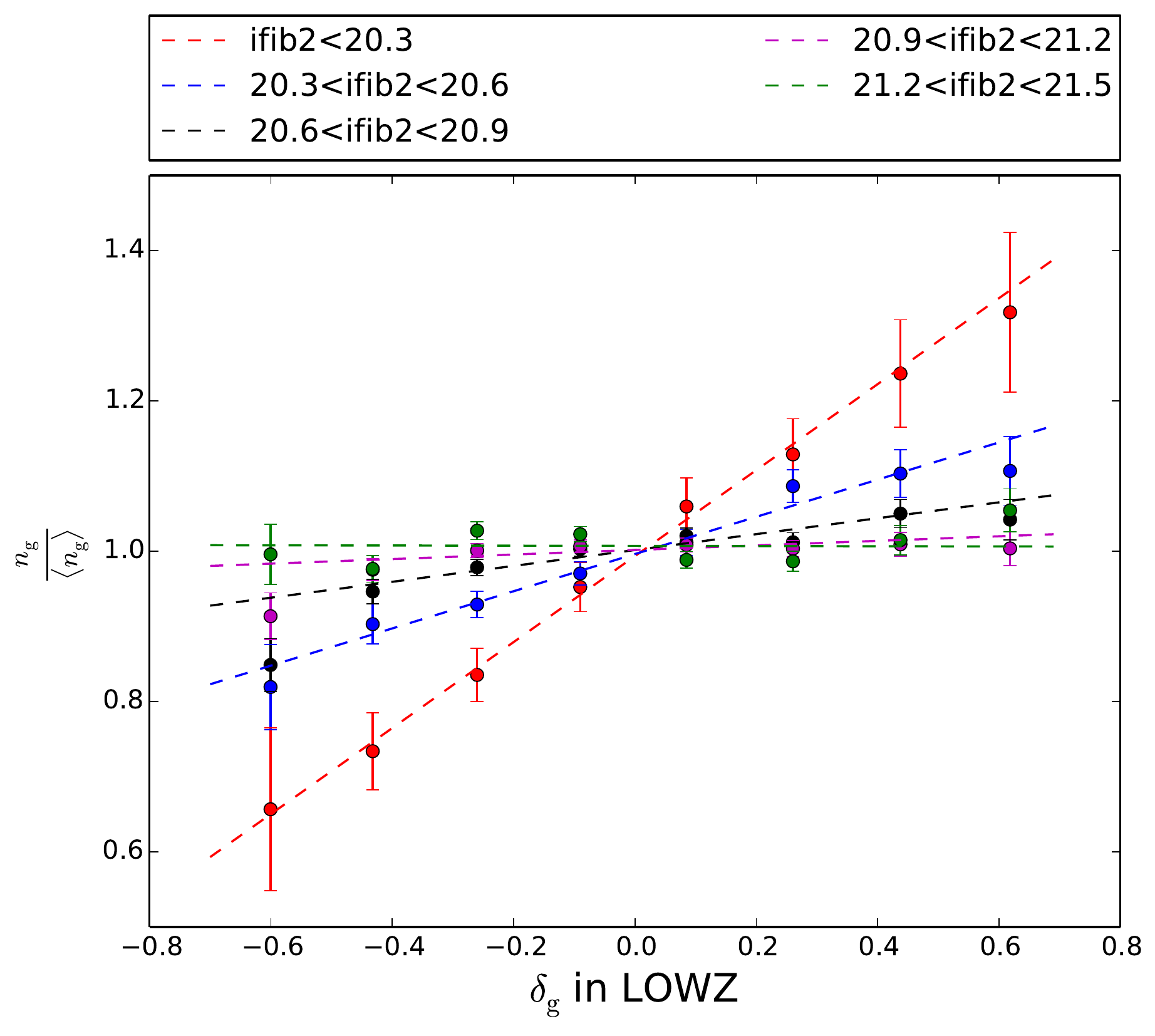}
    \includegraphics[width=\columnwidth]{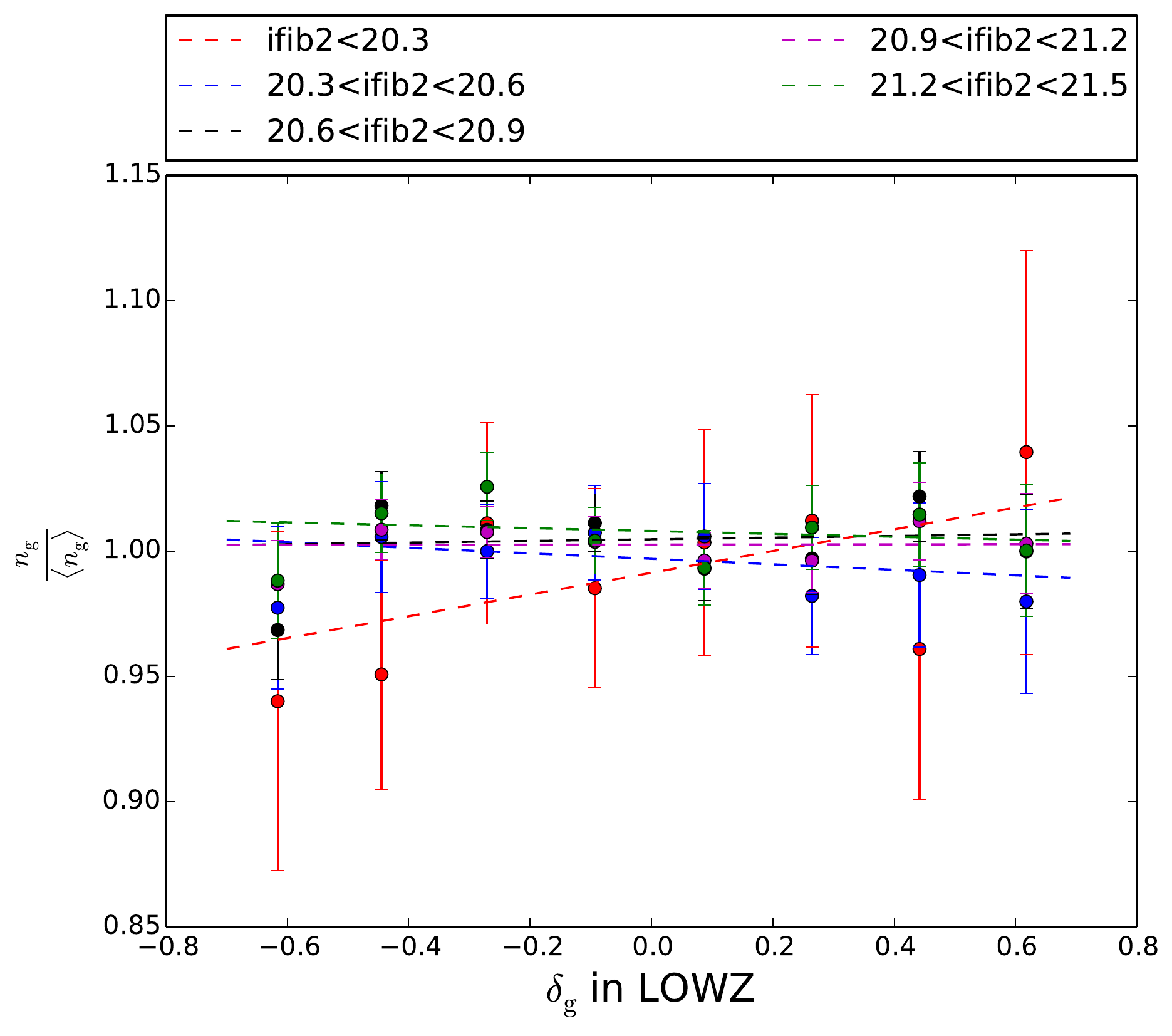}
    \caption{Plots similar to Fig. \ref{fig:chi2fit} but
    with the LOWZ over-density field as the foreground.
    In the top panel, we used the whole LOWZ sample as
    the foreground, whereas in the bottom panel, only
    galaxies at redshifts $z<0.29$ have been considered.}
    \label{fig:LOWZ}
\end{figure}

In Fig. \ref{fig:LOWZ} we show the ratio between the observed and the expected number counts as function of the number over-density in the LOWZ sample. We  
split the CMASS data again by $i_\mathrm{fib2}$, 
even though it is regarded as a measure of surface 
brightness, which is unaffected by lensing. However,
as the aperture covers most of the galaxy flux (both 
before and after it is lensed), $i_\mathrm{fib2}$
will catch extra photons from the 
magnification, i.e. even though surface brightness 
is conserved, the number of subpixels illuminated by 
this surface brightness increases. Indeed,
we find that significantly fewer 
very bright galaxies with $i$-fibre 
magnitudes $i_\mathrm{fib2}<20.6$ are observed behind
under-dense regions in LOWZ, and significantly more 
behind over-densities. However, such bright galaxies are
rare and have no effect on any template because the
average magnitude of even the closest CMASS galaxies (cf.
Fig. \ref{fig:ifibvsredplot}) is
fainter than the galaxies showing the effect. Due to 
their rareness, we also do not see any sizeable effect
when applying the classic galaxy-by-galaxy
weighting scheme. 
Moreover, the brightest galaxies in CMASS are the closest ones to us, so it is likely that this 
positive correlation has a clustering origin rather than being due to magnification. This is 
further supported by the fact that, if we restrict the LOWZ foreground to $z<0.29$, the 
cross-correlation is insignificant, suggesting that the significant cross-correlation visible in
the top panel of Fig. \ref{fig:LOWZ} is due to clustering between LOWZ galaxies at $z>0.29$ and
CMASS galaxies at low redshifts.

In order to rule out cosmic magnification as the origin of the excess in the power spectrum, a more comprehensive analysis comparing the results using photometric and spectroscopic catalogues as well as using different galaxy populations as foreground lenses is required. However, this study is beyond the scope of this work and is left for future research.

\section{Conclusions}

\begin{figure*}
	\centering
    \includegraphics[width=\textwidth]{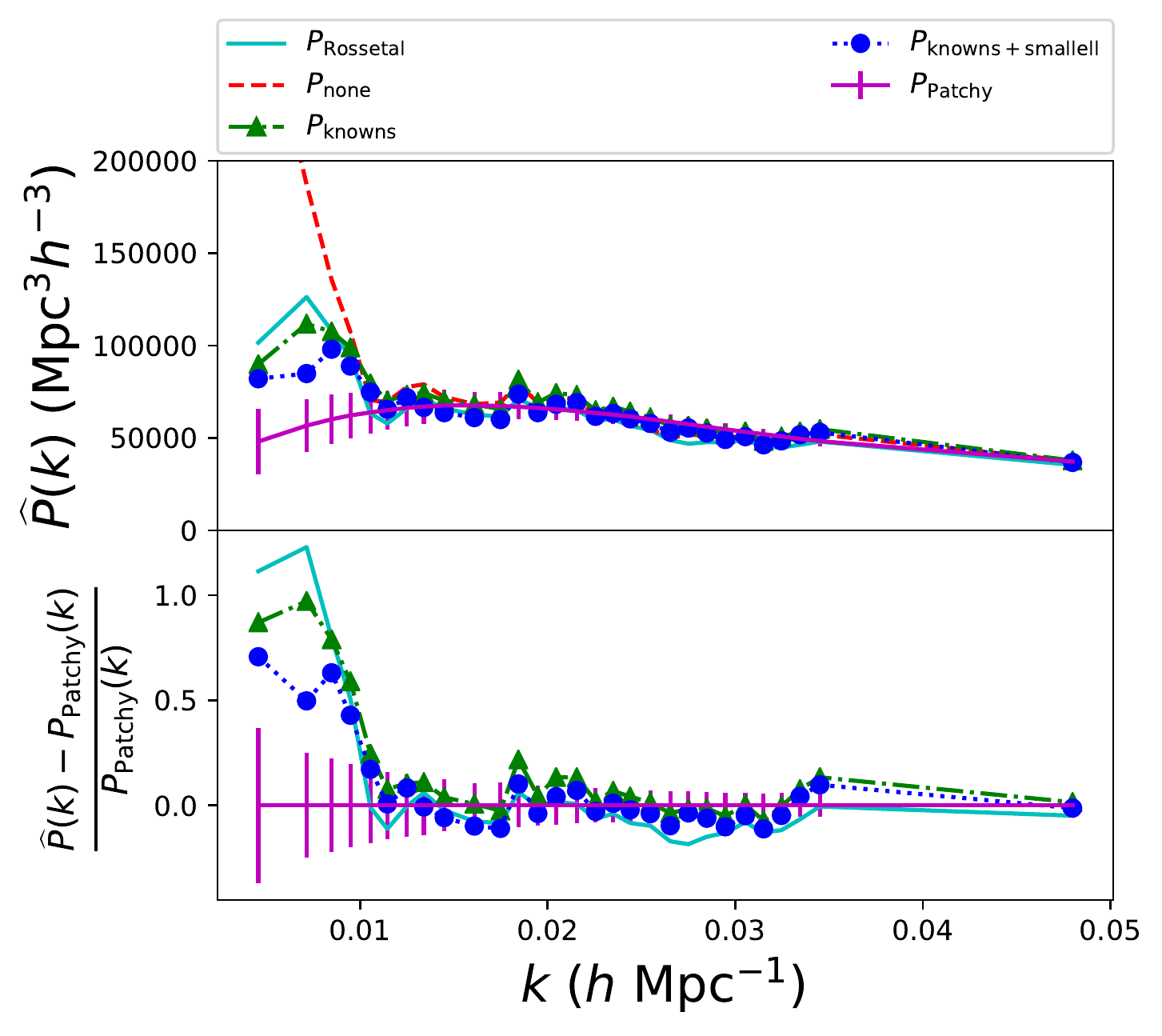}
    \caption{Power spectrum of BOSS DR 12 CMASS NGC data 
    after applying weights for all known systematics
    (stellar density, seeing, airmass, extinction,
    scanning stripes; to third order where 
    applicable) presented in Sec. 
    \ref{sec:application} (dot-dashed 
    green), as well as after additionally applying the 
    template removing all angular modes at $\ell<12$ (dotted
    blue). These
    are compared to the power spectrum we obtain using the
    same data but applying \citet{Rossinprep}'s corrective
    weights for stars and seeing (solid cyan) and not 
    applying any correction for systematic modes (dashed 
    red). The magenta line represents the average power 
    spectrum of 2049 MultiDark-Patchy realisations.}
    \label{fig:conclusionplot}
\end{figure*}

We have presented a practical approach to decontamination
using mode subtraction (cf. \ref{sec:practicalapprsingle} and 
\ref{sec:practicalapprmulti}). In Sec. \ref{sec:application}, we
generated templates to mitigate against
the effect of foreground stars, seeing, airmass,
galactic extinction and the SDSS scanning stripes. We 
applied these to the final SDSS-III BOSS CMASS NGC 
sample. We have found that mode subtraction mitigates
against systematic contaminants at least as well as
deriving and applying corrective weights to the observed
objects. As with the corrective weighting, we measure a
large-scale excess beyond the power spectrum expected
from the standard $\Lambda$CDM cosmology with Gaussian
initial density fluctuations. This excess is only 
present at scales that are much larger than the BAO 
scale, thus leaving the main results of BOSS unaffected.
In Fig. \ref{fig:conclusionplot}, we show that after
applying our template based approach the power spectrum is 
slightly reduced compared to after applying corrective 
weights. This is because
of a small correlation of the observed galaxy density field
with the scanning stripes of SDSS, which has previously not
been addressed.

We further tested our methodology by building a range of
phenomenological templates. Templates built on a comparison 
of a spherical
harmonics decomposition of the data with the distribution
of the decomposed mock catalogue data reduce the 
large-scale power offset (cf. Section \ref{sec:alm}).
However, after applying these
phenomenological templates, our large-scale power 
spectrum measurements are still discrepant with the 
average mock power spectrum (cf. Fig. \ref{fig:conclusionplot}). 
Thus, the excess signal is not only coming from
angular modes, but there might be
a contaminant, or a physical
effect, that amplifies the power spectrum along the LOS.
The power spectra presented in Fig. 
\ref{fig:P_stars+blobs} and \ref{fig:conclusionplot} have 
only been computed to test the Mode Subtraction method; they 
should not be
interpreted as a true measurement, as by using the mock
catalogues to generate our templates, we have already partially
assumed what we expect. 

In Sec. \ref{sec:shells}, we generate further
phenomenological templates based on cross-correlations 
between the CMASS and LOWZ, the other BOSS galaxy sample at 
lower redshifts, and between redshift shells within the CMASS 
sample. None of these templates have any sizeable effect on
the resulting power spectrum. We therefore do not yet have a satisfactory explanation for the 
large scale power spectrum excess, which we leave for future work.

In this work, we have shown that MOSES provides power 
spectrum measurements that are consistent with measurements
obtained using corrective weights. The computationally most
expensive part of MOSES is the generation of the templates.
This is done in the same way that \citet{Rossinprep} generate
their corrective weights, and when already having weights
at hand, Eq. \eqref{eq:farbitrary} provides a straightforward conversion of the weights into templates. As MOSES
is built to ensure an unbiased estimate of the power 
spectrum, and as one can easily introduce more templates to
explore more functional shapes of the contaminant effect, we 
encourage the use of MOSES.

\section*{Acknowledgements}

The authors would like to thank Hector Gil-Mar\'in for his 
support when using the MultiDark-Patchy mock
catalogues, and Michael Brown and Rob Crittenden for their valuable comments.

We acknowledge support from European Union's Horizon 2020 research and innovation programme ERC (BePreSySe, grant agreement 725327), Spanish MINECO under projects AYA2014-58747-P AEI/FEDER, UE, and MDM-2014-0369 of ICCUB (Unidad de Excelencia Mar\'ia de Maeztu). WJP acknowledges support from the European Research Council through the Darksurvey grant 614030. DB is supported by STFC consolidated grant ST/N000668/1. LS acknowledges support from NASA grant
12-EUCLID11-0004, and the DOE grant DE-SC0011840. JLB is supported by the
Spanish MINECO under grant BES-2015-071307, co-funded by the ESF.

We use publicly available SDSS-III data. Funding for SDSS-III has been provided by the Alfred P. Sloan Foundation, the Participating Institutions, the National Science Foundation, and the U.S. Department of Energy Office of Science. The SDSS-III web site is http://www.sdss3.org/.

SDSS-III is managed by the Astrophysical Research Consortium for the Participating Institutions of the SDSS-III Collaboration including the University of Arizona, the Brazilian Participation Group, Brookhaven National Laboratory, Carnegie Mellon University, University of Florida, the French Participation Group, the German Participation Group, Harvard University, the Instituto de Astrofisica de Canarias, the Michigan State/Notre Dame/JINA Participation Group, Johns Hopkins University, Lawrence Berkeley National Laboratory, Max Planck Institute for Astrophysics, Max Planck Institute for Extraterrestrial Physics, New Mexico State University, New York University, Ohio State University, Pennsylvania State University, University of Portsmouth, Princeton University, the Spanish Participation Group, University of Tokyo, University of Utah, Vanderbilt University, University of Virginia, University of Washington, and Yale University.

Some of the results in this paper have been derived using 
the HEALPix \citep{Gorski:2004by} package.
We used matplotlib \citep{Hunter:2007ouj}
to generate plots. We made use of the facilities and
staff of the UK Sciama High Performance Computing 
cluster supported
by the ICG, SEPNet and the University of Portsmouth.




\bibliographystyle{mnras}
\bibliography{Refs} 

\begin{thebibliography}{}
\makeatletter
\relax
\def\mn@urlcharsother{\let\do\@makeother \do\$\do\&\do\#\do\^\do\_\do\%\do\~}
\def\mn@doi{\begingroup\mn@urlcharsother \@ifnextchar [ {\mn@doi@}
  {\mn@doi@[]}}
\def\mn@doi@[#1]#2{\def\@tempa{#1}\ifx\@tempa\@empty \href
  {http://dx.doi.org/#2} {doi:#2}\else \href {http://dx.doi.org/#2} {#1}\fi
  \endgroup}
\def\mn@eprint#1#2{\mn@eprint@#1:#2::\@nil}
\def\mn@eprint@arXiv#1{\href {http://arxiv.org/abs/#1} {{\tt arXiv:#1}}}
\def\mn@eprint@dblp#1{\href {http://dblp.uni-trier.de/rec/bibtex/#1.xml}
  {dblp:#1}}
\def\mn@eprint@#1:#2:#3:#4\@nil{\def\@tempa {#1}\def\@tempb {#2}\def\@tempc
  {#3}\ifx \@tempc \@empty \let \@tempc \@tempb \let \@tempb \@tempa \fi \ifx
  \@tempb \@empty \def\@tempb {arXiv}\fi \@ifundefined
  {mn@eprint@\@tempb}{\@tempb:\@tempc}{\expandafter \expandafter \csname
  mn@eprint@\@tempb\endcsname \expandafter{\@tempc}}}

\bibitem[\protect\citeauthoryear{Abazajian et~al.}{Abazajian
  et~al.}{2004}]{Abazajian:2004aja}
Abazajian K.,  et~al., 2004, \mn@doi [Astron. J.] {10.1086/421365}, 128, 502

\bibitem[\protect\citeauthoryear{Agarwal et~al.}{Agarwal
  et~al.}{2014}]{Agarwal:2013ajb}
Agarwal N.,  et~al., 2014, \mn@doi [JCAP] {10.1088/1475-7516/2014/04/007},
  1404, 007

\bibitem[\protect\citeauthoryear{Alam et~al.}{Alam et~al.}{2015}]{Alam:2015mbd}
Alam S.,  et~al., 2015, \mn@doi [Astrophys. J. Suppl.]
  {10.1088/0067-0049/219/1/12}, 219, 12

\bibitem[\protect\citeauthoryear{Dawson et~al.}{Dawson
  et~al.}{2013}]{Dawson:2012va}
Dawson K.~S.,  et~al., 2013, \mn@doi [Astron. J.] {10.1088/0004-6256/145/1/10},
  145, 10

\bibitem[\protect\citeauthoryear{Eisenstein et~al.}{Eisenstein
  et~al.}{2011}]{Eisenstein:2011sa}
Eisenstein D.~J.,  et~al., 2011, \mn@doi [Astron. J.]
  {10.1088/0004-6256/142/3/72}, 142, 72

\bibitem[\protect\citeauthoryear{Elsner, Leistedt  \& Peiris}{Elsner
  et~al.}{2016}]{Elsner:2015aga}
Elsner F.,  Leistedt B.,   Peiris H.~V.,  2016, \mn@doi [Mon. Not. Roy. Astron.
  Soc.] {10.1093/mnras/stv2777}, 456, 2095

\bibitem[\protect\citeauthoryear{Elsner, Leistedt  \& Peiris}{Elsner
  et~al.}{2017}]{Elsner:2016bvs}
Elsner F.,  Leistedt B.,   Peiris H.~V.,  2017, \mn@doi [Mon. Not. Roy. Astron.
  Soc.] {10.1093/mnras/stw2752}, 465, 1847

\bibitem[\protect\citeauthoryear{Feldman, Kaiser  \& Peacock}{Feldman
  et~al.}{1994}]{Feldman:1993ky}
Feldman H.~A.,  Kaiser N.,   Peacock J.~A.,  1994, \mn@doi [Astrophys. J.]
  {10.1086/174036}, 426, 23

\bibitem[\protect\citeauthoryear{Gil-Marín et~al.}{Gil-Marín
  et~al.}{2016}]{Gil-Marin:2015sqa}
Gil-Marín H.,  et~al., 2016, \mn@doi [Mon. Not. Roy. Astron. Soc.]
  {10.1093/mnras/stw1096}, 460, 4188

\bibitem[\protect\citeauthoryear{Gorski, Hivon, Banday, Wandelt, Hansen,
  Reinecke  \& Bartelman}{Gorski et~al.}{2005}]{Gorski:2004by}
Gorski K.~M.,  Hivon E.,  Banday A.~J.,  Wandelt B.~D.,  Hansen F.~K.,
  Reinecke M.,   Bartelman M.,  2005, \mn@doi [Astrophys. J.] {10.1086/427976},
  622, 759

\bibitem[\protect\citeauthoryear{Gunn et~al.}{Gunn et~al.}{2006}]{Gunn:2006tw}
Gunn J.~E.,  et~al., 2006, \mn@doi [Astron. J.] {10.1086/500975}, 131, 2332

\bibitem[\protect\citeauthoryear{Hivon, Gorski, Netterfield, Crill, Prunet  \&
  Hansen}{Hivon et~al.}{2002}]{Hivon:2001jp}
Hivon E.,  Gorski K.~M.,  Netterfield C.~B.,  Crill B.~P.,  Prunet S.,   Hansen
  F.,  2002, \mn@doi [Astrophys. J.] {10.1086/338126}, 567, 2

\bibitem[\protect\citeauthoryear{Ho, Hirata, Padmanabhan, Seljak  \&
  Bahcall}{Ho et~al.}{2008}]{Ho:2008bz}
Ho S.,  Hirata C.,  Padmanabhan N.,  Seljak U.,   Bahcall N.,  2008, \mn@doi
  [Phys. Rev.] {10.1103/PhysRevD.78.043519}, D78, 043519

\bibitem[\protect\citeauthoryear{Ho et~al.}{Ho et~al.}{2012}]{Ho:2012vy}
Ho S.,  et~al., 2012, \mn@doi [Astrophys. J.] {10.1088/0004-637X/761/1/14},
  761, 14

\bibitem[\protect\citeauthoryear{Hunter}{Hunter}{2007}]{Hunter:2007ouj}
Hunter J.~D.,  2007, \mn@doi [Comput. Sci. Eng.] {10.1109/MCSE.2007.55}, 9, 90

\bibitem[\protect\citeauthoryear{Jasche \& Lavaux}{Jasche \&
  Lavaux}{2017}]{Jasche:2017rze}
Jasche J.,  Lavaux G.,  2017, \mn@doi [Astron. Astrophys.]
  {10.1051/0004-6361/201730909}, 606, A37

\bibitem[\protect\citeauthoryear{Jasche \& Wandelt}{Jasche \&
  Wandelt}{2013}]{Jasche:2013lwa}
Jasche J.,  Wandelt B.~D.,  2013, \mn@doi [Astrophys. J.]
  {10.1088/0004-637X/779/1/15}, 779, 15

\bibitem[\protect\citeauthoryear{Jasche, Kitaura, Wandelt  \& Ensslin}{Jasche
  et~al.}{2010}]{Jasche:2009hx}
Jasche J.,  Kitaura F.~S.,  Wandelt B.~D.,   Ensslin T.~A.,  2010, \mn@doi
  [Mon. Not. Roy. Astron. Soc.] {10.1111/j.1365-2966.2010.16610.x}, 406, 60

\bibitem[\protect\citeauthoryear{Kalus, Percival, Bacon  \& Samushia}{Kalus
  et~al.}{2016}]{Kalus:2016cno}
Kalus B.,  Percival W.~J.,  Bacon D.,   Samushia L.,  2016, \mn@doi [Mon. Not.
  Roy. Astron. Soc.] {10.1093/mnras/stw2008}, 463, 467

\bibitem[\protect\citeauthoryear{Kitaura, Yepes  \& Prada}{Kitaura
  et~al.}{2014}]{Kitaura:2013cwa}
Kitaura F.-S.,  Yepes G.,   Prada F.,  2014, \mn@doi [Mon. Not. Roy. Astron.
  Soc.] {10.1093/mnrasl/slt172}, 439, 21

\bibitem[\protect\citeauthoryear{Kitaura, Gil-Mar\'in, Scoccola, Chuang,
  M{\"u}ller, Yepes  \& Prada}{Kitaura et~al.}{2015}]{Kitaura:2014mja}
Kitaura F.-S.,  Gil-Mar\'in H.,  Scoccola C.,  Chuang C.-H.,  M{\"u}ller V.,
  Yepes G.,   Prada F.,  2015, \mn@doi [Mon. Not. Roy. Astron. Soc.]
  {10.1093/mnras/stv645}, 450, 1836

\bibitem[\protect\citeauthoryear{Kitaura et~al.}{Kitaura
  et~al.}{2016}]{Kitaura:2015uqa}
Kitaura F.-S.,  et~al., 2016, \mn@doi [Mon. Not. Roy. Astron. Soc.]
  {10.1093/mnras/stv2826}, 456, 4156

\bibitem[\protect\citeauthoryear{Klypin, Yepes, Gottlober, Prada  \&
  Hess}{Klypin et~al.}{2016}]{Klypin:2014kpa}
Klypin A.,  Yepes G.,  Gottlober S.,  Prada F.,   Hess S.,  2016, \mn@doi [Mon.
  Not. Roy. Astron. Soc.] {10.1093/mnras/stw248}, 457, 4340

\bibitem[\protect\citeauthoryear{Leistedt \& Peiris}{Leistedt \&
  Peiris}{2014}]{Leistedt:2014wia}
Leistedt B.,  Peiris H.~V.,  2014, \mn@doi [Mon. Not. Roy. Astron. Soc.]
  {10.1093/mnras/stu1439}, 444, 2

\bibitem[\protect\citeauthoryear{Leistedt, Peiris, Mortlock, Benoit-L\'evy  \&
  Pontzen}{Leistedt et~al.}{2013}]{Leistedt:2013gfa}
Leistedt B.,  Peiris H.~V.,  Mortlock D.~J.,  Benoit-L\'evy A.,   Pontzen A.,
  2013, \mn@doi [Mon. Not. Roy. Astron. Soc.] {10.1093/mnras/stt1359}, 435,
  1857

\bibitem[\protect\citeauthoryear{Leistedt, Peiris  \& Roth}{Leistedt
  et~al.}{2014}]{Leistedt:2014zqa}
Leistedt B.,  Peiris H.~V.,   Roth N.,  2014, \mn@doi [Phys. Rev. Lett.]
  {10.1103/PhysRevLett.113.221301}, 113, 221301

\bibitem[\protect\citeauthoryear{Pullen \& Hirata}{Pullen \&
  Hirata}{2010}]{Pullen:2010zy}
Pullen A.~R.,  Hirata C.~M.,  2010, \mn@doi [JCAP]
  {10.1088/1475-7516/2010/05/027}, 1005, 027

\bibitem[\protect\citeauthoryear{Pullen \& Hirata}{Pullen \&
  Hirata}{2013}]{Pullen:2012rd}
Pullen A.~R.,  Hirata C.~M.,  2013, \mn@doi [Publ. Astron. Soc. Pac.]
  {10.1086/671189}, 125, 705

\bibitem[\protect\citeauthoryear{Reid et~al.}{Reid et~al.}{2016}]{Reid:2015gra}
Reid B.,  et~al., 2016, \mn@doi [Mon. Not. Roy. Astron. Soc.]
  {10.1093/mnras/stv2382}, 455, 1553

\bibitem[\protect\citeauthoryear{Rodríguez-Torres et~al.}{Rodríguez-Torres
  et~al.}{2016}]{Rodriguez-Torres:2015vqa}
Rodríguez-Torres S.~A.,  et~al., 2016, \mn@doi [Mon. Not. Roy. Astron. Soc.]
  {10.1093/mnras/stw1014}, 460, 1173

\bibitem[\protect\citeauthoryear{Ross et~al.}{Ross et~al.}{2011}]{Ross:2011cz}
Ross A.~J.,  et~al., 2011, \mn@doi [Mon. Not. Roy. Astron. Soc.]
  {10.1111/j.1365-2966.2011.19351.x}, 417, 1350

\bibitem[\protect\citeauthoryear{Ross et~al.}{Ross
  et~al.}{2013}]{Ross:2012sxfNL}
Ross A.~J.,  et~al., 2013, \mn@doi [Mon. Not. Roy. Astron. Soc.]
  {10.1093/mnras/sts094}, 428, 1116

\bibitem[\protect\citeauthoryear{Ross et~al.}{Ross et~al.}{2017}]{Rossinprep}
Ross A.~J.,  et~al., 2017, \mn@doi [Mon. Not. Roy. Astron. Soc.]
  {10.1093/mnras/stw2372}, 464, 1168

\bibitem[\protect\citeauthoryear{Rybicki \& Press}{Rybicki \&
  Press}{1992}]{Rybicki:1992jz}
Rybicki G.~B.,  Press W.~H.,  1992, \mn@doi [Astrophys. J.] {10.1086/171845},
  398, 169

\bibitem[\protect\citeauthoryear{Schlafly \& Finkbeiner}{Schlafly \&
  Finkbeiner}{2011}]{Schlafly:2010dz}
Schlafly E.~F.,  Finkbeiner D.~P.,  2011, \mn@doi [Astrophys. J.]
  {10.1088/0004-637X/737/2/103}, 737, 103

\bibitem[\protect\citeauthoryear{Schlegel, Finkbeiner  \& Davis}{Schlegel
  et~al.}{1998}]{Schlegel:1997yv}
Schlegel D.~J.,  Finkbeiner D.~P.,   Davis M.,  1998, \mn@doi [Astrophys. J.]
  {10.1086/305772}, 500, 525

\bibitem[\protect\citeauthoryear{Slosar, Seljak  \& Makarov}{Slosar
  et~al.}{2004}]{Slosar:2004fr}
Slosar A.,  Seljak U.,   Makarov A.,  2004, \mn@doi [Phys. Rev.]
  {10.1103/PhysRevD.69.123003}, D69, 123003

\bibitem[\protect\citeauthoryear{Slosar, Hirata, Seljak, Ho  \&
  Padmanabhan}{Slosar et~al.}{2008}]{Slosar:2008hx}
Slosar A.,  Hirata C.,  Seljak U.,  Ho S.,   Padmanabhan N.,  2008, \mn@doi
  [JCAP] {10.1088/1475-7516/2008/08/031}, 0808, 031

\bibitem[\protect\citeauthoryear{Smee et~al.}{Smee et~al.}{2013}]{Smee:2012wd}
Smee S.,  et~al., 2013, \mn@doi [Astron. J.] {10.1088/0004-6256/146/2/32}, 146,
  32

\bibitem[\protect\citeauthoryear{Stoughton et~al.}{Stoughton
  et~al.}{2002}]{Stoughton:2002ae}
Stoughton C.,  et~al., 2002, \mn@doi [Astron. J.] {10.1086/324741}, 123, 485

\bibitem[\protect\citeauthoryear{Tegmark}{Tegmark}{1997}]{Tegmark:1996qtQML}
Tegmark M.,  1997, \mn@doi [Phys. Rev.] {10.1103/PhysRevD.55.5895}, D55, 5895

\bibitem[\protect\citeauthoryear{Zhao, Kitaura, Chuang, Prada, Yepes  \&
  Tao}{Zhao et~al.}{2015}]{Zhao:2015jga}
Zhao C.,  Kitaura F.-S.,  Chuang C.-H.,  Prada F.,  Yepes G.,   Tao C.,  2015,
  \mn@doi [Mon. Not. Roy. Astron. Soc.] {10.1093/mnras/stv1262}, 451, 4266

\makeatother
\end{thebibliography}



\appendix

\section{Maps}
\label{sec:maps}

\begin{figure}
 \centering
\includegraphics[width=\columnwidth]{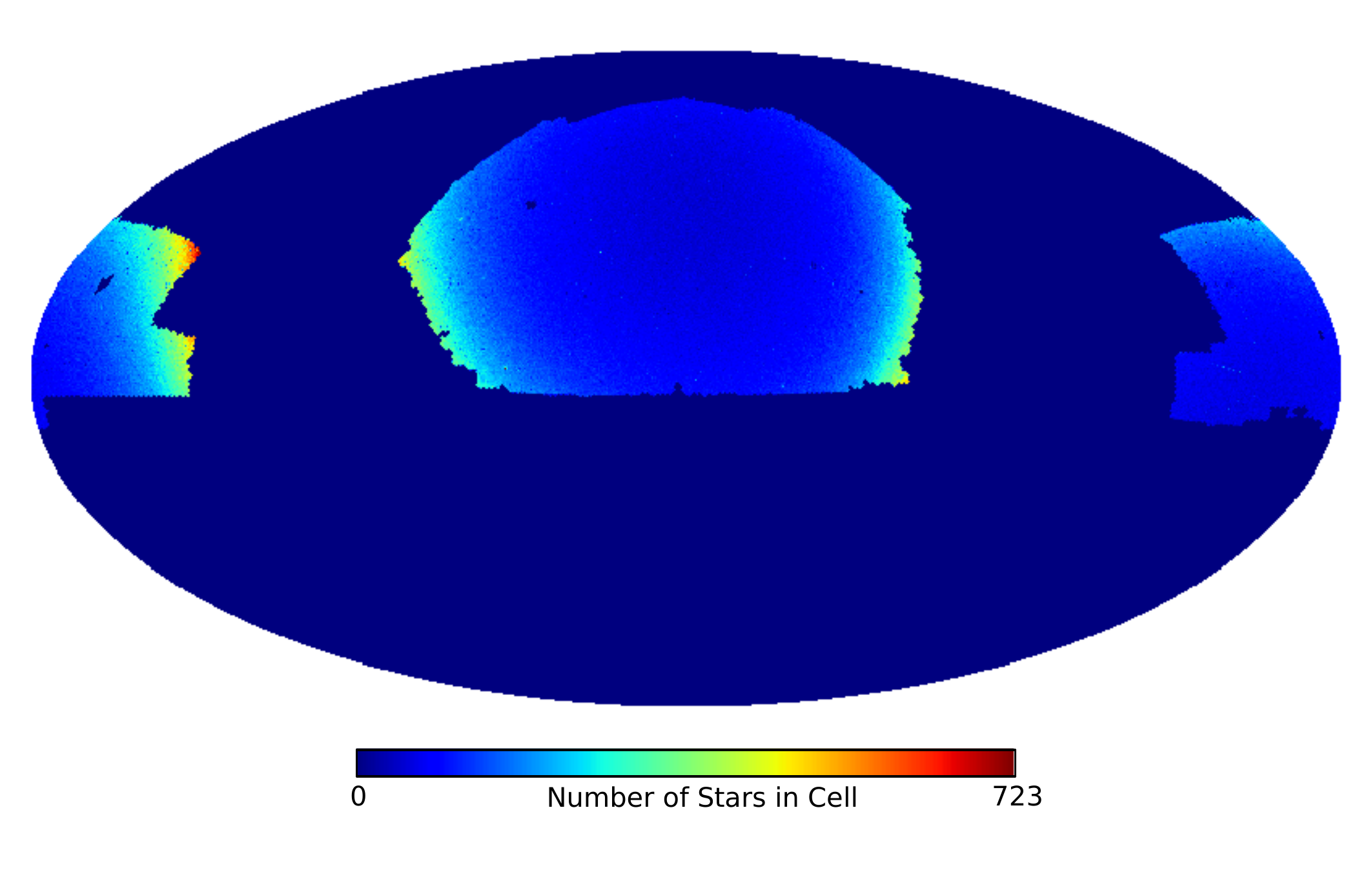}
\caption{The distribution of stars in the 8$^\mathrm{th}$ SDSS data 
release in HEALPix. The map is presented in Mollweide projection, 
equatorial coordinates, astronomical orientation, i.e. east is 
left, and it has been rotated by $180^\circ$ 
to show the NGC in the centre. The resolution is 
$N_\mathrm{side}=256$. The
catalogue includes stars in areas that were not targeted 
by BOSS. These are masked out 
in the relevant cells.
\label{fig:starsNSIDE256}}
 \centering
 \includegraphics[width=\linewidth]{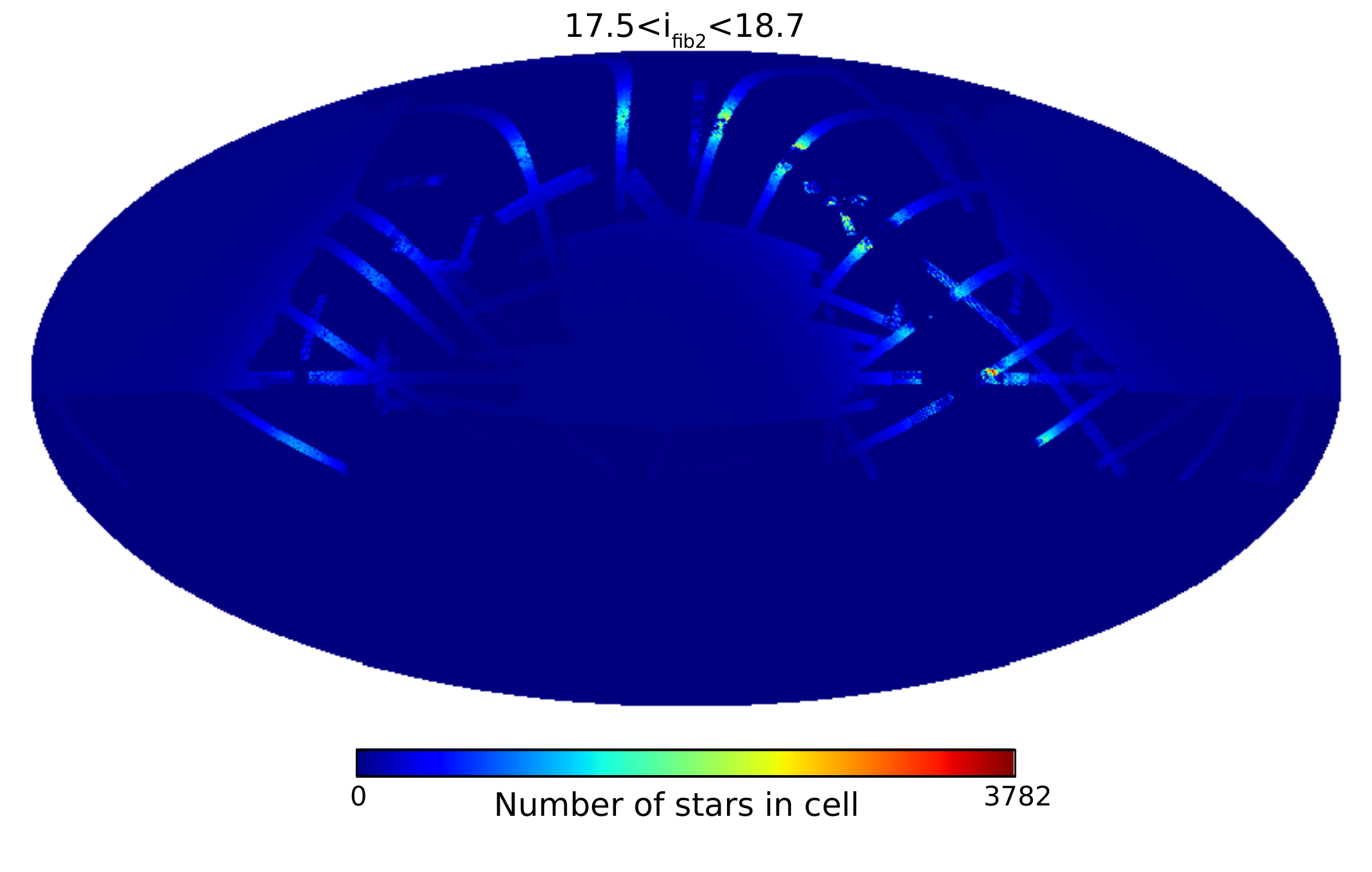}
 \includegraphics[width=\linewidth]{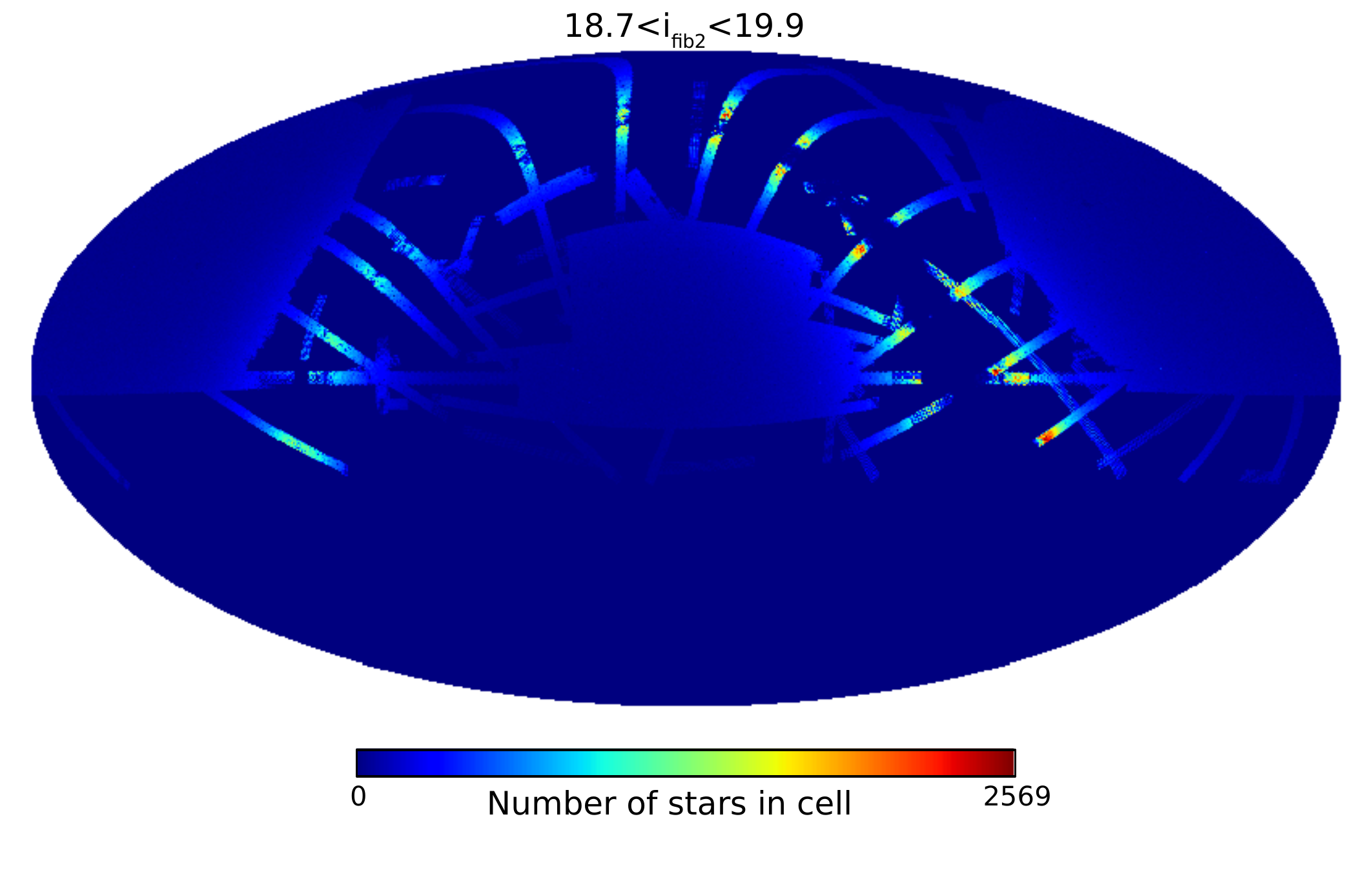}
 \caption{Maps of two sub-samples of the SDSS DR8 star 
 	catalogue. The upper panel shows the distribution of
    bright stars with $17.5<i<18.7$ and the
    lower one faint stars with $18.7<i<19.9$.
    The plot is in Mollweide projection and in equatorial
    coordinates with astronomical orientation.}
 \label{fig:starmagsplit2}
\end{figure}

\begin{figure}
 \centering
 \includegraphics[width=0.49\linewidth]{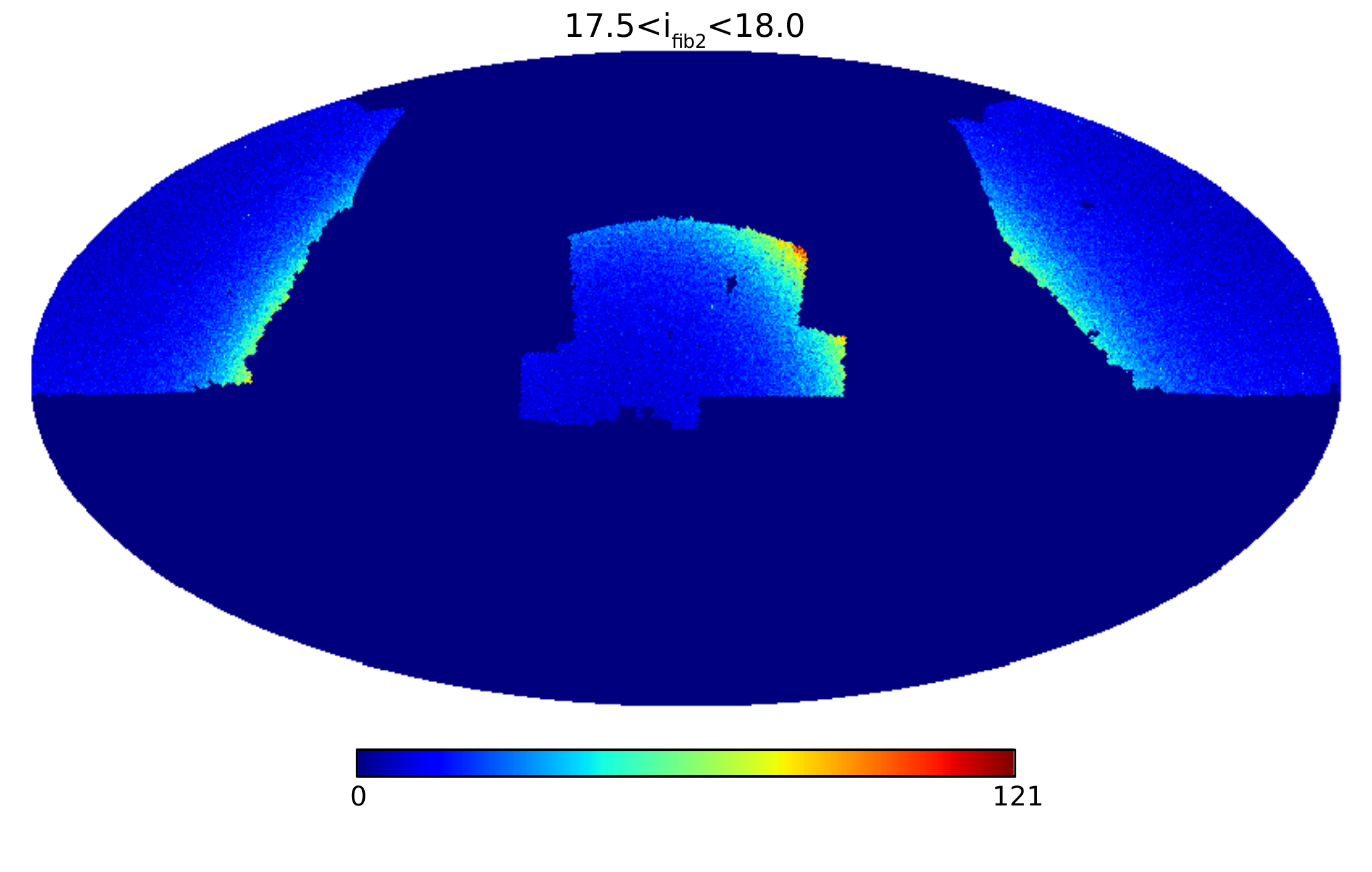}
 \includegraphics[width=0.49\linewidth]{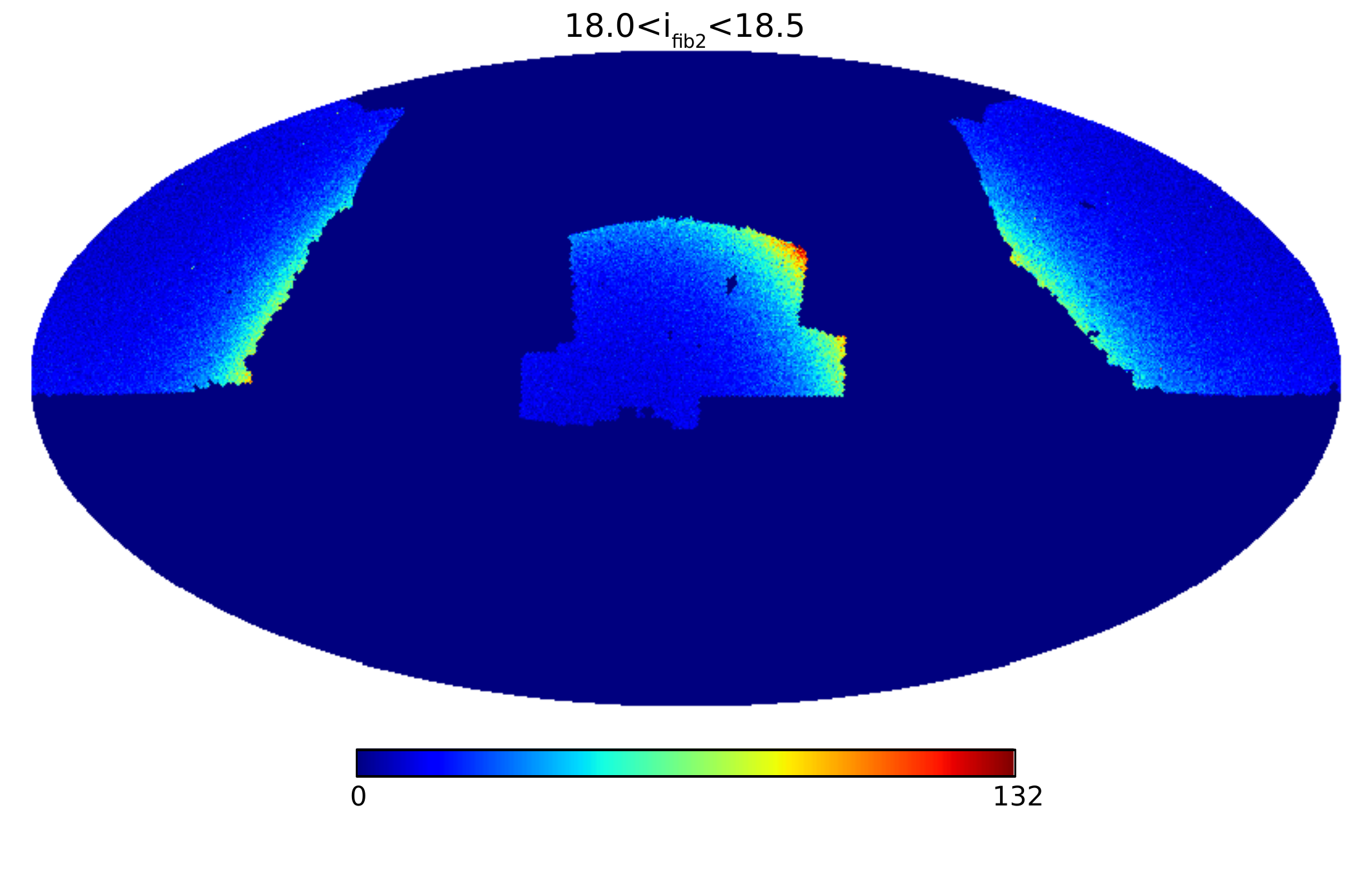}
 \includegraphics[width=0.49\linewidth]{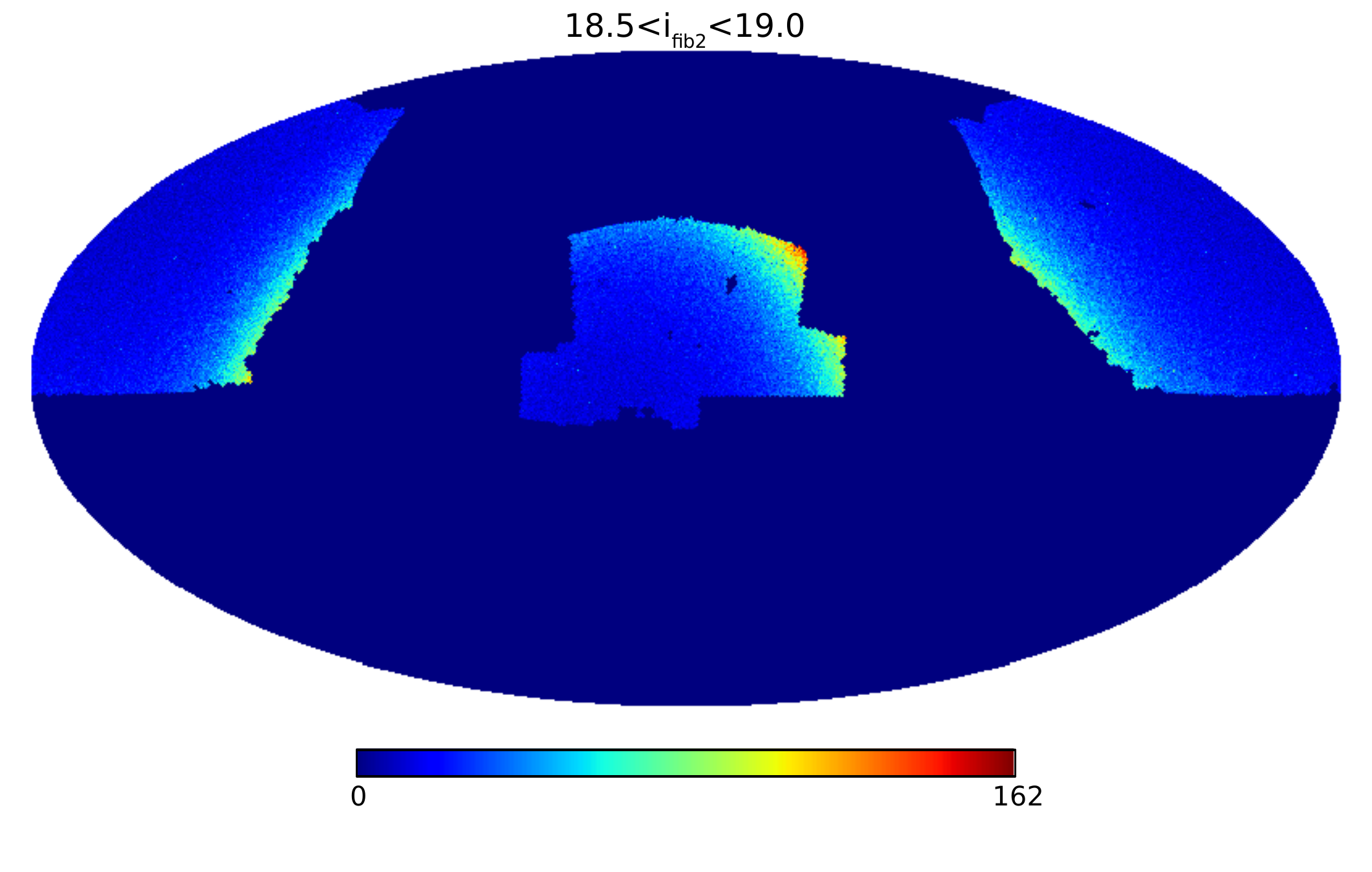}
 \includegraphics[width=0.49\linewidth]{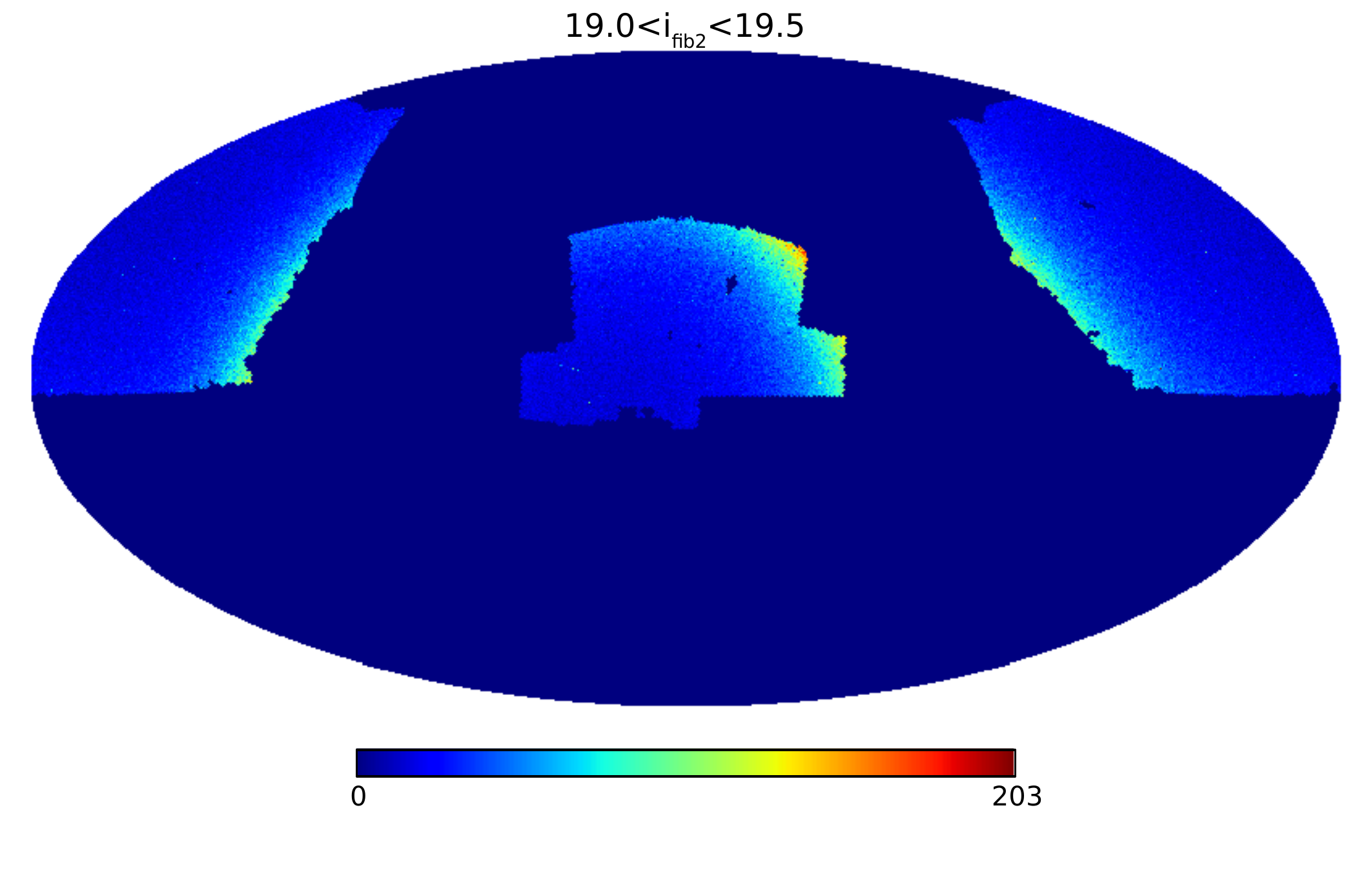}
 \includegraphics[width=0.49\linewidth]{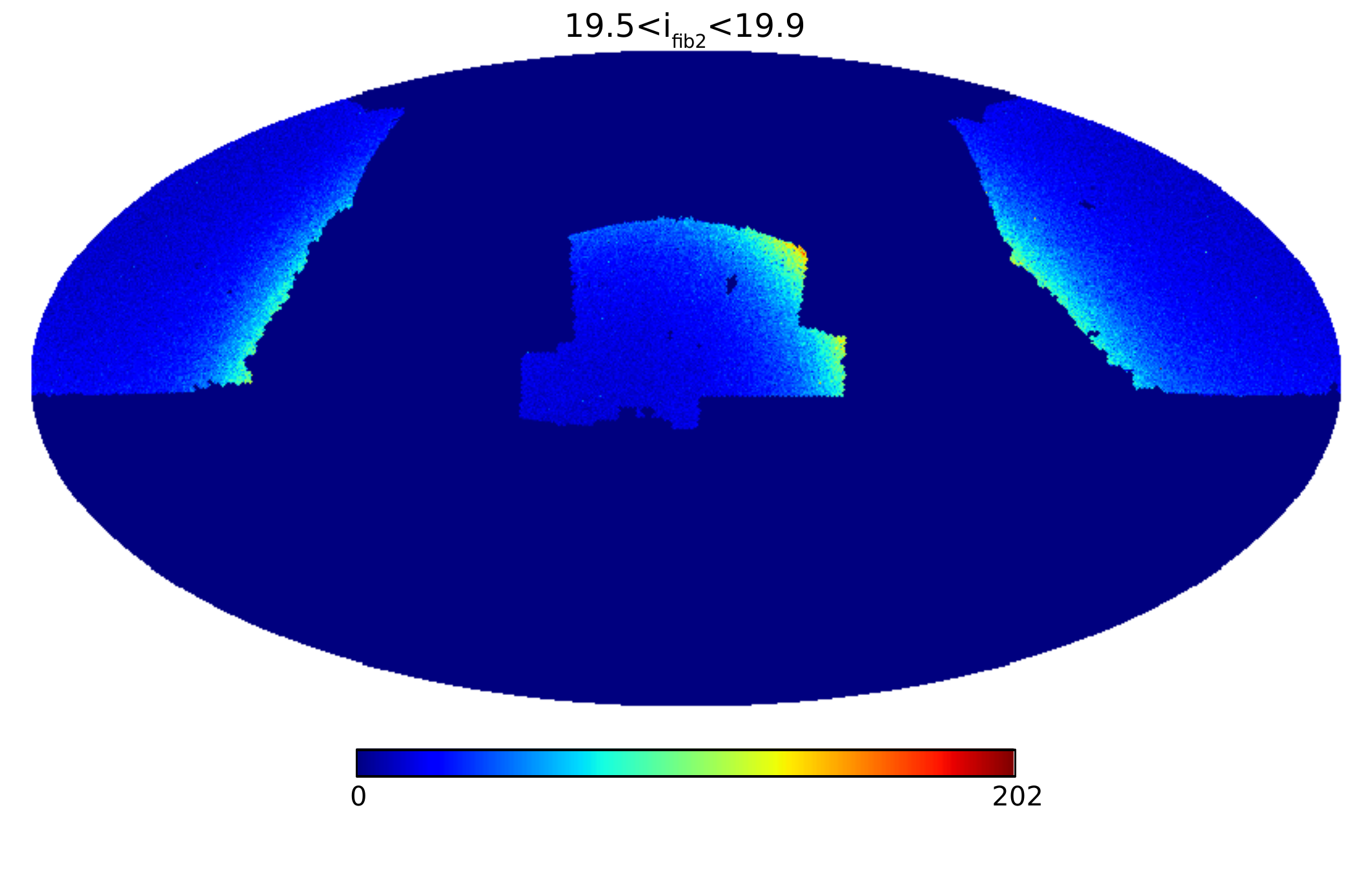}
 \caption{Maps of five sub-samples of the SDSS DR8 star 
 	catalogue. The panels show the distribution of
    stars with $17.5<i<18.0$ (top left), $18.0<i<18.5$ (top right), $18.5<i<19.0$ (centre left), $19.0<i<19.5$ (centre right), and $19.5<i<19.9$ (bottom).
    The plot is in Mollweide projection and in equatorial
    coordinates with astronomical orientation.}
 \label{fig:starmagsplit5}
\end{figure}

\begin{figure}
 \centering
 \includegraphics[width=\linewidth]{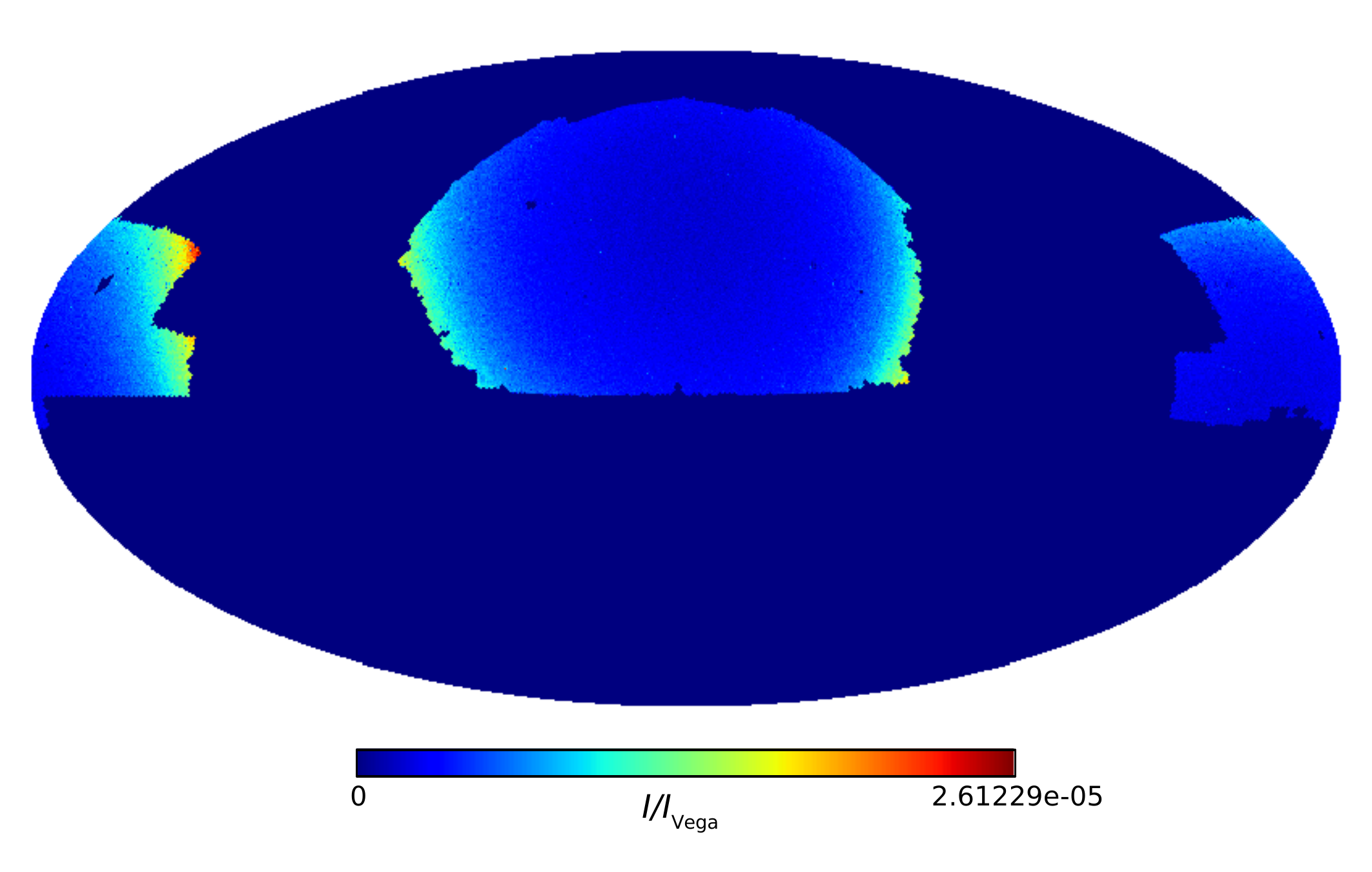}
 \caption{Map of the brightness distribution
 	of the SDSS DR8 star 
 	catalogue. 
    The plot is in Mollweide projection and in equatorial
    coordinates with astronomical orientation. The map is 
    rotated by $180^\circ$ to feature the NGC in the centre. The brightness is given
    in units of the brightness of the star 
    Vega.}
 \label{fig:intmap}
\end{figure}

\begin{figure}
 \centering
\includegraphics[width=\columnwidth]{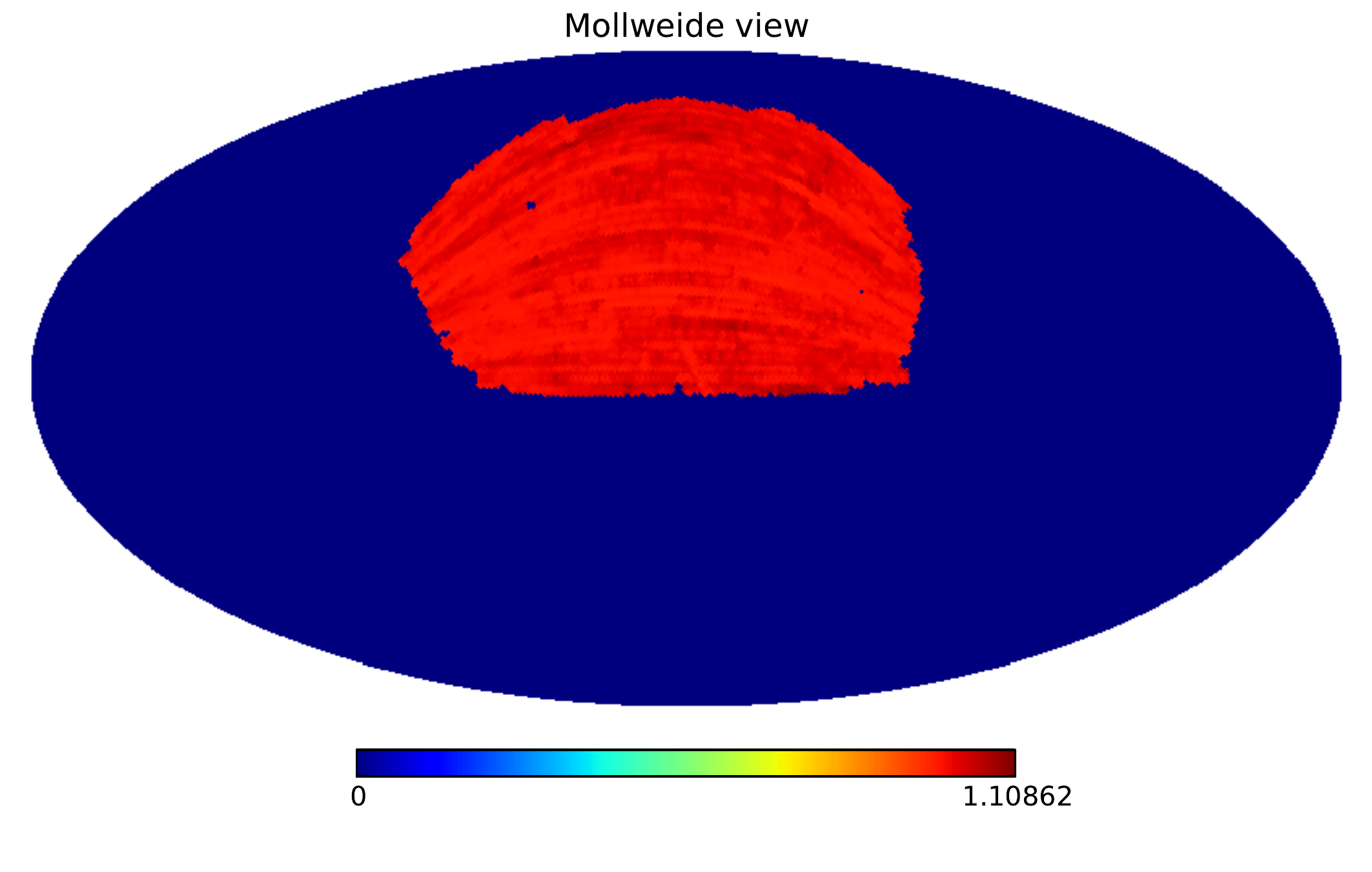}
\caption{The seeing condition weights of BOSS DR12 CMASS NGC in HEALPix. The map is presented in 
Mollweide projection, 
equatorial coordinates and astronomical orientation, 
but it is rotated by $180^\circ$ such that the
region observed is in the centre of the map.
\label{fig:Seeingmap}}
 \centering
\includegraphics[width=\columnwidth]{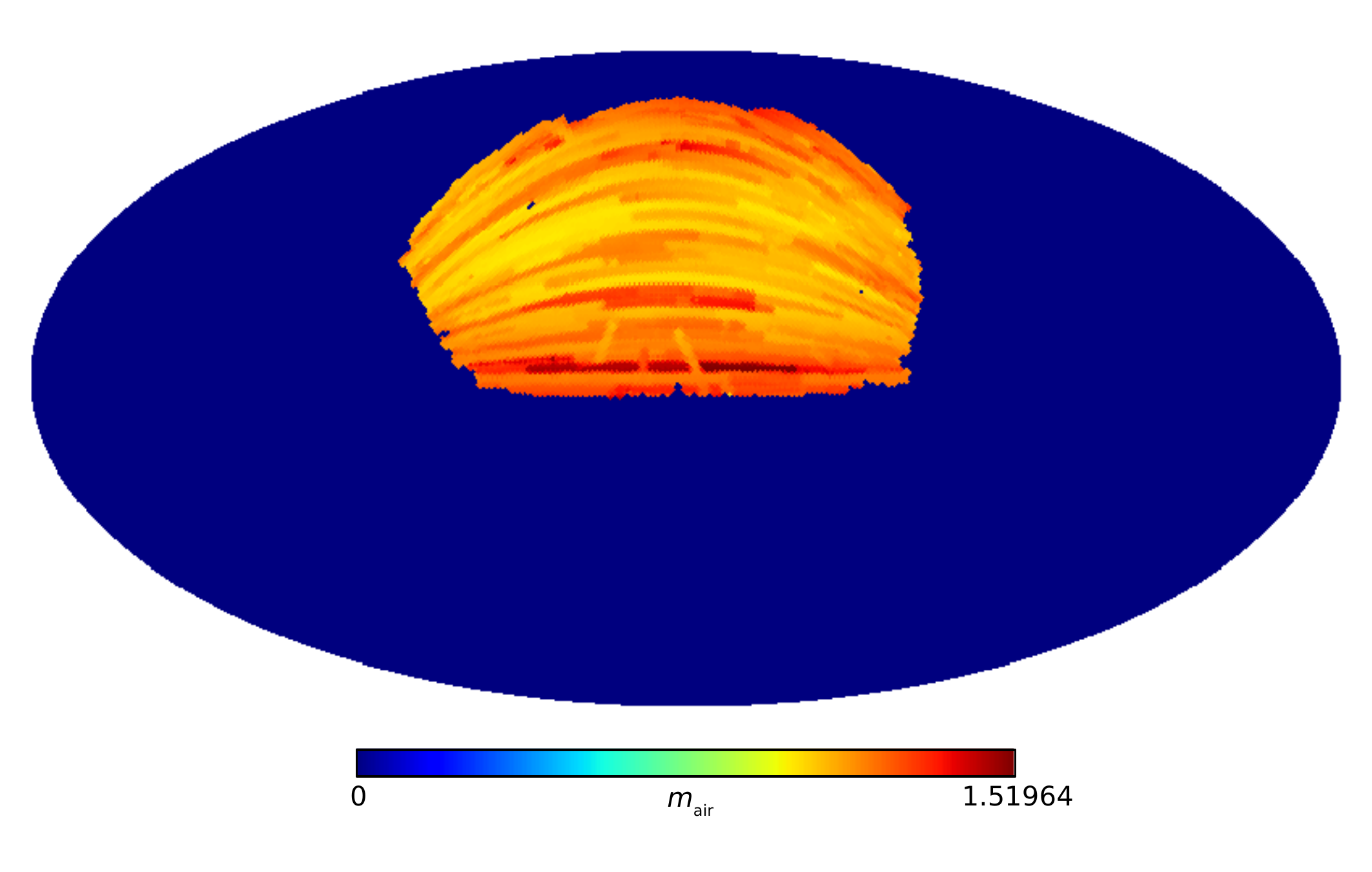}
\caption{The airmass $m_\mathrm{air}$ (cf.
	Eq. \ref{eq:airmass}) in the NGC sub-sample of BOSS 
DR12 CMASS in HEALPix. The map is presented in 
Mollweide projection, 
equatorial coordinates and astronomical orientation, 
but it is rotated by $180^\circ$ such that the
region observed is in the centre of the map.
\label{fig:Airmassmap}}
 \centering
\includegraphics[width=\columnwidth]{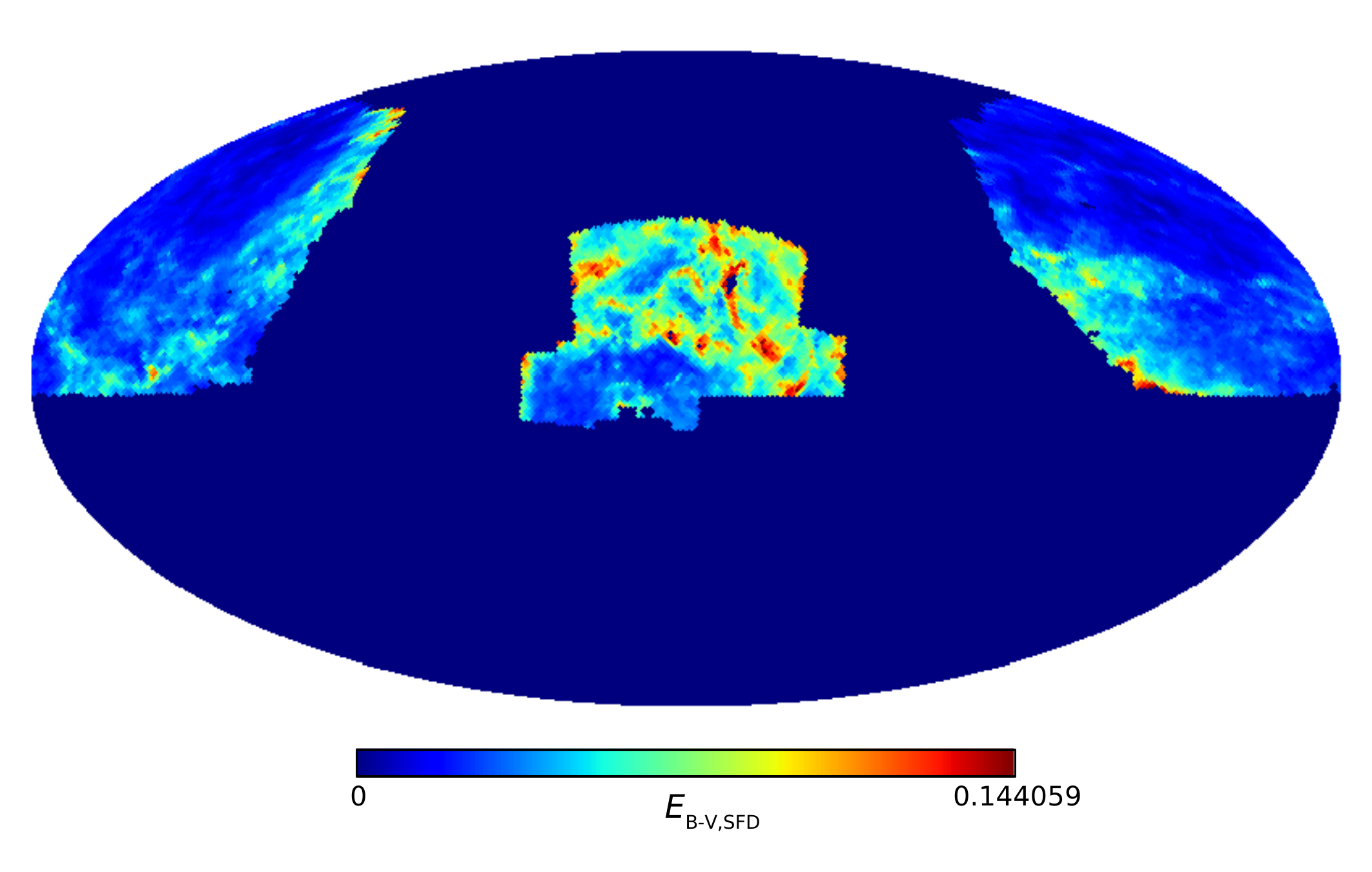}
\caption{The values of $E_\mathrm{E-V,SFD}$
used to correct for extinction in the BOSS 
targetting in HEALPix. The map is presented in 
Mollweide projection, 
equatorial coordinates and astronomical orientation.
\label{fig:Extinctionmap}}
\end{figure}

\begin{figure}
	\centering
    \includegraphics[width=\columnwidth]{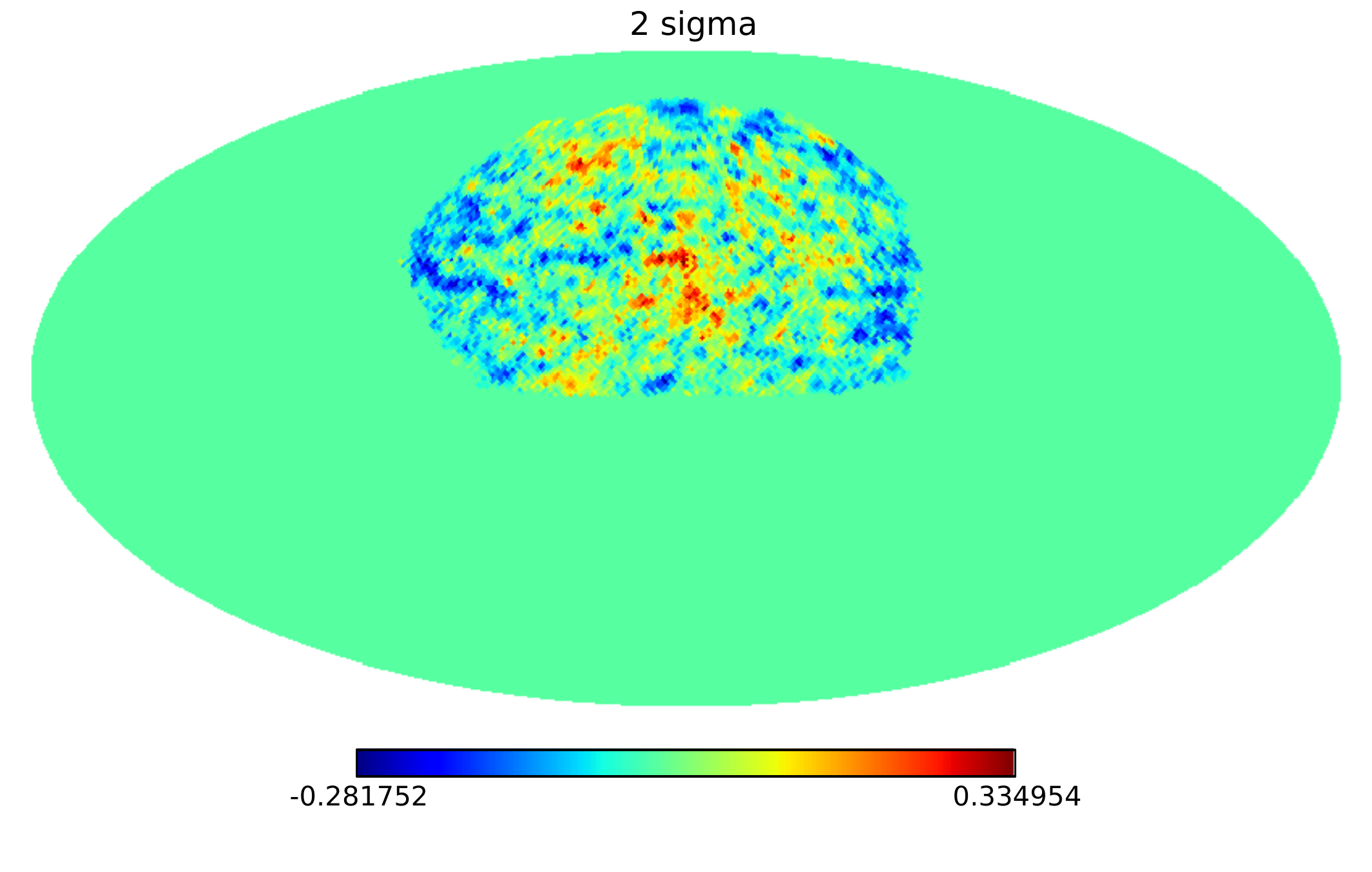}
    \includegraphics[width=\columnwidth]{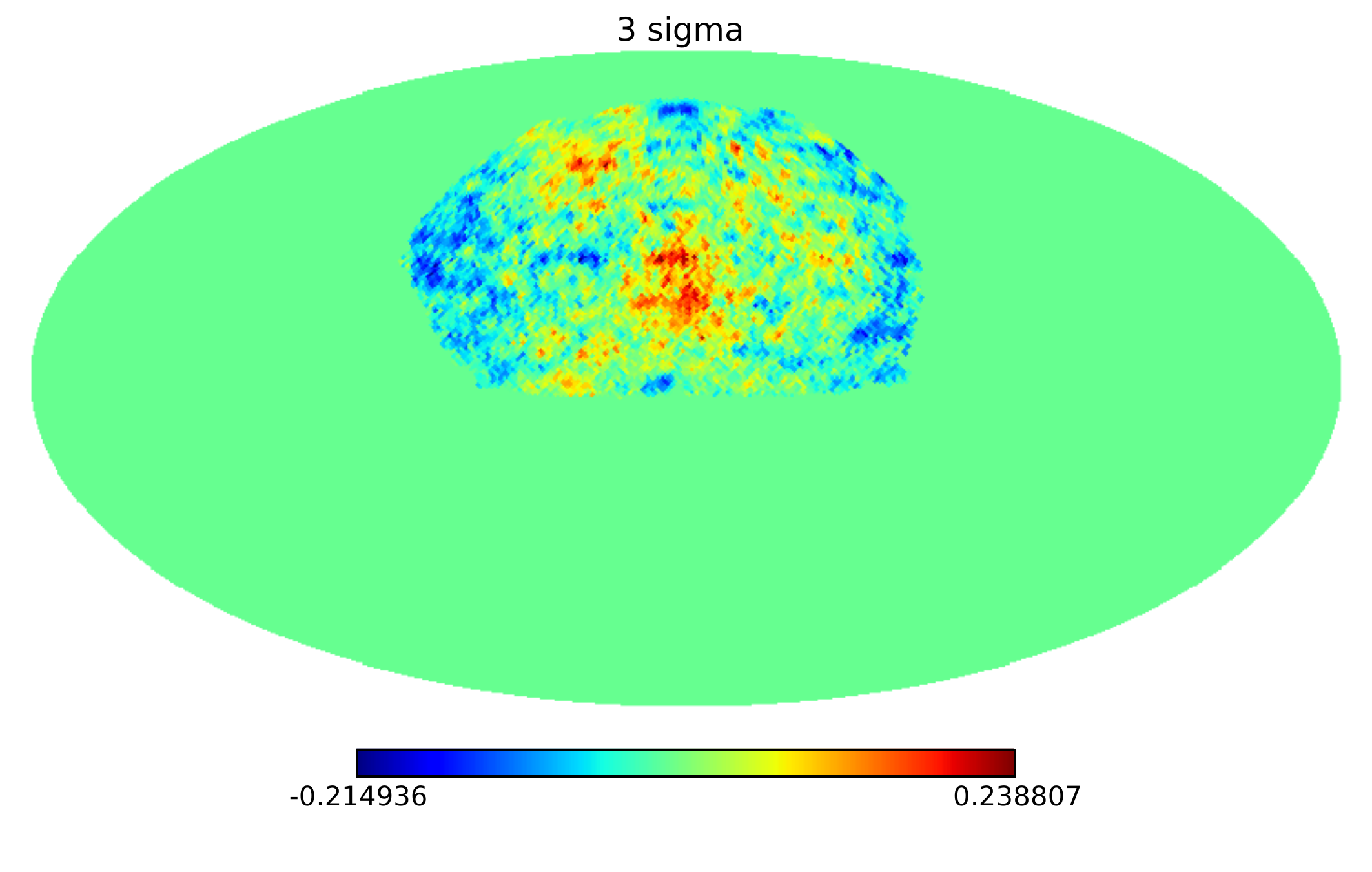}
    \includegraphics[width=\columnwidth]{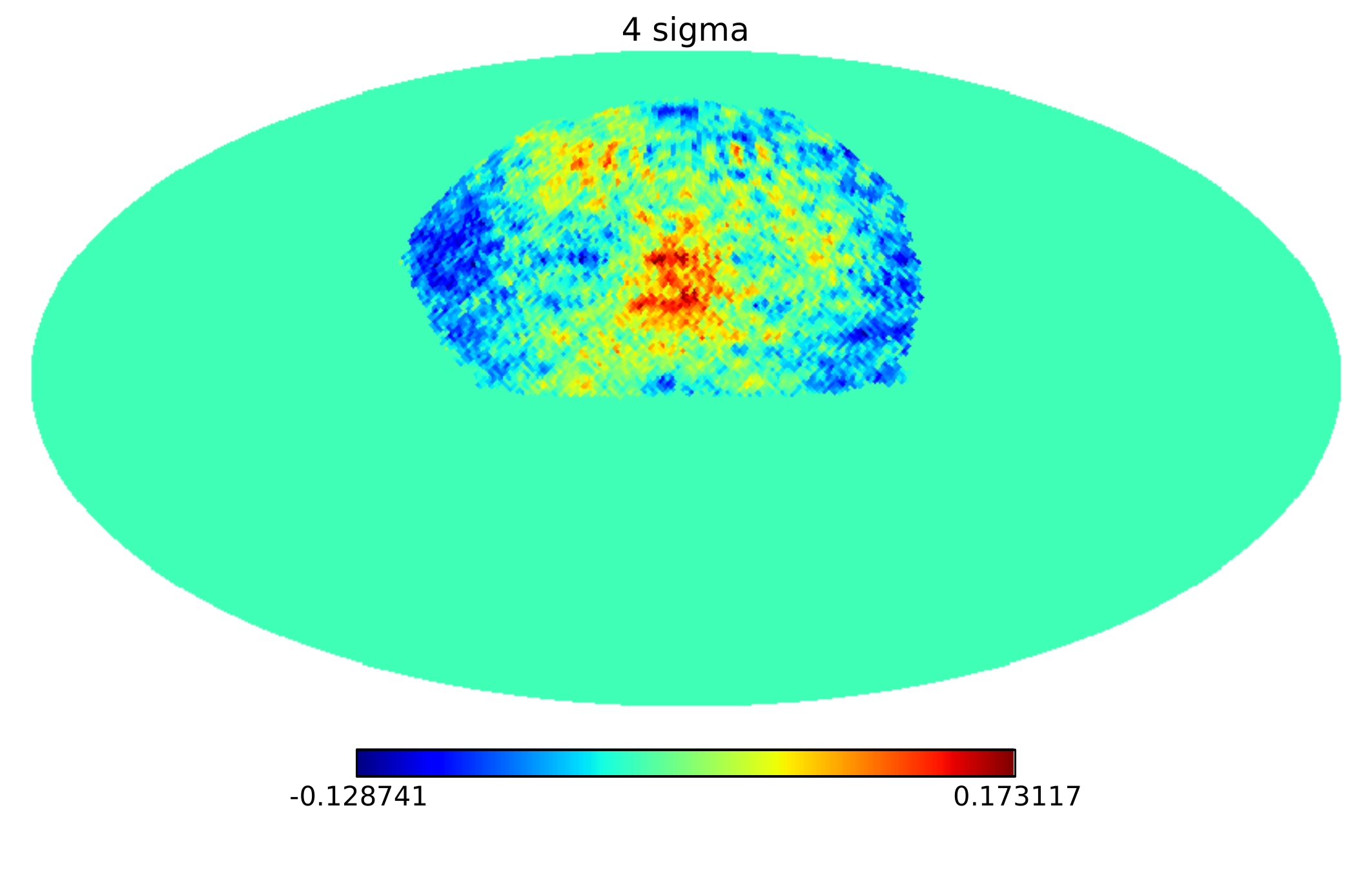}
    \caption{Phenomenological ``contaminant'' maps for
    a $2\sigma$ (top), $3\sigma$ (centre) and
	$4\sigma$ (bottom) threshold as
    defined by Eq. \eqref{eq:blobmap} presented in
    Mollweide projection, equatorial coordinates and
    astronomical orientation. The maps have been rotated
    to centre the BOSS NGC footprint. We masked out 
    regions outside of the BOSS NGC footprint since they
    are unphysical and do not contribute to the 
    templates as the number density of the randoms is
    zero.}
    \label{fig:blobmaps}
\end{figure}



\bsp	
\label{lastpage}
\end{document}